\documentclass[longauth]{aa} % for the long lists of affiliations 
\usepackage{graphicx}
\usepackage{txfonts}
\usepackage{enumerate}
\usepackage[english]{babel}
\usepackage[T1]{fontenc}
\usepackage{amsmath}
\usepackage{natbib}
\usepackage[pagebackref,
                 colorlinks = true,  
                 linkcolor = magenta,
                 urlcolor  = magenta,
                 citecolor = blue,
                 anchorcolor = green,
                 hyperindex = true,
                 hyperfigures]{hyperref}

\begin{document}

   \title{Expected signatures from hadronic emission processes\\ in the TeV spectra of BL Lac objects}
%   \subtitle{Hadronic signatures}

   \author{A. Zech\inst{1}
          \and
          M. Cerruti\inst{2}
          \and
          D. Mazin\inst{3}
          }

   \institute{LUTH, Observatoire de Paris, CNRS, Universit\'e Paris Diderot, PSL Research University, 5 Place Jules Janssen, 92190 Meudon, France\\
              \email{andreas.zech@obspm.fr}
         \and
             LPNHE, Universit\'e Pierre et Marie Curie Paris 6, Universit\'e Denis Diderot Paris 7, CNRS/IN2P3, 4 Place Jussieu, F-75252, Paris Cedex 5, France\\
             \email{mcerruti@lpnhe.in2p3.fr}
             \and
             Institute for Cosmic Ray Research, University of Tokyo, Japan
%             \email{mazin@icrr.u-tokyo.ac.jp}
             }

   \date{Received ...; accepted ...}

% \abstract{}{}{}{}{} 
% 5 {} token are mandatory
 
  \abstract
  % context heading (optional)
  % {} leave it empty if necessary  
   {The wealth of recent data from imaging air Cherenkov telescopes (IACTs), ultra-high energy cosmic-ray experiments 
   and neutrino telescopes have fuelled a renewed interest in hadronic emission models for $\gamma$-loud blazars.}
  % aims heading (mandatory)
   {We explore physically plausible solutions for a lepto-hadronic interpretation of the stationary emission
   from high-frequency peaked BL Lac objects (HBLs). The modelled spectral energy distributions are then searched
   for specific signatures at very high energies that could help to distinguish the hadronic origin of the emission from a standard
   leptonic scenario.}
  % methods heading (mandatory)
   {By introducing a few basic constraints on parameters of the model, such as assuming the co-acceleration of electrons and protons, we 
   significantly reduced the number of free parameters. We then systematically explored the parameter space of the size of the emission region and its magnetic field for
   two bright $\gamma$-loud HBLs, PKS\,2155-304 and Mrk\,421.
   For all solutions close to equipartition between the energy densities of protons and of the 
   magnetic field, and with acceptable jet power and light-crossing timescales, we inspected the spectral hardening in the multi-TeV domain from proton-photon induced cascades and muon-synchrotron emission inside the source.
   Very-high-energy spectra simulated with the available instrument functions from the future Cherenkov Telescope Array (CTA) were evaluated for detectable features
   as a function of exposure time, source redshift, and flux level.}
  % results heading (mandatory)
    {A range of hadronic scenarios are found to provide satisfactory solutions for the broad band emission of the sources under study. The TeV spectrum can be dominated 
    either by proton-synchrotron emission or by muon-synchrotron emission. The solutions for HBLs cover a parameter space that is distinct from the one found for the most 
    extreme BL Lac objects in an earlier study. Over a large range of model parameters, the spectral hardening due to internal synchrotron-pair cascades, the ``cascade bump'', 
    should be detectable for acceptable exposure times with the future CTA for a few nearby and bright HBLs.}
  % conclusions heading (optional), leave it empty if necessary 
   {}

   \keywords{astroparticle physics; radiative transfer; radiation mechanisms: non-thermal; BL Lacertae objects: individual: PKS 2155-304, Mrk 421; gamma rays: galaxies}

   \maketitle
%
% ________________________________________________________________
\section{Introduction}

The still-open question on the origin of ultra-high-energy cosmic rays (UHECRs) and astrophysical neutrinos on the one hand, and the wealth of available data from 
$\gamma$-ray emitting blazars on the other \citep[from MeV to TeV energies, see e.g.][]{Ackermann11b,Senturk2013}, has led to renewed interest in hadronic emission models for 
those sources. Radiative emission in the most common scenarios either comes from a region inside the relativistic jet \citep[e.g.][]{Mannheim93,Dermer01,Mucke01,Dimitrakoudis2012,Bosch2012,Boettcher2013,Mastichiadis2013,Petropoulou2015,Diltz2015} or from interactions of escaping hadrons along the path 
from the source to Earth \citep[e.g.][]{Essey2010,Essey2011,Dermer12,Murase2012,Tavecchio2014a}. Contrary to the more commonly assumed leptonic scenarios, 
in which the two characteristic broad bumps of the non-thermal spectral energy distribution (SED) of blazars are described with electron-synchrotron and Inverse Compton emission 
\citep{Konigl1981,Sikora1994}, hadronic scenarios introduce relativistic protons to explain the high-energy bump that is generally seen from keV to GeV energies in flat-spectrum 
radio quasars (FSRQs) and in the MeV to TeV range for BL Lac objects. In the hadronic framework, this high-energy component can be attributed to either proton-synchrotron emission 
or radiation from secondary products of proton-photon or proton-proton interactions. These kinds of scenarios thus admit the possibility of a direct link between UHECRs and electromagnetic emission 
from blazars. They also lead necessarily to the emission of high-energy neutrinos from the decay of proton-induced pions and muons. This suggests that blazars, or more generally radio-loud
active galactic nuclei (AGNs), of which blazars are a sub-class of objects with their jets assumed to be closely aligned to the line of sight, are potential sources of the PeV neutrinos recently 
detected with IceCube \citep{Aartsen13Ta}.

Even though hadronic emission models in general require higher jet powers and face more difficulties to account for short-term variability than leptonic models, they still present a viable
and very intriguing alternative within the available constraints from current observational data. Future instruments, such as the Cherenkov Telescope Array (CTA)~\citep{Actis2011,Acharya13}, will be 
able to probe blazar spectra above a few tens of GeV and cover the whole very-high-energy range (VHE, energies above 100\,GeV) to above 100\,TeV, with much better sensitivity 
and spectral resolution than current Imaging Air Cherenkov telescopes (IACTs). This motivates a search for potential signatures in the VHE $\gamma$-ray spectrum that would help distinguish hadronic scenarios from the 
simpler leptonic models.

\defcitealias{Cerruti2015}{C15}\citet[][hereafter \citetalias{Cerruti2015}]{Cerruti2015} have recently characterised the SEDs of a distinct class of so-called ultra-high-frequency peaked 
BL Lac objects (UHBLs) with a stationary one-zone model that provides a complete treatment of all relevant emission processes for relativistic electrons and protons. This model, which 
will also be used for the current study, allows us to treat leptonic, hadronic and mixed scenarios with a single code.
The authors have shown that hadronic and mixed lepto-hadronic scenarios provide an interesting alternative for the interpretation of UHBL SEDs, compared to purely leptonic synchrotron 
self-Compton (SSC) models. The latter are found to require extreme parameters for such sources.
Two distinct regions were identified in the parameter space spanned by the source extension and magnetic field strength, leading to 
interpretations of the high-energy bump either as proton-synchrotron dominated (``hadronic'' solution; for high magnetic fields of the order of a few 10\,G) or as consisting of
a combination of SSC radiation and emission from synchrotron-pair cascades triggered by proton-photon interactions and the subsequent decay of the generated pions and other mesons 
(mixed ``lepto-hadronic'' solution; for magnetic fields of the order of a few 0.1\,G). In each of these parameter regions, solutions were found with jet powers below the Eddington luminosity, 
distinguishing these objects from the more luminuous blazar classes studied by~\citet{Zdziarski2015}, which require very high jet powers.

Although UHBLs are located at the most extreme end of the ``blazar sequence'' \citep{Fossati1998}, high-frequency peaked BL Lac objects (HBLs) with a high-energy peak located in the 
0.1 to 1\,TeV range are far more numerous in the current sample of blazars detected with IACTs. 
They clearly outnumber all other types of blazars detected at these energies\footnote{For an up-to-date catalogue of TeV sources, see \url{http://tevcat.uchicago.edu}}. 
Hadronic solutions are routinely proposed as alternatives to the standard SSC scenario for these kinds of  sources, but a systematic exploration of \mbox{(lepto-)}hadronic solutions 
for a given dataset has not yet been attempted, to the best of our knowledge. A more general study of the impact of the magnetic field strength and density of the target photon field of
the proton-synchrotron model is presented by \citet{Muecke2003}. In most of the current literature, very few a priori physical constraints are imposed and only exemplary solutions 
are presented. 

After a short description of the physical constraints we impose on our model to reduce the number of free parameters (Sect.~\ref{sec:model}), 
we explore the characteristics of different hadronic solutions for HBLs with respect to the source extension, magnetic field strength, jet
power, and equipartition between the magnetic and kinetic energy density in Sect.~\ref{sec:scenarios}. Then we apply the model to SEDs of the two HBLs \object{PKS\,2155-304} 
and \object{Mrk\,421} from low flux states in Sect.~\ref{sec:application}. We describe the different sets of solutions, and briefly discuss the jet power, deviation from equipartition and variability 
timescales they imply. In Sect.~\ref{sec:UHBL}, we compare these solutions to a previous study of UHBLs. We then search the modelled SEDs for signatures of
spectral hardening in the VHE range, which are caused by the emission from synchrotron-pair cascades and from muon-synchrotron radiation, and compare these to the expected 
sensitivity of CTA in Sect.~\ref{sec:signatures}. The relation of the detectability of such features to the flux level and redshift are explored.
Finally, we present a critical discussion of these results, their limitations, and implications for UHECR and astro-neutrino searches in Sect.~\ref{sec:discussion}.

Throughout this work, the absorption on the extragalactic background light (EBL) is computed using the model by \citet{Franceschini2008}. In Sect.~\ref{sec:discussion}, we discuss 
how this particular choice affects our results. For the transformations between the two nearby blazars under study and the observer on Earth, we have adopted a standard cosmology 
with $\Omega_m = 0.3$, $\Omega_{\Lambda} = 0.7,$ and $H_0 = 68$\,km\,s$^{-1}$\,Mpc$^{-1}$. 
 
%__________________________________________________________________

\section{Physical constraints of model parameters}
\label{sec:model}

LEHA \citepalias[see][]{Cerruti2015} is a code that was developed recently to simulate the stationary emission from BL Lac objects for leptonic, hadronic, and mixed scenarios. 
In this ``blob in jet'' code, relativistic populations of electrons and protons are confined within a plasma blob of radius $R$ with a tangled magnetic field of strength $B$ and
a bulk Lorentz factor $\Gamma$. The primary particle distributions follow power laws with self-consistent synchrotron cooling breaks.

In addition to synchrotron emission from primary protons and electrons, and SSC emission, the code also treats proton-photon 
interactions with meson-production and subsequent decay by use of the SOPHIA Monte Carlo package \citep{Muecke2000}. Secondary particles from such interactions trigger 
pair-synchrotron cascades that are followed for several generations. In addition, the muon-synchrotron spectrum is extracted following slight modifications of the SOPHIA code. 
In the LEHA code, Bethe-Heitler and photon-photon pair production are calculated and the synchrotron-pair cascades they trigger are evaluated as well. Only proton-proton interactions 
are not considered, because to be efficient this kind of a process would require very high target proton densities inside the jet in our framework.

In our application to high-frequency peaked BL Lac objects, the code provides a simple description for a continuous plasma flow through a stationary or slowly-moving acceleration region inside the jet, without specification of the acceleration mechanism. A flow of accelerated particles is continuously injected into a stationary or slowly moving radiation zone and then continues down the jet with bulk Lorentz factor $\Gamma$, while expanding adiabatically. This picture is motivated by the fact that very long baseline interferometry (VLBI) observations of a large sample of blazars show that 
although luminous blazar types often exhibit rapidly-moving radio knots, in HBLs they tend to be stationary \citep[e.g.][]{Hervet2016}.
The homogeneous, spherical emission zone of constant radius that we are modelling represents the radiation region into which particles are injected. Radiative emission is strongly dominated 
by emission from this zone, because the plasma flow rapidly loses energy through adiabatic expansion farther down the jet. For example, when imposing that the magnetic flux be conserved, the magnetic field strength scales as $B \propto R^{-2}$, implying an energy output through synchrotron emission that is 
proportional to $R^{-4}$. The decreasing particle density (proportional to $R^{-3}$) together with the decreasing target photon density also leads to a rapid reduction in the proton-photon interaction rate. In addition, radiative losses will further reduce the emission level, especially at the highest energies where our study is focused. Therefore only the single acceleration and emission zone is modelled and emission from the plasma flow farther down the jet is neglected. 

To reduce the number of free parameters and to focus on physically plausible solutions from the onset, we impose a number of constraints that are based on simplifying 
assumptions about the nature of the emission region and the acceleration process. These constraints correspond for the most part to those applied by \citetalias{Cerruti2015}:

\begin{itemize}

\item Leptons and protons are supposed to be co-accelerated and to follow power laws with the same intrinsic index $n_1$, before cooling. This index 
is assumed to be close to two, consistent with Fermi-like acceleration mechanisms. However, no specific assumptions are made about the underlying
mechanism.\newline

\item To account for synchrotron cooling of the primary particles, the stationary spectra of primary electrons and protons, which are used as an input to the code, are 
characterised by broken power laws with exponential cut-offs \citepalias[see][Eq. 1]{Cerruti2015} . The Lorentz factors corresponding to the break energies $\gamma_{(e;p) , break}$ 
are determined from a comparison of radiative and adiabatic cooling timescales. The second spectral slope above these breaks is given by $n_2 = n_1 + 1$. 
For the application to HBLs, the electron population is generally completely cooled,  whereas the proton spectrum is not impacted. 
We verify that for all the solutions presented here, proton energy loss is indeed always dominated by adiabatic cooling and thus there is no cooling 
break in the proton spectra.
Comparisons of the relevant acceleration and cooling timescales for a few exemplary models are shown in Sects.~\ref{app:2155} and~\ref{app:421} and are discussed below.   \newline

\item The minimum proton Lorentz factor $\gamma_{p,min}$ is not constrained and is set to one, whereas for the electron spectrum we generally need to adjust 
the minimum Lorentz factor $\gamma_{e,min}$ to match constraints from low-energy optical and radio data. This choice for $\gamma_{p;min}$ maximises the 
proton energy density $u_p$, which represents a major contribution to the jet power. It is thus a conservative choice when considering the energetics of the models.\newline

\item The maximum proton Lorentz factor $\gamma_{p;max}$, is derived from the equilibrium between particle acceleration and energy loss, meaning radiative and adiabatic cooling, through 
a comparison of their characteristic timescales. The acceleration timescale as a function of the Lorentz factor is simply described by $\tau_{acc} (\gamma) = \gamma / \psi \cdot m c / (e B) $ 
with an assumed efficiency of 
$\psi = 0.1$. In the regime where adiabatic cooling dominates over radiative cooling, which is the regime of interest for HBLs in our framework, the maximum proton Lorentz factor is 
\begin{equation}
\gamma_{p;max} \propto B \cdot R,
\label{equ:gpmax}
\end{equation} 
as can be seen from Eq. 18 in \citetalias{Cerruti2015}. The same behaviour is found when simply applying the Hillas criterion \citep{Hillas1984} to constrain $\gamma_{p;max}$.

The maximum energy of the electron spectrum, defined by the Lorentz factor $\gamma_{e,max}$, is left as a free parameter that we constrain with X-ray data. For the solutions found here,
the ratio of $\gamma_{p,max}$ over $\gamma_{e,max}$ is consistent with the values one might expect in diffusive acceleration for a Kraichnan turbulence spectrum, but a specific investigation of the
acceleration mechanism is beyond the scope of the present work (cf.\ \citet{Biermann1987, Mucke01} and \citetalias{Cerruti2015} for more discussion).\newline

\item The bulk Doppler factor of the emission region is fixed at $\delta=$ 30, a typical value for bright VHE-detected BL Lac objects \citep{Tavecchio2010}. This  allows for the 
observed variability in those sources when using reasonable source parameters, except for the most extreme minute-scale flares \citep{Gaidos1996,Aharonian2007,Albert07b}, the origin of which are
not yet clearly understood. The impact on our study of the value of this parameter is discussed briefly in Sect.~\ref{sec:discussion}.\newline

\item We also assume that the viewing angle between the jet axis and the line of sight to the observer is very small, leading to a relation between bulk Doppler factor and bulk 
Lorentz factor of 
\begin{equation}
\delta \approx 2 \Gamma.
\label{equ:delta}
\end{equation}
This small-angle approximation seems justified when considering that we are investigating particularly bright sources in this study.

\end{itemize}

Given these constraints, we are left with: two quantities, $B$ and $R$, that we treat as free parameters; and five quantities, $n_1$, $\gamma_{e, min}$, $\gamma_{e, max}$, and the normalisation 
of the electron and proton spectra $K_{(e;p)}$, that are more or less constrained by the SED for a given choice of the free parameters. To find a complete set of solutions for a given source, we first 
varied $n_1$ for values close to two that provided a satisfactory representation of the optical and X-ray data. Then, for a small set of the selected $n_1$ values, we scanned $B$ and $R$ while adjusting 
the particle densities for electrons and protons with $K_{(e;p)}$, and the minimum and maximum electron energies to the SED, using the usual ``fit-by-eye'' method. 
The flux level of the electron- and proton-synchrotron peaks in the SED can be kept constant during these scans by varying the particle densities as
\begin{equation}
K_{e;p} \propto R^{-3} \cdot B^{-2} \cdot f(\gamma_{(e;p) , break}, \gamma_{(e;p) , max}, n_1),
\end{equation}
where the last term is a function of the maximum particle Lorentz factor, the Lorentz factors of the cooling breaks, and the spectral index of the particle distributions, which all influence the 
particle content of the source.

We also constrained the parameter space of models for a given SED by requiring that the size of the emission region and the overall jet power be physically acceptable. A characteristic size 
scale for AGNs is given by the Schwarzschild radius $R_S$ of the supermassive black hole. Although for blob-in-jet models one generally expects emission regions of a size at least an order of 
magnitude larger than the Schwarzschild radius, as a minimum requirement we discarded only those solutions with $R < R_S$. The maximum size of the emission region is  usually limited by comparing 
the light-crossing time to the observed variability timescale of the source, but for the persistent flux states we are interested in, the latter information is generally not available
or not very constraining. This will be discussed more in Sect.~\ref{sec:discussion}.

The jet power for a two-sided jet was estimated as 
\begin{equation}
L_{j} \approx 2 \pi R^2 \beta c \Gamma^2 (u_B + u_e + u_p) + 2 L_r,
\label{equ:Ljet}
\end{equation}
where $u_B$, $u_e$, $u_p$ are the co-moving stationary energy densities of the magnetic field, electrons and protons, respectively, and $L_r$ corresponds to the power in the radiation field 
\citep[e.g.][and references therein]{Dermer2014}. In hadronic scenarios, $u_e$ and $L_r$ are generally very small compared to the other components and thus we neglected them. 
Any additional component due to cold protons inside the jet were also neglected. As is usually the case in one-zone hadronic scenarios, the jet power was assumed to be largely dominated by the power 
from the ``blob''. An additional extensive jet could have been added, as is done for certain leptonic models \citep[see e.g.][]{Katarzynski2001, Abramowski12f, Hervet2015} if one wants to account for 
electron synchrotron emission in the radio band, but its contribution to the total jet power would be much smaller than that from the ``blob''.

The value of $L_j$ is not constrained from first principles, but can be compared to the Eddington luminosity $L_{edd}$ as an order-of-magnitude reference. Based on measurements of the jet power of radio-loud AGNs from X-ray cavity data \citep[e.g.][]{Cavagnolo2010}, highly super-Eddington powers are generally not expected. 
We conservatively required that acceptable solutions should have $L_{j} < L_{edd}$ for a given source.
 
We also derived the ratio $\eta$ between the kinetic energy density of the relativistic particles, which is largely dominated by $u_p$ for all our solutions,
and the energy density of the magnetic field $u_B$:
\begin{equation}
\eta = \frac{u_p} {u_B}.
\end{equation}
Solutions can then be characterised by their proximity to equipartition between these components ($\eta=1$). For stationary solutions, having a system close to equipartition is physically appealing 
\citep[e.g. discussion by][]{Boettcher2013} and models close to equipartition are in general also energetically favoured, such that $\eta$ can be used as an order-of-magnitude reference to select physically plausible solutions. We required $0.1 <  \eta  < 10$ in the following. It should however be noted that deviation from equipartition between kinetic electron-energy density and magnetic energy 
density, even by more than an order of magnitude, is frequently encountered in the standard one-zone SSC models for these type of sources \citep[e.g.][]{Cerruti2013a}.  

%______________________________________________________________
\section{Hadronic scenarios for high-frequency peaked BL\ Lac objects }
\label{sec:scenarios}

When modelling the SEDs of high-frequency peaked BL Lac objects within our hadronic scenario, we distinguish two limiting cases. These depend on whether the VHE spectrum
is dominated by proton-synchrotron emission from the primary protons or by muon-synchrotron emission from muons generated in pion decays that follow proton-photon interactions. 
In general, both components contribute at least to some extent to the VHE spectrum. In all the hadronic scenarios discussed here, the low-energy bump of the SED that covers the
optical to X-ray range is interpreted as electron-synchrotron radiation\footnote{An alternative definition of ``hadronic'' models, where the whole SED is dominated by emission linked
to hadrons, is used by \citet{Mastichiadis2013}. Such scenarios seem to lead to very high jet powers and are not considered here.}. 
 
\subsection{Proton-synchrotron dominated very-high-energy spectrum} 

In this scenario, proton-synchrotron emission is responsible for the entire high-energy bump. A contribution from
muon-synchrotron or cascade emission only appears above several TeV or tens of TeV, where it can lead to small spectral features as will be discussed below.
When the high-energy bump is ascribed to proton-synchrotron emission, the proton-synchrotron peak frequency is fixed to the peak frequency of the bump.

\begin{figure}
   \centering
        \includegraphics[width=\columnwidth]{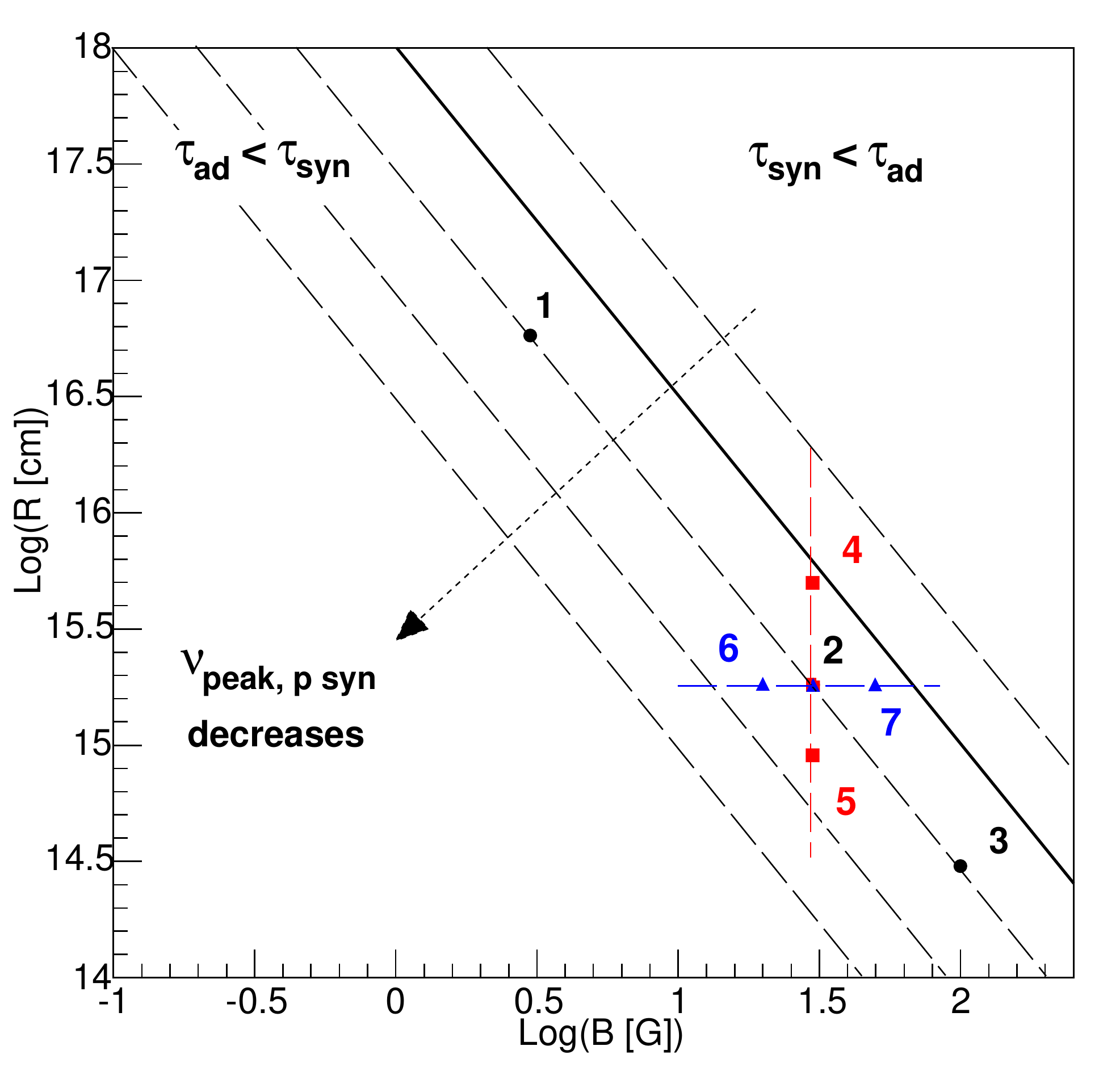}
      \caption{Location of hadronic solutions in the $\log{R}$-$\log{B}$ parameter plane. The frequency of the proton synchrotron emission peak remains constant along
             the dashed lines. The solid diagonal line separates the adiabatic cooling dominated domain from the radiative cooling dominated domain. The location
             of the typical solutions for HBLs shown in Figs.~\ref{fig:ps_diagonal},~\ref{fig:varyR} and ~\ref{fig:varyB} are indicated with markers and numbers. The dotted 
             arrow points in the direction of decreasing proton synchrotron peak frequency.}
         \label{fig:rb_schema}
   \end{figure}

Solutions with a constant synchrotron peak frequency lie on diagonal lines in the $\log{B}$-$\log{R}$ parameter plane that satisfy
\begin{equation}
\log{R} \propto -3/2 \cdot \log{B}
\label{equ:rb_diag}
\end{equation}
for a given $n_1$ and $\delta$. This can be seen from Eqs.\ 20 and 29 in \citetalias{Cerruti2015}. Solutions with higher or lower peak
frequencies lie on parallels towards higher or lower values of $R$, respectively. This is shown schematically in Fig.~\ref{fig:rb_schema}.
All the solutions for the HBLs we have studied lie in the adiabatic cooling dominated regime, meaning to the left of the bold line in Fig.~\ref{fig:rb_schema}.

\begin{figure*}
   \centering
        \includegraphics[width=0.33\textwidth]{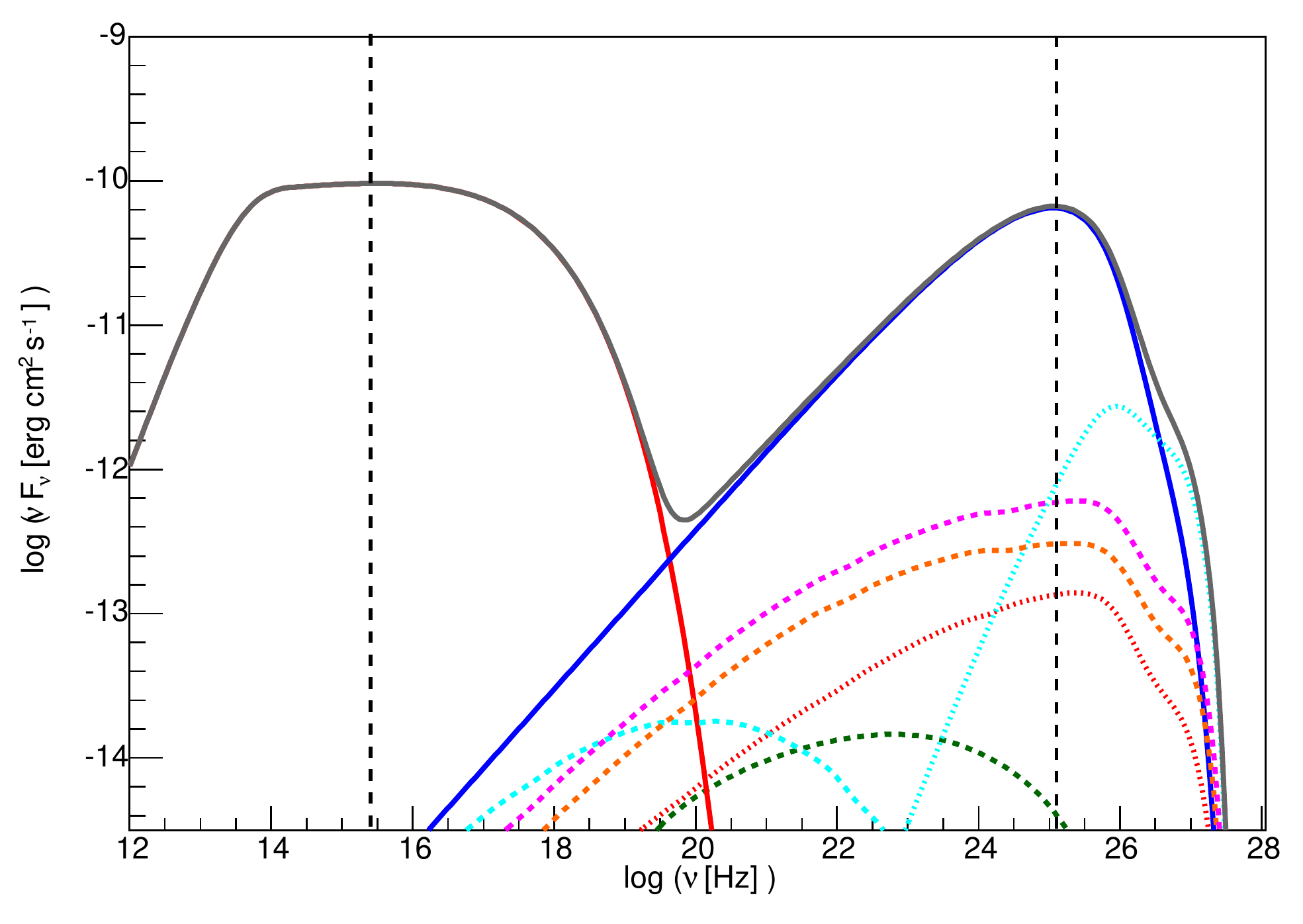}
         \includegraphics[width=0.33\textwidth]{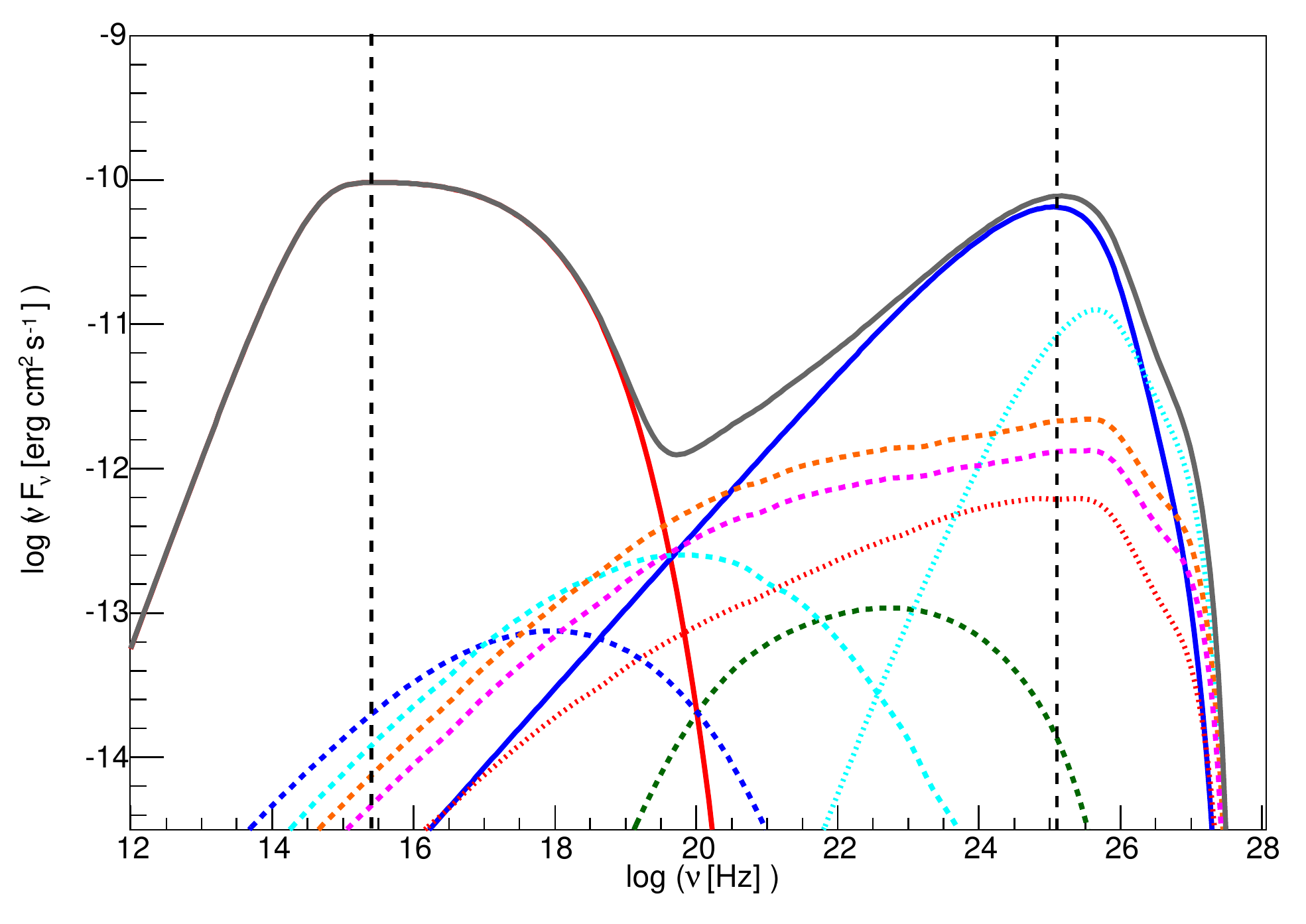}
          \includegraphics[width=0.33\textwidth]{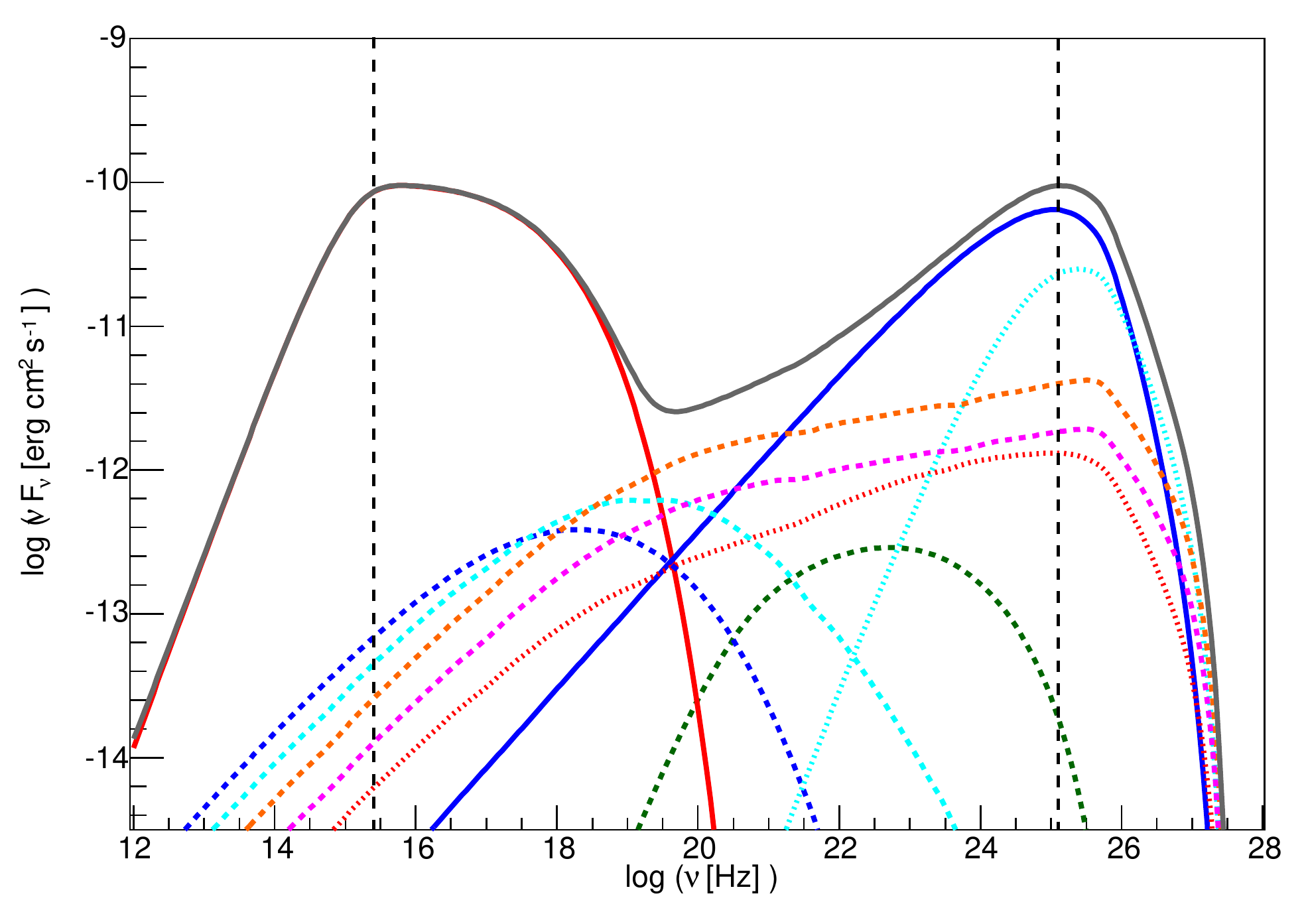}
          \includegraphics[width=\textwidth]{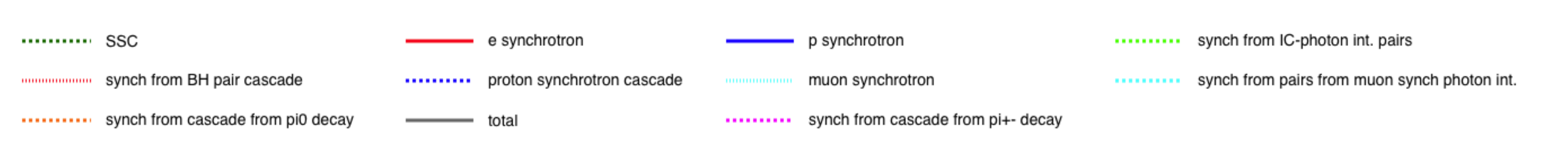}
      \caption{Variation in the modelled SED when moving along the diagonal in $\log{R}$-$\log{B}$ space. From left to right, models with $\log{B [G]} =0.5$,$1.5$,$2.0$ are shown.
                   Particle densities are adjusted to maintain the same overall flux level between the different models.
                   Solid red and blue lines indicate the electron and proton synchrotron emission. The dotted and dashed lines show muon-synchrotron emission at the highest energies,
                   and the SSC and muon-synchrotron cascade components at lower energies, as indicated in the legend. The VHE spectrum is absorbed by the EBL, assuming a redshift of 0.116,  corresponding to the
                   source PKS\,2155-304. The spectral index $n_1$ is chosen to be 1.9. The two dashed vertical lines are there to guide the eye by marking the approximate
                   peak postitions of the model in the central figure. }
         \label{fig:ps_diagonal}
   \end{figure*}

When moving along the diagonal line of constant peak frequency towards higher $B$ and smaller $R$ while keeping the flux level constant, the equipartition ratio
$\eta$ increases. The magnetic energy density $u_B$ increases as $B^2$. The proton energy density $u_p$ increases roughly as $B^3$ due to the relation given in Eq.~\ref{equ:rb_diag} and the dependence of $\gamma_{p,max}$ on $R$ and $B$
(cf. Eq.~\ref{equ:gpmax}). As $\eta$ increases, proton-photon interactions become more frequent, leading to a more significant contribution 
from subsequent synchrotron-pair cascades and from muon-synchrotron emission. This can be seen in Fig.~\ref{fig:ps_diagonal}, in which model SEDs
are shown for three different locations on a diagonal in the $\log{B}$-$\log{R}$ plane. The proton-synchrotron component remains dominant as long as the peak frequency is
sufficiently high and the particle density remains sufficiently low. This is particularly the case for all the hadronic solutions found for UHBLs, as will be discussed in Sect.~\ref{sec:UHBL}.

Increasing $B$ moves the low energy cut-off in the electron-synchrotron spectrum to higher frequencies, as shown in Fig.~\ref{fig:ps_diagonal}, whereas solutions with large $R$ and small $B$ require large $\gamma_{e;min}$, so as not to violate
constraints from optical and radio data. Very large values of $\gamma_{e;min}$ are however physically difficult to account for, except in very specific scenarios 
\citep[see for example the discussion by][]{Katarzynski2006}.

\begin{figure}
   \centering
        \includegraphics[width=\columnwidth]{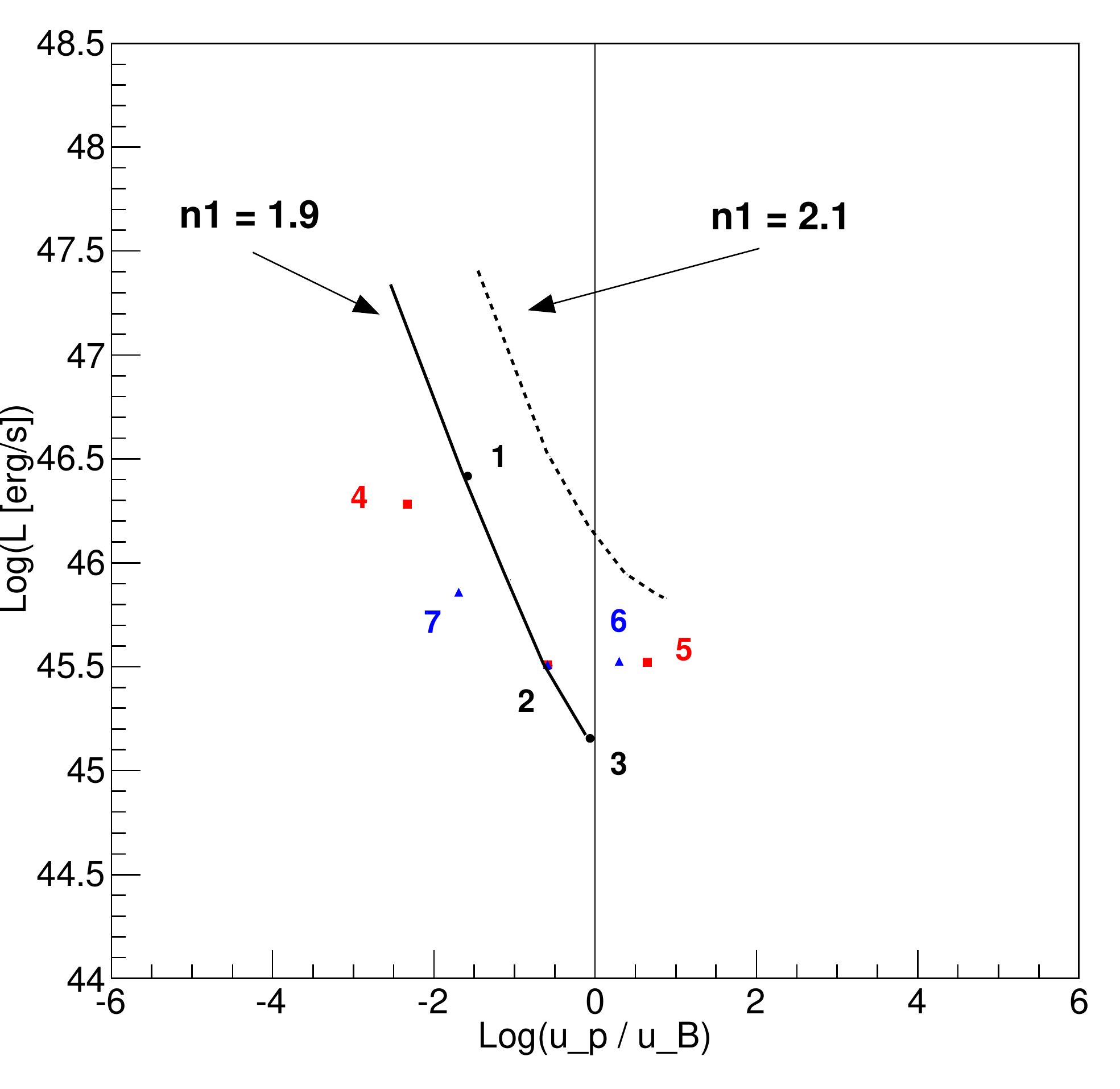}
      \caption{Location of hadronic solutions in $\log{L_j}$-$\log{\eta}$ parameter space. The markers indicated the same solutions as shown in Fig.~\ref{fig:rb_schema}.
                   The dashed curve represents the location of the proton-synchrotron solutions for a higher spectral index of the particle population. The vertical line
                    marks equipartition between energy density of the magnetic field and relativistic protons.}
         \label{fig:Leta_schema}
   \end{figure}

From Eqs.~\ref{equ:Ljet} and~\ref{equ:rb_diag} it can be seen that the jet power decreases with $1/B$ when moving on the diagonal towards 
higher $B$ and lower $R$, as long as the contribution from the magnetic energy density dominates, meaning  $\eta << 1$. Once the proton energy density
dominates, $L_j$ remains approximately constant for $n_1 \sim
 2$. The location of the different solutions in the $\log{L_j}$-$\log{\eta}$ plane is 
shown in Fig.~\ref{fig:Leta_schema}.

In the same figure, a second set of solutions is given for a different, larger value of $n_1$. When increasing $n_1$ for a given $B$ and $R$, the proton energy 
density $u_p$ increases because a given population of protons close to $\gamma_{p;max}$ implies a larger population of protons at lower energy. This also 
leads to an increase in $L_j$ and $\eta$.

\subsection{Muon-synchrotron dominated  very-high-energy spectrum}

Starting from a given proton-synchroton dominated solution, one can find an alternative scenario when moving to lower values of $B$ or $R$
(cf. Fig.~\ref{fig:rb_schema}, red and blue lines) or both. This will shift the proton-synchrotron peak to a lower frequency, but will also lead to an increase of $\eta$,
due to an increase in particle density that compensates the smaller emission region or magnetic field strength. In the modelled SED, this results in a stronger presence of 
emission from cascades and from the muon-synchrotron component at high energies. In this scenario, the high-energy bump is thus represented by a combination of 
different proton-induced components. 

\begin{figure*}
   \centering
                  \includegraphics[width=0.33\textwidth]{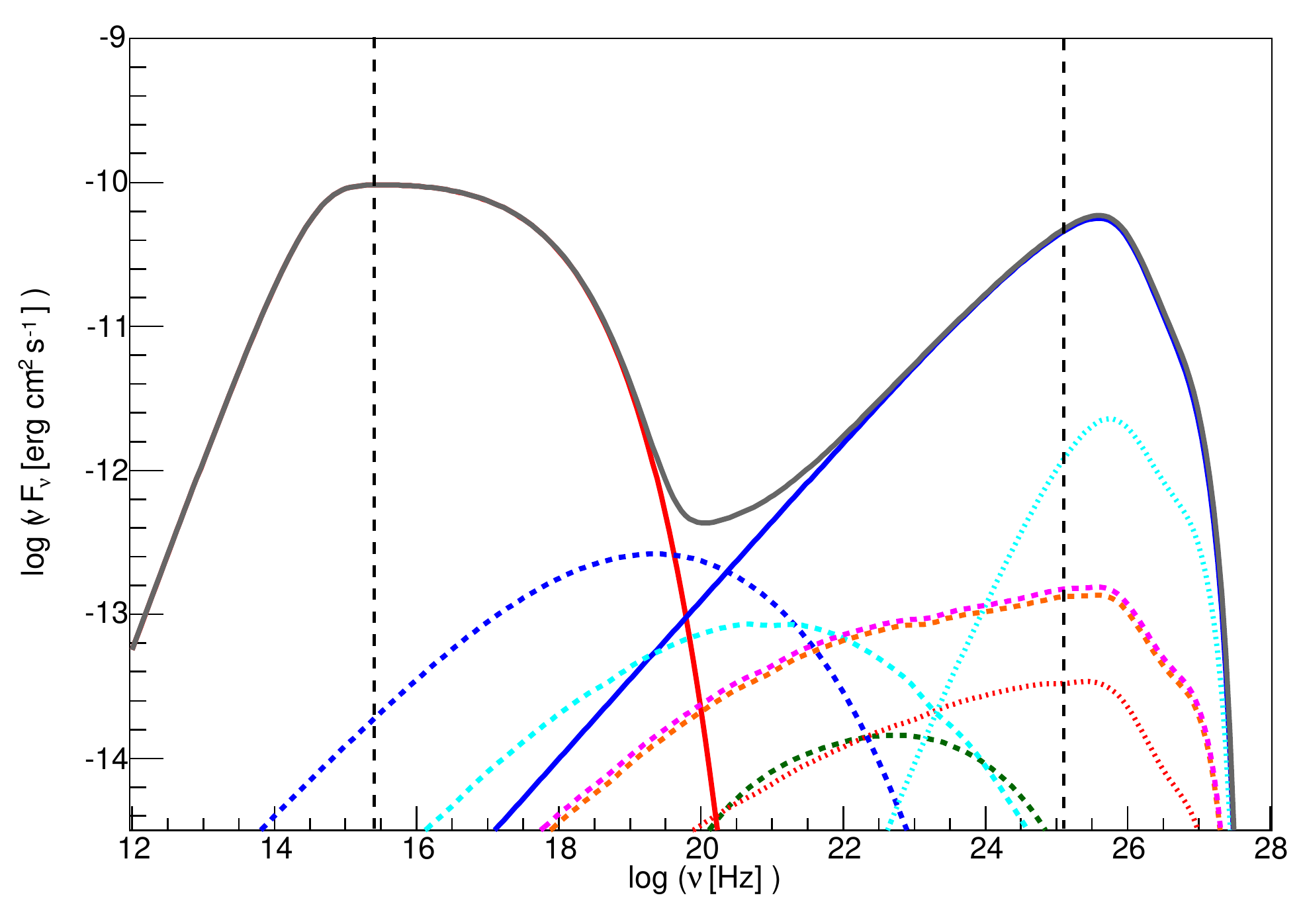}
         \includegraphics[width=0.33\textwidth]{sed_1_5_fignice}
          \includegraphics[width=0.33\textwidth]{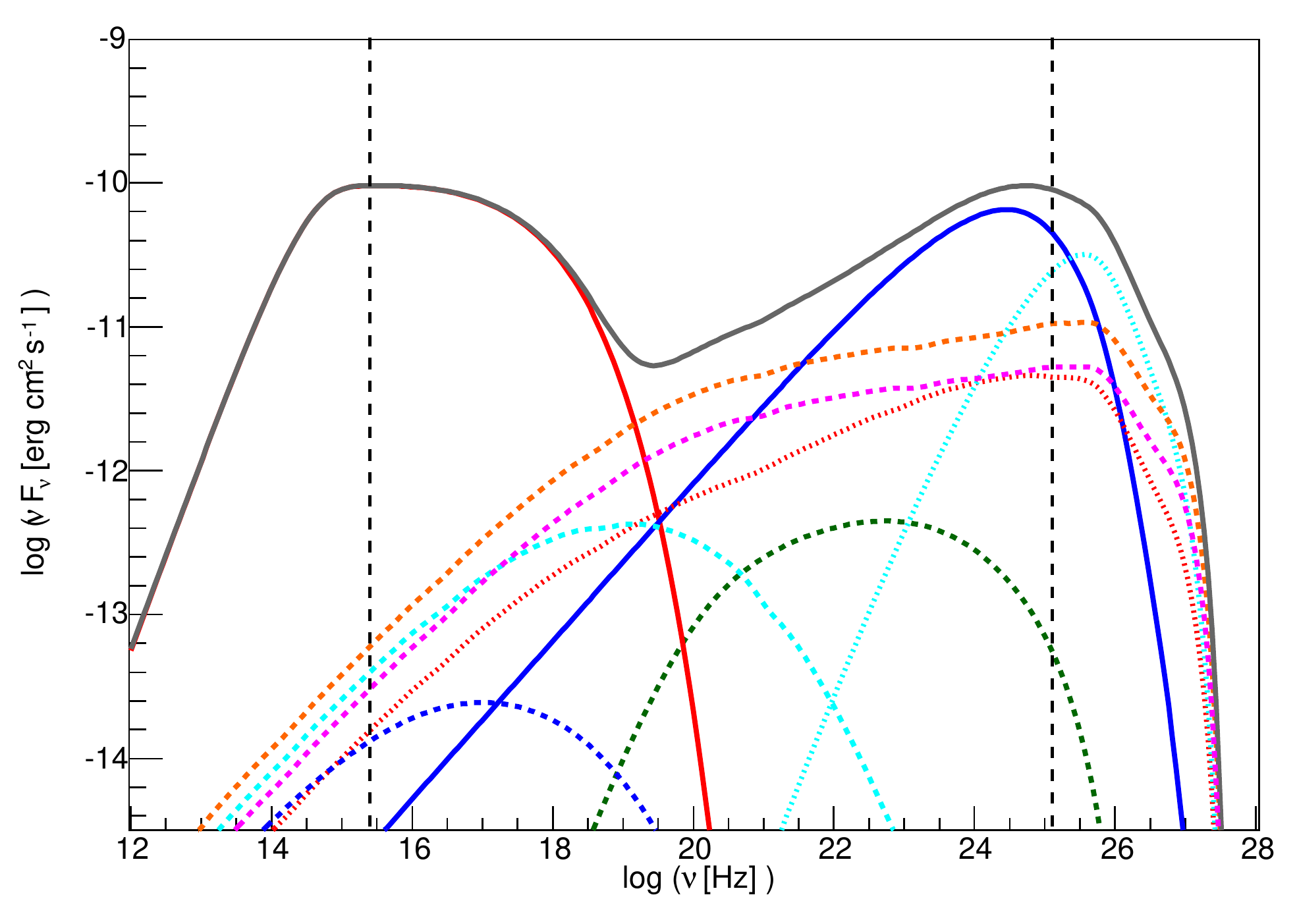}
      \caption{Variation in the modelled SED when varying $R$, i.e.\ moving along the red line in $\log{R}$-$\log{B}$ space. From left to right, models with $\log{R [cm]}=15.7$, $15.3$, and$15.0$ are shown. Particle densities are adjusted to maintain the same overall flux level between the different models. Definition of the different curves as in Fig.~\ref{fig:ps_diagonal}. }
         \label{fig:varyR}
   \end{figure*}

\begin{figure*}
   \centering
     \includegraphics[width=0.33\textwidth]{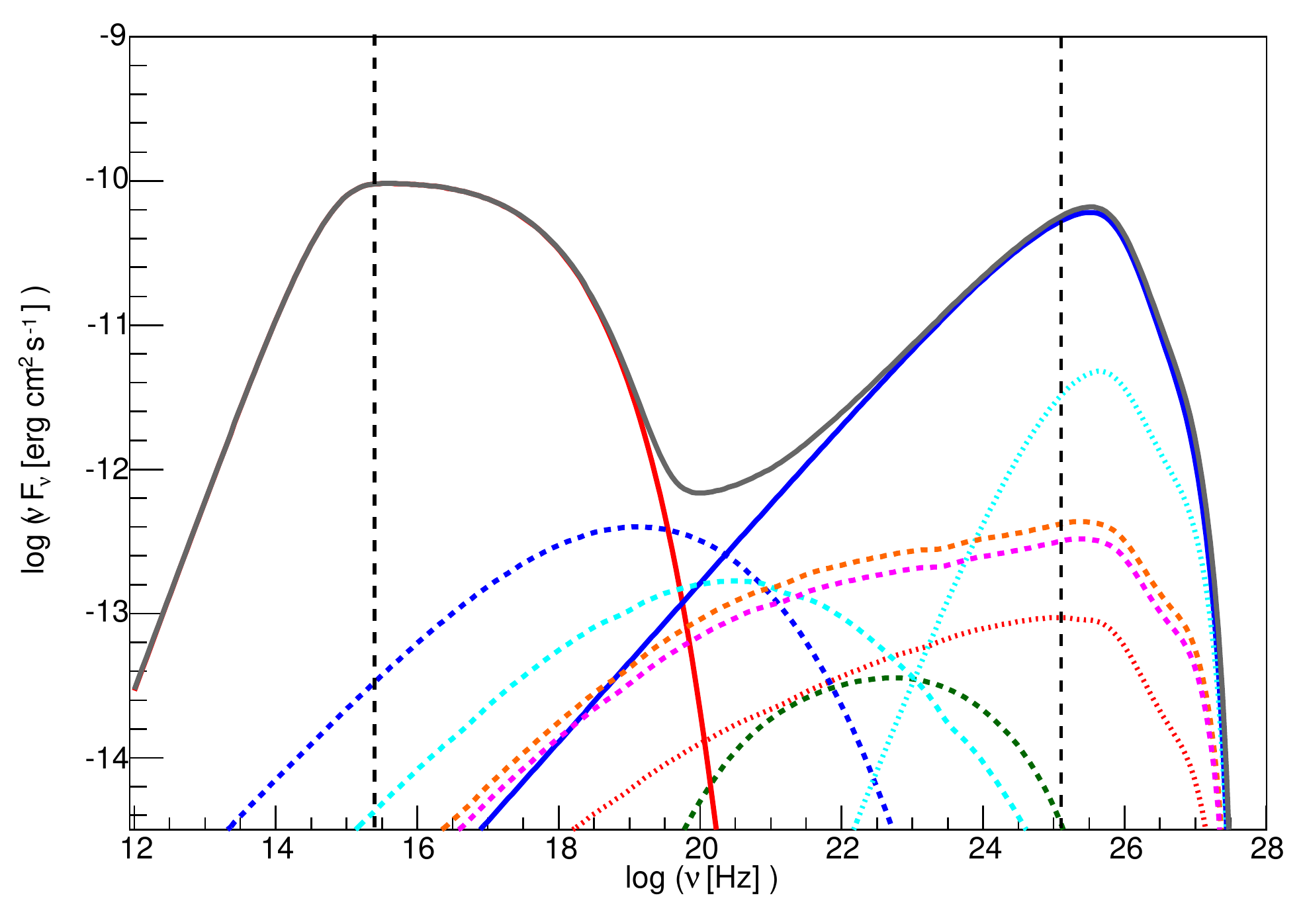}
     \includegraphics[width=0.33\textwidth]{sed_1_5_fignice}
    \includegraphics[width=0.33\textwidth]{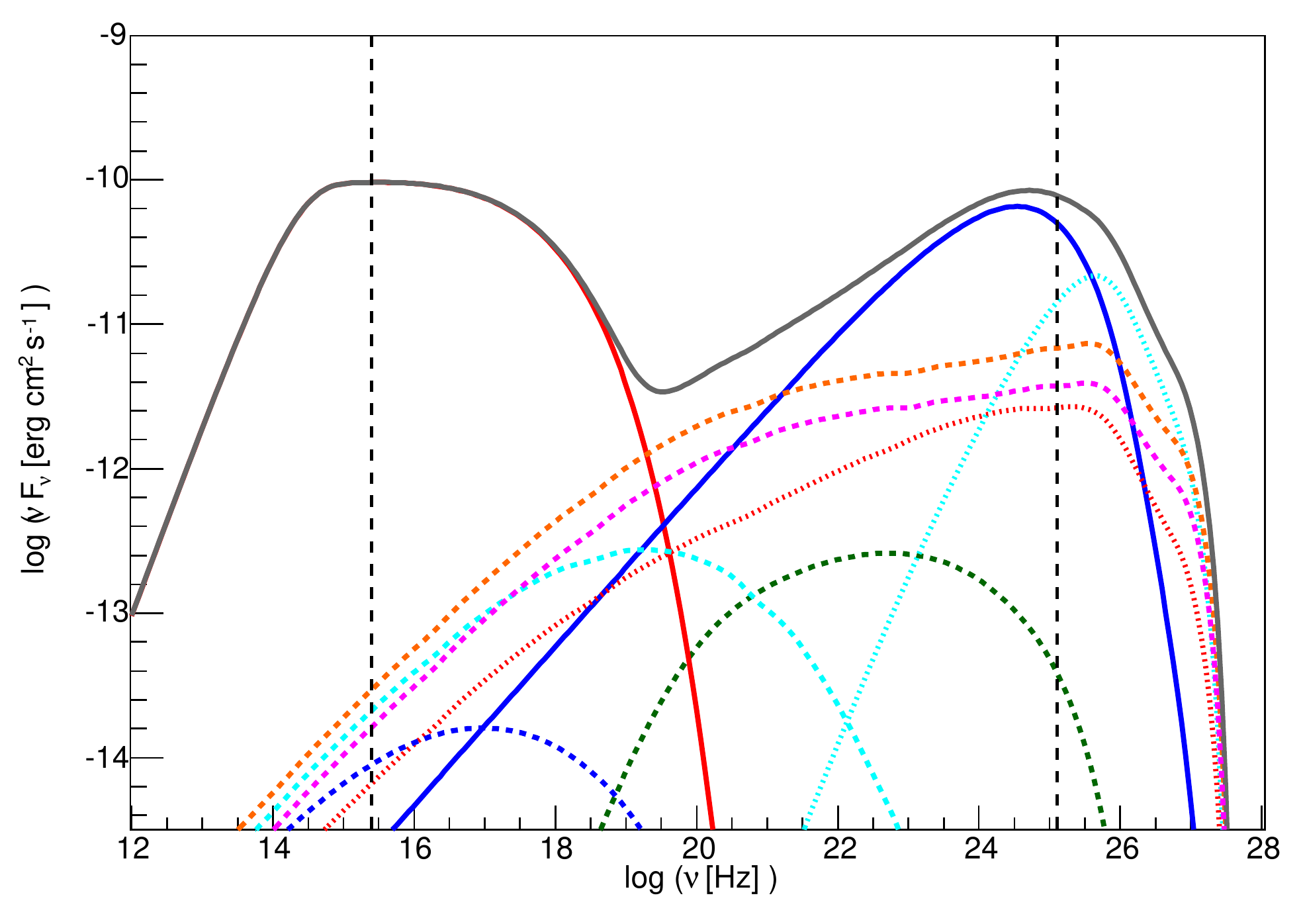}
       \caption{Variation in the modelled SED when varying $B$, i.e.\ moving along the blue line in $\log{R}$-$\log{B}$ space. From left to right, models with $\log{B [G]}=1.7$,
                  $1.5$,$1.3$ are shown. Particle densities are adjusted to maintain the same overall flux level between the different models. Definition of the different curves as in Fig.~\ref{fig:ps_diagonal}. }
         \label{fig:varyB}
   \end{figure*}

An example for the transition from a proton-synchrotron to a muon-synchrotron dominated VHE spectrum is shown in Figs.~\ref{fig:varyR} and~\ref{fig:varyB}.
The muon-synchrotron and cascade emission becomes dominant in the TeV range in Fig.~\ref{fig:varyR} in the right panel. 
Variations in either $R$ or $B$ lead to very similar models, except that by decreasing only $B$ one lowers the energy cut-off in the electron-synchrotron spectrum.
Note the small shift in the left slope of the low-energy bump in Fig.~\ref{fig:varyB} between the three panels.

When lowering $B$ or $R$ from a given initial value while keeping the flux level constant, the jet power $L_j$ decreases as long as $u_B$ dominates over $u_p$. 
As $\eta$ increases, $u_p$ will eventually become dominant and $L_j$ will increase, as can be derived from Eq.~\ref{equ:Ljet}. 
The muon-synchrotron dominated scenario provides an energetic advantage for another reason: due to the combination of several components in 
the high-energy bump, the spectral index $n_1$ is no longer constrained by the often relatively flat high-energy spectrum in the MeV and GeV range, in the $\nu F_{\nu}$ representation.
This scenario can accommodate broad high-energy bumps with smaller $n_1$ than the proton-synchrotron scenario, leading to a lower $L_j$ due to the
smaller proton number density.

\subsection{Appearance of a ``cascade bump'' spectral feature}

The combination of spectral components from proton-synchrotron emission and from proton-photon interactions can lead to spectral hardening in the 
VHE spectrum, as is seen in some of the exemplary models in Figs.~\ref{fig:ps_diagonal},~\ref{fig:varyR}, and~\ref{fig:varyB}.
The necessary condition for this kind of a feature to occur is a sufficiently high value of $\eta$, such that proton-photon interactions are non-negligible against
proton-synchrotron emission. For the proton-synchrotron scenario this is the case for solutions on the diagonal towards large $B$ and small $R$, where
the muon-synchrotron component starts becoming visible at the highest energies.

This ``cascade bump'' feature can also be seen in certain muon-synchrotron solutions. When the muon-synchrotron component is very prominent, the 
transition between muon and proton synchrotron emission occurs below the TeV spectrum and could lead to distinctive features between the {\it Fermi}-LAT 
and TeV energy ranges.
In other cases, both the proton and muon-synchrotron emission contribute to the TeV spectrum, which can lead to spectral features from the muon and cascade
emission at the highest observable energies.
It should be mentioned that this feature is also seen in other hadronic models \citep[e.g.][]{Muecke2003, Boettcher2013}, but there has been no systematic study
of its dependence on the model parameters and of its detectability.
Because this feature depends on the relative contributions from several components and thus on the exact parameter set used for a given solution, the only feasible approach 
to evaluate its prevalence seems to be a case study of a large number of solutions for a few given sources, which is attempted in the following section.

%______________________________________________________________

\section{Application to PKS\,2155-304 and Mrk\,421}
\label{sec:application}

\subsection{Datasets and constraints}

We selected two prominent TeV emitting HBLs as exemplary sources for our 
study, because their broad-band SEDs have been well measured in several multi-wavelength campaigns: Mrk\,421 in the northern hemisphere and PKS\,2155-304 in the southern hemisphere.
Because we are interested in the persistent emission from these blazars and do not try to interpret emission from flares, we focus here on datasets 
corresponding to low flux states. 

The nearby blazar Mrk\,421 has a redshift of $z=0.031$. The black hole at its centre has a mass of about $1.7 \times 10^8 M_{\odot}$ \citep{Woo2005}, and thus a Schwarzschild radius $R_{S, Mrk421} \sim 5.0 \times 10^{13} \mathrm{cm}$. This corresponds to an Eddington luminosity of $L_{edd, Mrk421} \sim 2.1 \times 10^{46} \mathrm{erg} \mathrm{s}^{-1}$. 

The SED for Mrk\,421 taken from \citet{Abdo11c}, from a multi-wavelength campaign in 2009 in which the source was found in a low flux state, includes: data points from several telescopes in the radio and optical band, data from {\it SWIFT} (X-rays and UV), {\it RXTE} (X-rays), 
{\it Fermi}-LAT ($\gamma$-rays), and MAGIC (VHE). The published MAGIC spectrum had been de-absorbed by the authors using the EBL model by \citet{Franceschini2008}. To compare the spectral points with our model curve, we absorbed the published VHE data points using the same model. The high-energy bump of the SED peaks around 100\,GeV with a peak energy flux of roughly $8 \times 10^{-11} \mathrm{\,erg\,cm^{-2}\,s^{-1}}$.

The HBL\ PKS\,2155-304 is more distant than Mrk\,421, with a redshift of $z=0.116$. Its black-hole mass is still subject to discussion. 
Based on measurements of the host-galaxy luminosity by \citet{Kotilainen1998} and on the relations given by \citet{Bettoni2003}, a mass of $1 - 2 \times 10^9 M_{\odot}$ is estimated by\citet{Aharonian2007}. However, when accounting for the scatter in the relation between bulge luminosity and black-hole mass, its mass could be as low as $2 \times 10^8 M_{\odot}$  
\citep{Rieger2010,McLure2002}. In addition, as pointed out by \citet{Aharonian2007}, the host galaxy luminosity might require further confirmation.
We thus considered a range for the Schwarzschild radius of $R_{S, PKS2155-304} \sim 3 \times 10^{13} - 3 \times 10^{14} \mathrm{cm}$, corresponding to 
$L_{edd, PKS2155-304} \sim 3 \times 10^{46} - 3 \times 10^{47} \mathrm{erg} \ \mathrm{s}^{-1}$.

The SED for PKS\,2155-304, taken from \citet{Aharonian09a}, includes data from the ATOM telescope (optical), {\it SWIFT}, {\it RXTE}, {\it Fermi}-LAT, and H.E.S.S. (VHE), all taken during a multi-wavelength campaign in 2008, in which PKS\,2155-304 was found in a relatively low, although not quiescent state. The published VHE data in this case correspond to the absorbed fluxes. 
The high-energy bump of the SED peaks around a few 10\,GeV, with a peak energy flux of roughly the same level as for Mrk\,421.

\subsection{Hadronic solutions for PKS\,2155-304}

When we interpreted the high-energy bump as proton synchrotron emission, the spectral index of the proton population was constrained by the 
{\it Fermi}-LAT spectrum to $n_1 \gtrsim 2.0$. This also fixed the spectral index of the electron population following our simple co-acceleration 
scenario. Given the high magnetic field strengths in this scenario, the entire electron population is cooled by synchrotron radiation and has thus a 
photon index of $n_1 + 1$.
The parameters $\gamma_{e;min}$ and $\gamma_{e;max}$ had to be adjusted to obey the constraints from the optical and X-ray emission, which 
we assumed to stem from the same emission region as the high-energy radiation.  

The high-energy bump could also be modelled as a combination of a proton-synchrotron peak in the {\it Fermi}-LAT band and a muon-synchrotron 
component that dominates the TeV spectrum. In this case, the spectral index of the proton spectrum was smaller, meaning that the spectrum was steeper in the
high-energy range than in the proton-synchrotron dominated scenario. 

We found good solutions for $1.9 \lesssim n_1 \lesssim 2.1$. Models with smaller $n_1$ are no longer compatible with the optical and X-ray data, whereas 
larger $n_1$ provide still acceptable representations of the SED, but lead to very large jet powers $L_j > L_{edd, PKS2155-304}$.
These solutions correspond to a wide range of magnetic field strengths. For $B$ smaller than a few Gauss, the source extension and jet power become
very large. On the other hand, $B \gtrsim 100$\,G corresponds to solutions with very small $R$, close to $R_{S, PKS2155-304}$.

\begin{figure*}
   \centering
        \includegraphics[width=\columnwidth]{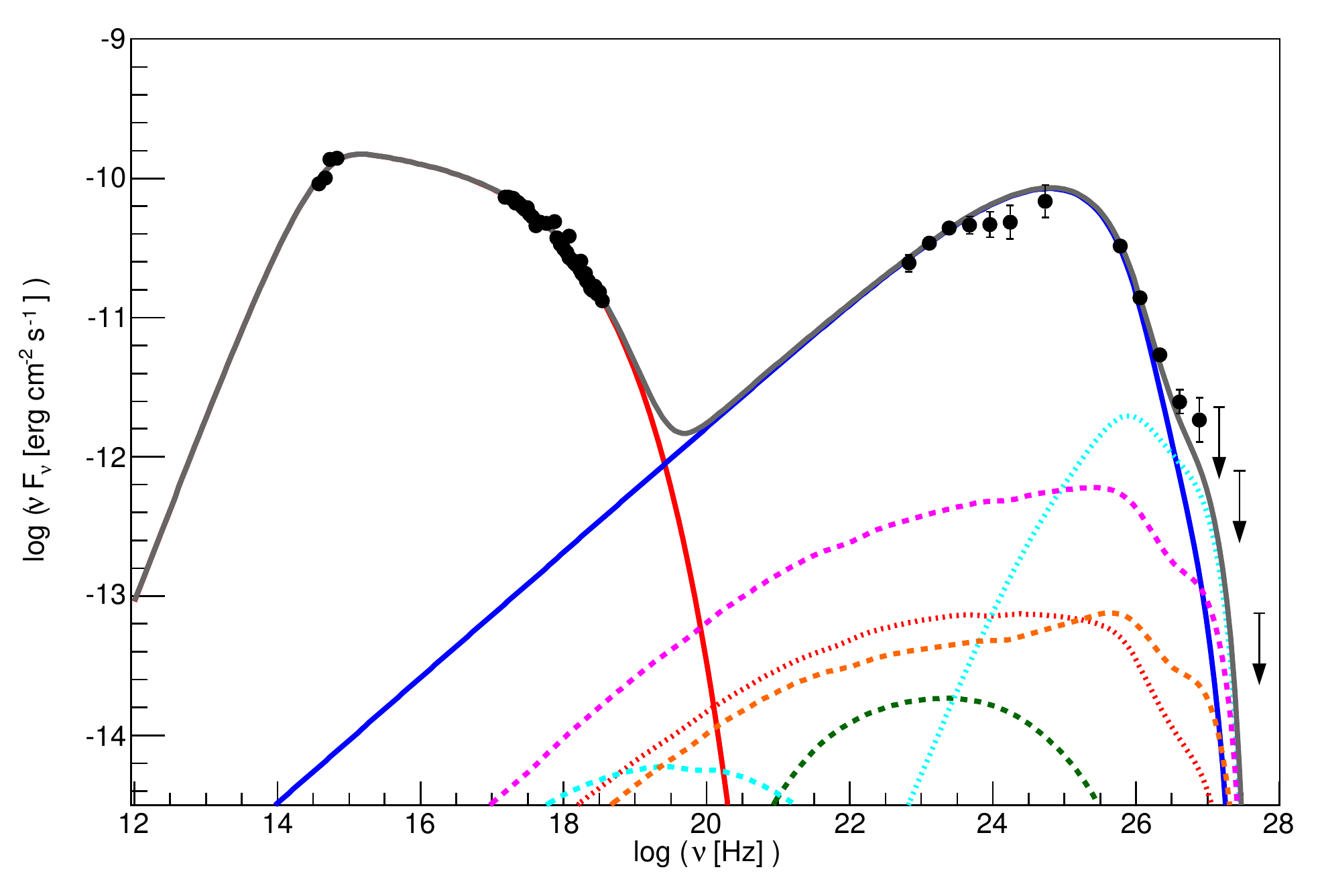} 
        \includegraphics[width=\columnwidth]{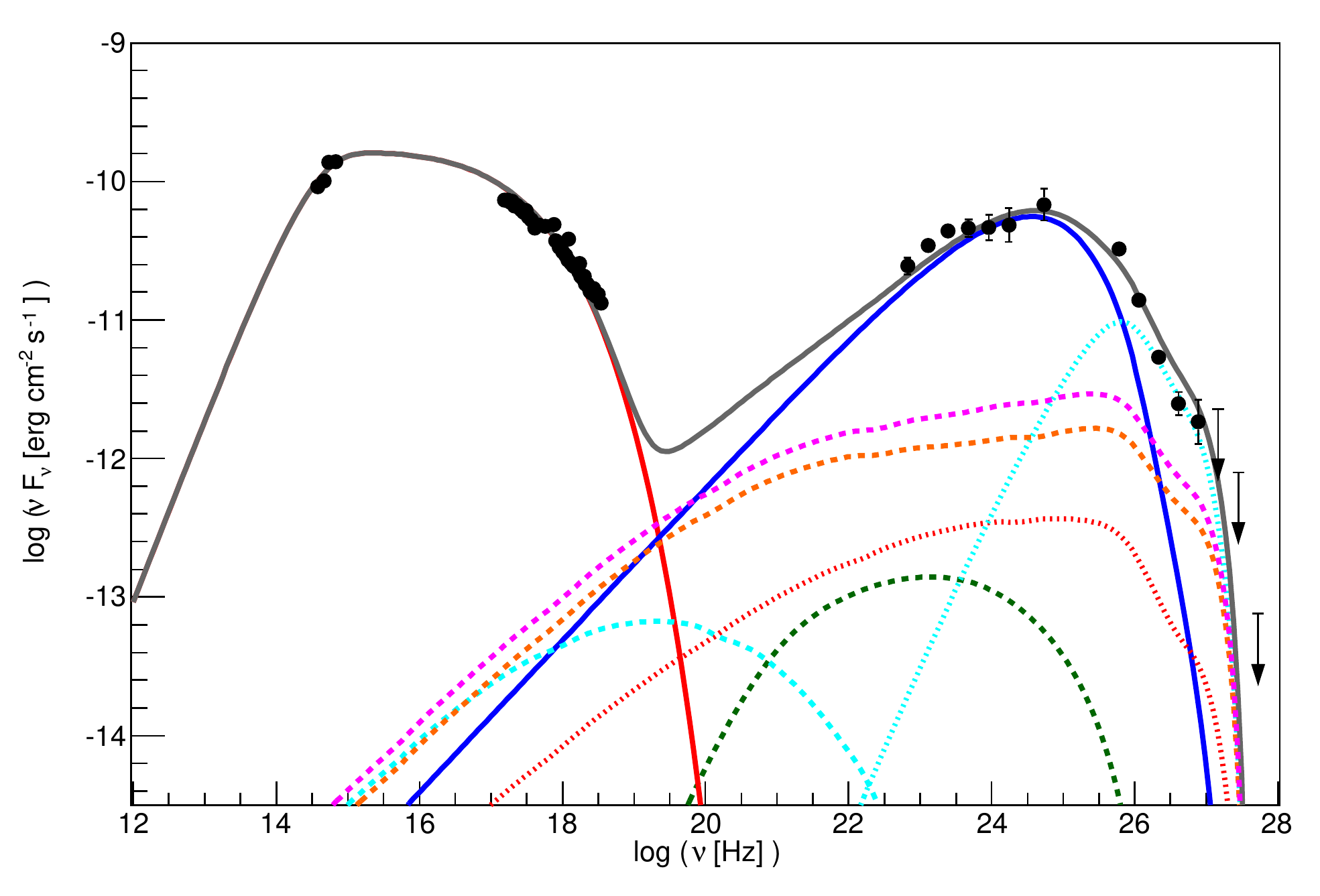}  
      \caption{SEDs for PKS\,2155-304 with two hadronic models where proton synchrotron emission (left figure) or muon-synchrotron emission (right figure) dominates the 
      TeV spectrum. These solutions correspond to a spectral index of $n_1=2.1$ and $1.9$, a magnetic field with $\log{B [G]}=0.3$ and $0.7$, and an emission region of size 
       $\log{R [cm]}=9 \times 10^{16}$ and $1.5 \times 10^{16}$, respectively. See Fig.~\ref{fig:ps_diagonal} for a description of the different curves. The dataset is described
       in the text. }
         \label{fig:sed_2155}
   \end{figure*}

% example solutions and time scales
Two examples of such solutions for PKS\,2155-304 are shown in Fig.~\ref{fig:sed_2155}. A visible ``cascade bump'' appears in the VHE spectrum in both cases.
The relevant acceleration and cooling timescales of the particle populations for these two solutions are shown in Sect.~\ref{app:2155times}. 
In all cases, energy losses are clearly dominated by adiabatic losses, and photo-meson losses, which are dominated by photopion production, are the slowest processes. 
Differences between the two solutions for each source are relatively small. The photo-meson time scale is smaller for the solutions with smaller $R$ and higher particle
densities. The solutions for Mrk\,421 have smaller source extensions, leading to shorter timescales for adiabatic cooling and thus smaller maximum proton energies.

\begin{figure}
\centering
 \includegraphics[width=\columnwidth]{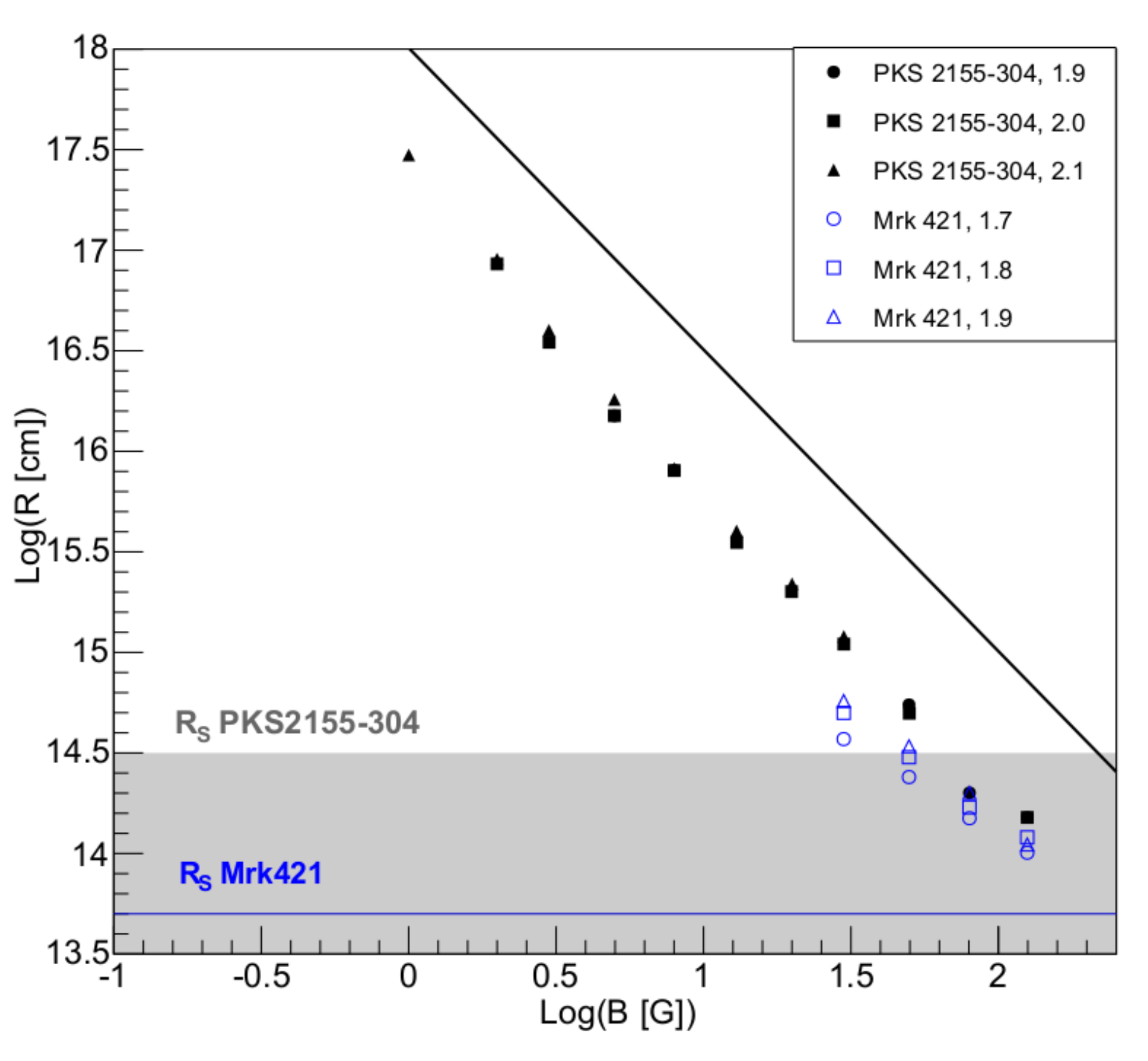}
 \caption{Source extension vs. magnetic field for HBL models. The different symbols correspond to the two different sources and to different spectral indices $n_1$. The grey 
 band indicates the range of values suggested for the Schwarzschild radius of PKS\,2155-304, whereas the blue line shows the Schwarzschild radius of Mrk\,421.}
  \label{fig:rb_hbl}
\end{figure}

% location of solutions in BR plane
The hadronic solutions for both sources are located inside the adiabatic-cooling dominated regime in $\log{B}$-$\log{R}$ space (cf. Fig.~\ref{fig:rb_hbl}). 
For PKS\,2155-304, solutions can be found over the whole physically acceptable range of source extensions. The contribution from muon-synchrotron and cascade emission is 
completely negligible for magnetic fields of a few Gauss but increases up to the same level as proton-synchrotron emission at the highest allowed magnetic fields, meaning approximately 100\,G. 

Large source extensions imply long variability timescales that might pose a problem for sources such as PKS\,2155-304 and Mrk\,421 that are known for 
their rapid variability, although rapid flares and the continuous component likely arise from different emission regions \citep[see e.g.][]{Abramowski12f}.
It should also be noted that the minimum Lorentz factor for the electron distribution can become large for very extended sources. Although for all the solutions
for Mrk\,421 that are discussed below $\gamma_{min}$ does not become larger than about 700, for the PKS\,2155-304 models with the largest extensions 
it can increase up to 1800.

\begin{figure}
\centering
 \includegraphics[width=\columnwidth]{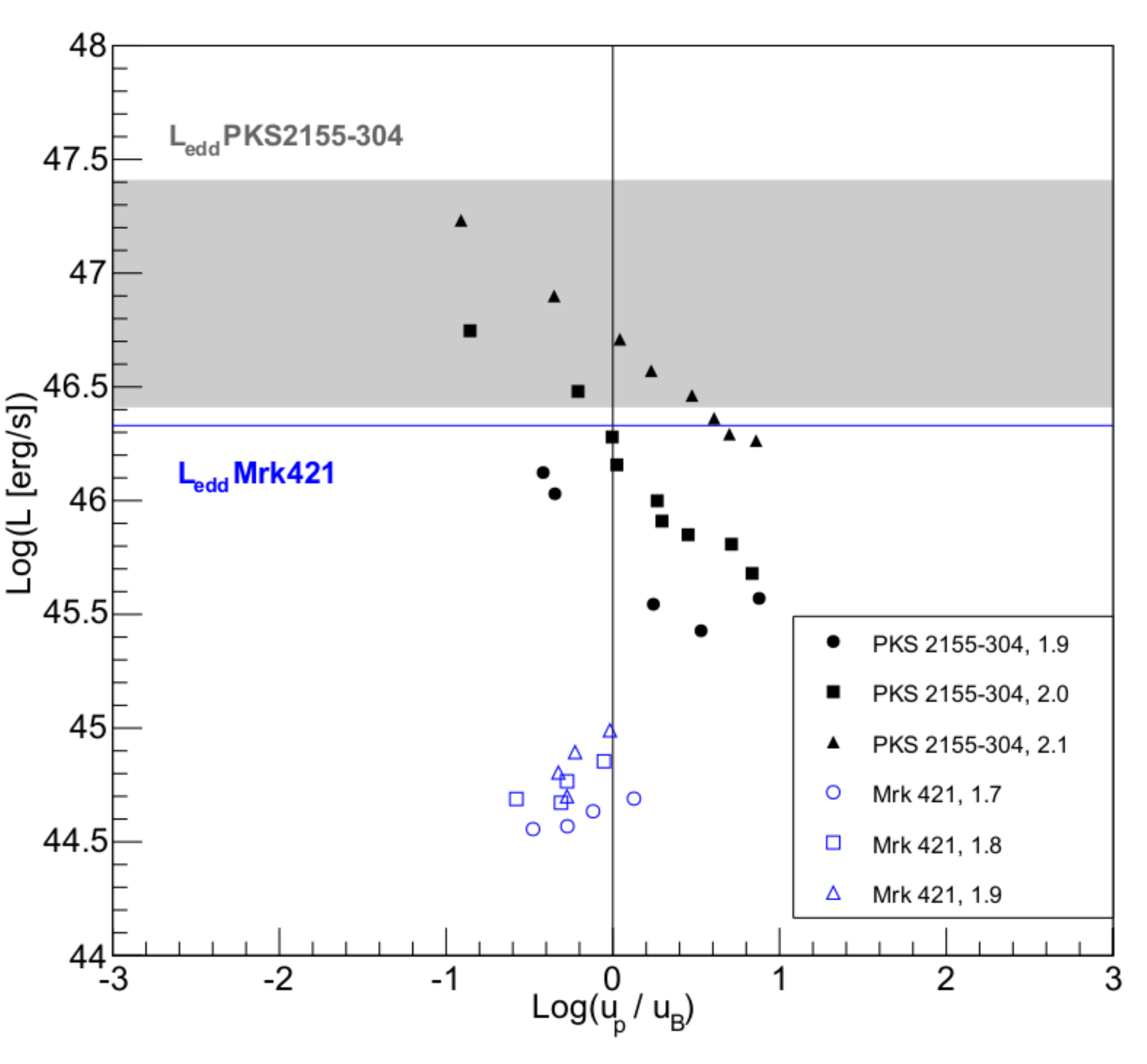}
 \caption{Jet power vs. equipartition ratio for HBL models. The different symbols correspond to 
 the two different sources and to different spectral indices $n_1$. The grey band indicates the range of values suggested for the Eddington luminosity of PKS\,2155-304, 
 whereas the blue line shows the Eddington luminosity of Mrk\,421. }
 \label{fig:Leta_hbl}
\end{figure}

% location of solutions in Leta plane
The energy budget of the solutions for PKS\,2155-304 can be dominated by the energy density of the protons or of the magnetic field, as seen in Fig.~\ref{fig:Leta_hbl}. 
The jet power becomes large for solutions with small magnetic fields and thus large emission regions. As discussed above, the value of $n_1$ has a strong influence
on the jet power. For the study presented in Sect.~\ref{sec:signatures}, we consider all the solutions shown in Figs.~\ref{fig:rb_hbl} and~\ref{fig:Leta_hbl}, but the most easily acceptable solutions
in terms of jet power and source extension are those with intermediate values of $R$ and $B$ and a relatively small $n_1$.

\subsection{Hadronic solutions for Mrk\,421}

\begin{figure*}
   \centering
              \includegraphics[width=\columnwidth]{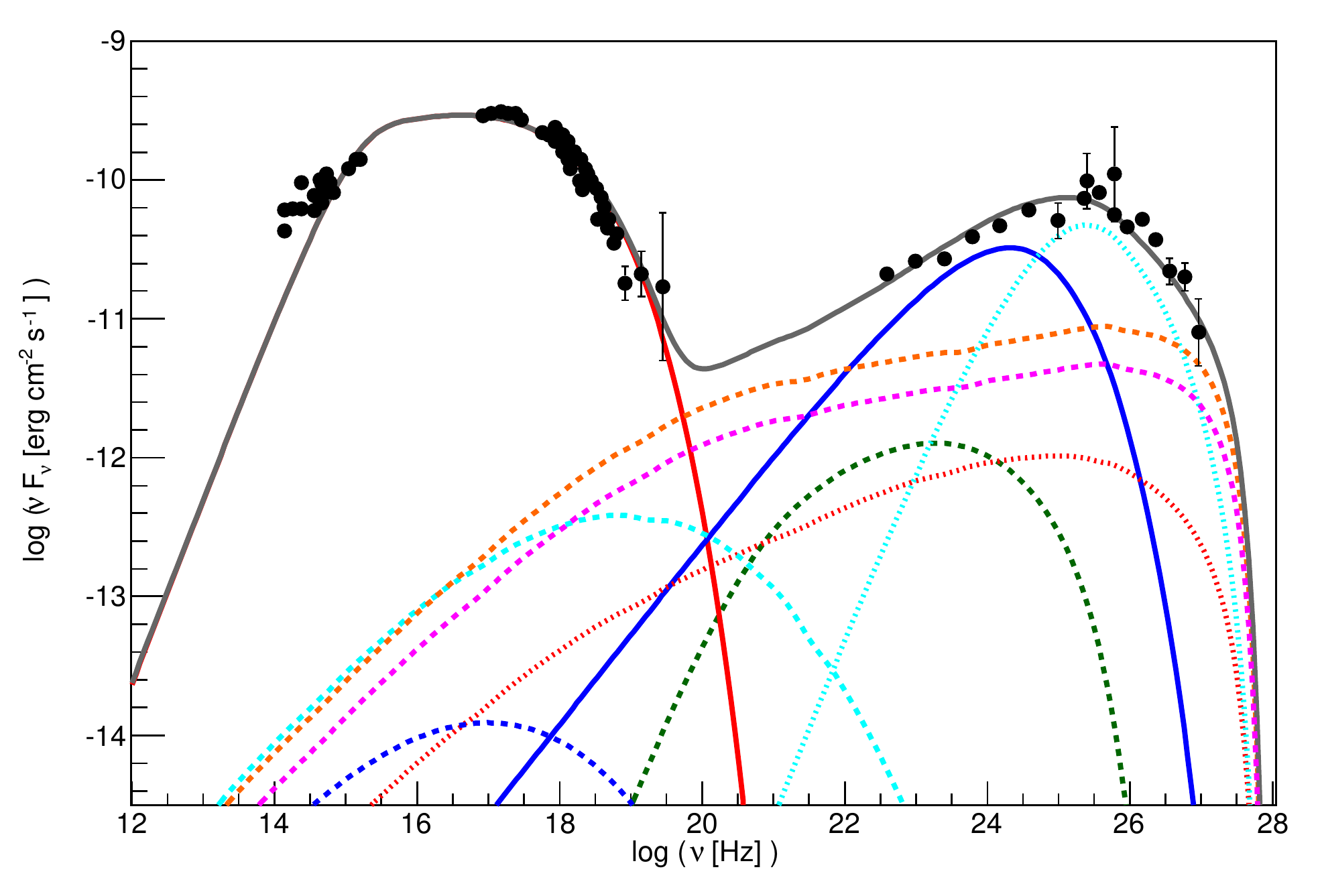}
        \includegraphics[width=\columnwidth]{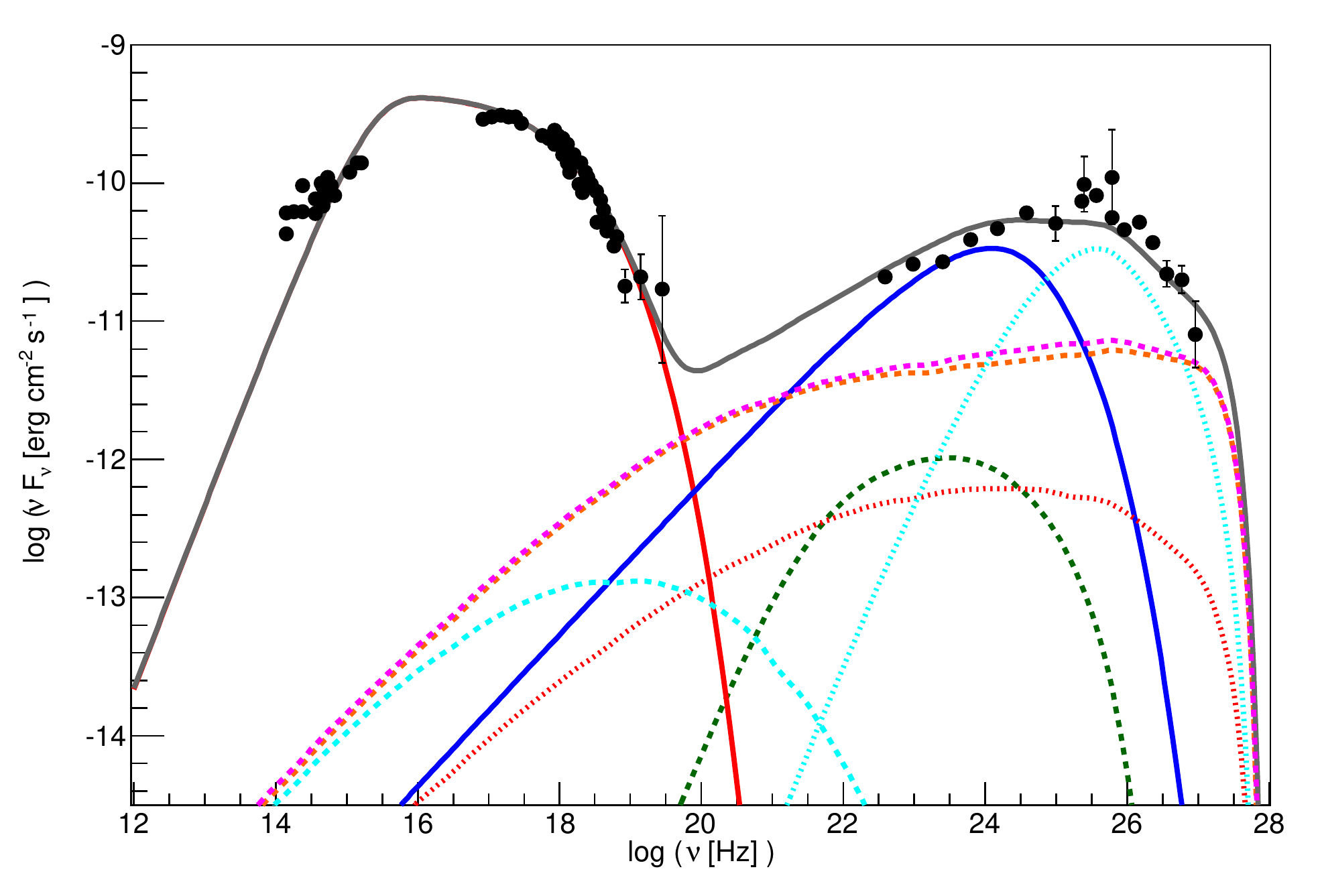}
      \caption{SEDs for Mrk\,421 with two hadronic models where muon-synchrotron emission dominates the TeV spectrum. 
         These solutions correspond to a spectral index $n_1=1.7$ and $1.9$,
        a magnetic field with $\log{B [G]}=1.9$ and $1.5$, respectively, and an emission region of size $\log{R [cm]}=1.5 \times 10^{14}$ and $5.7 \times 10^{14}$, respectively. 
        See Fig.~\ref{fig:ps_diagonal} for a description of the different curves. The dataset is described in the text.}
         \label{fig:sed_421}
   \end{figure*}

For the SED of Mrk\,421, the proton-synchrotron scenario does not provide an acceptable solution because the spectral index $n_1$ is strongly constrained by the optical data, 
ruling out solutions with a sufficiently flat shape of the high-energy bump to match the {\it Fermi}-LAT data. Good solutions are restricted to values of $1.7 \lesssim n_1 \lesssim 1.9$
and require a significant contribution from muon-synchrotron emission to reproduce the shape of the high-energy bump. 
Two examples of such solutions are shown in Fig.~\ref{fig:sed_421}.

% location of solutions in BR plane and Leta plane

These solutions occupy a restricted range of magnetic  field strengths, as seen in Fig.~\ref{fig:rb_hbl}. For $B$ smaller than a few 10\,G, the muon-synchrotron component is too low
compared to the proton-synchrotron component to match the given flux level, and the proton-synchrotron peak has shifted too far to lower energies, leading to a depression in the 
TeV spectrum. The hadronic solution proposed by~\citet{Abdo11c}, using the model by~\citet{Mucke01}, which served as an inspiration for the hadronic part of our code, falls into the same region as our solutions with $\log{B [G]} = 1.7$ and $\log{R [cm]} = 14.6$, even though their Doppler factor is much smaller with $\delta = 12$.

As is shown in Fig.~\ref{fig:Leta_hbl}, the solutions are characterised by an energy budget close to equipartition and by a total jet power that is lower than in the case of PKS\,2155-304. 
In fact, all solutions for Mrk\,421 have $L_j < 0.1 L_{edd, Mrk421}$. When comparing solutions with the same magnetic field strength and the same proton
index $n_1 = 1.9$ for the two sources, the jet power necessary for the PKS\,2155-304 model is about a factor of five larger than for Mrk\,421. Given the similar observed flux levels of the two
sources, to compensate for the difference in redshift between them a larger emission region and a higher particle density are needed for PKS\,2155-304, leading to an increased jet power and 
a higher maximum proton energy. Although the high-energy bump for these parameters is interpreted as muon-synchrotron emission for Mrk\,421, proton-synchrotron radiation dominates in 
PKS\,2155-304. 

When comparing our solutions for Mrk\,421 to the one proposed by~\citet{Abdo11c}, the jet powers are very similar, such that $\log{L_j [erg /s]} = 44.7$ for the latter. However, the smaller $\delta$ in their
solution requires compensation through a higher proton density and leads to a clear dominance of the proton energy with $\log(\eta) = 0.7$.

\subsection{Synchrotron self-Compton and mixed lepto-hadronic solutions}
\label{sec:SSC}

A standard one-zone SSC model cannot account for the observed SEDs if one assumes that the electron input spectrum is a
simple power law that is only modified by a synchrotron cooling break. One-zone SSC solutions require the introduction of an ad hoc 
broken-power-law shape for the electron distribution, in violation of our physical assumptions. 

\begin{figure*}
   \centering
       \includegraphics[width=\columnwidth]{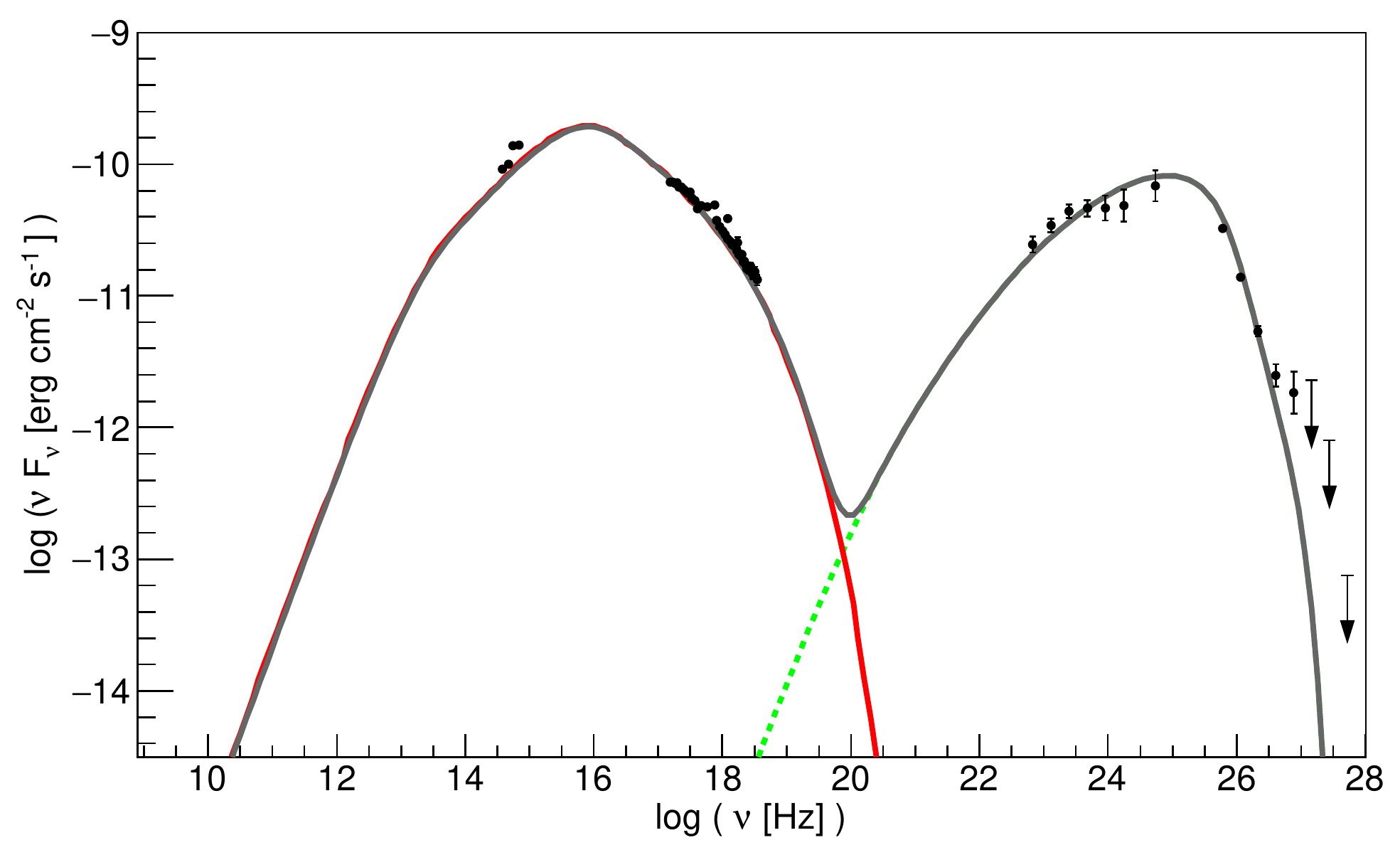}
        \includegraphics[width=\columnwidth]{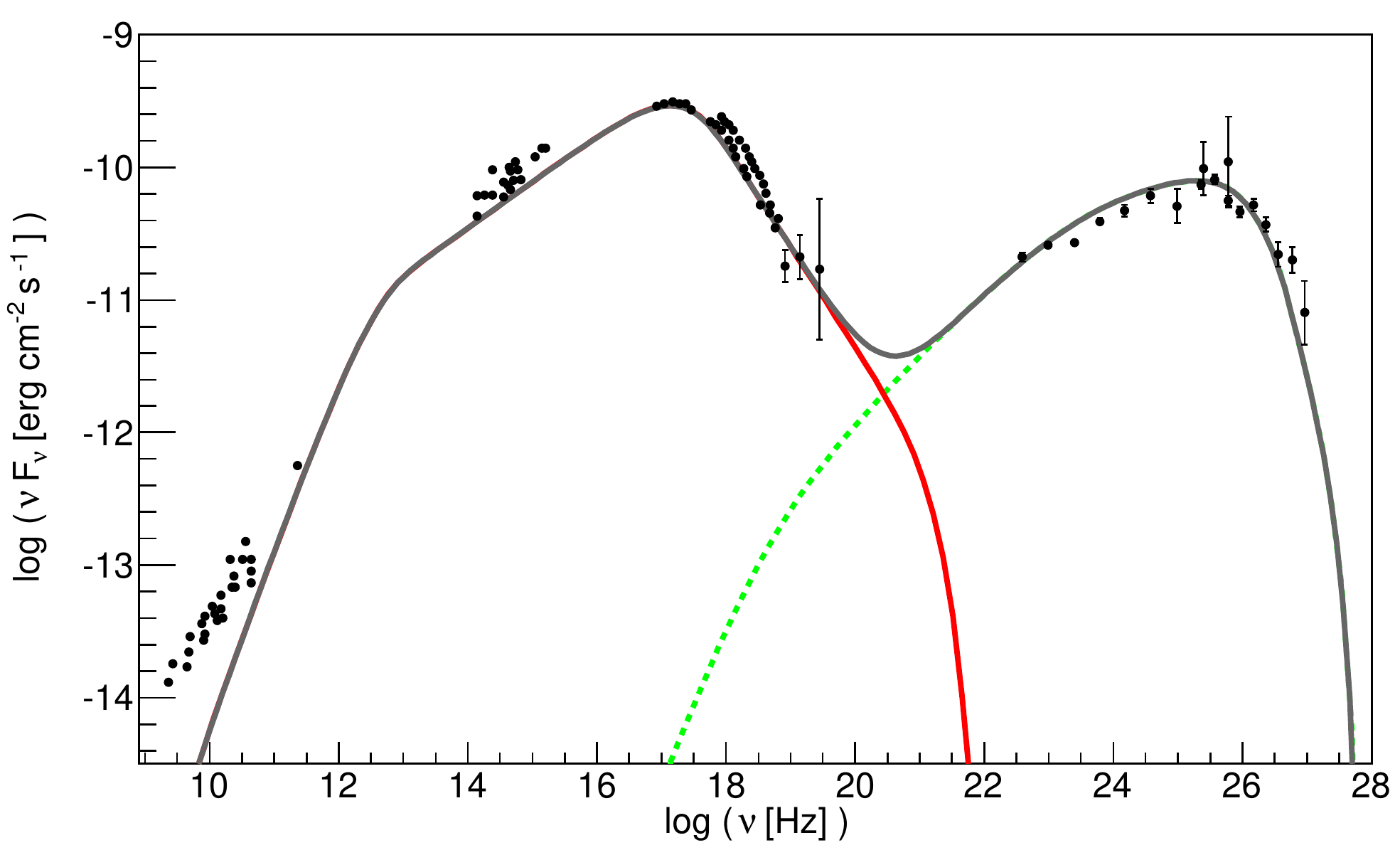}
      \caption{SEDs for PKS\,2155-304 (left figure) and Mrk\,421 (right figure) with an SSC model, where only electron synchrotron emission
       and SSC emission are considered. These solution correspond to
        a magnetic field with $\log{B [G]}=-1.4$ and $-1.1$, and an emission region of size $\log{ R [cm]}=16.8$ and $16.0$, respectively. The solid red line corresponds to the
        electron-synchrotron emission and the dotted green line to the SSC component. The datasets are described in the text.}
         \label{fig:sed_ssc}
   \end{figure*}

When loosening those assumptions for electron populations following a broken power law, accurate representations can be found for both SEDs with the exception of the optical
data from the ATOM telescope in the SED of PKS\,2155-304, which seem to require either an additional component \citep{Abramowski12f} or a more complicated electron 
distribution \citep{Aharonian09a}. When we adjusted the model to the X-ray, $\gamma$-ray, and VHE data while constraining the shape of the synchrotron bump to roughly account for 
the optical flux, and while keeping a fixed Doppler factor of $\delta = 30$, the remaining parameters of the leptonic SSC model were well constrained. The resulting solutions can be seen 
in Fig.~\ref{fig:sed_ssc}. They correspond to a size of the emission region of
approximately $\ 7 \times 10^{16} \mathrm{cm}$ and approximately $\ 10^{16}$\,cm, a magnetic field strength of $0.04$\,G and $0.08$\,G and a jet power of $ 8 \times 10^{43}$\, erg s$^{-1}$ and $1 \times 10^{43}$\, erg s$^{-1}$ for PKS\,2155-304 and Mrk\,421, respectively.
Additional solutions can be found when modifying $\delta$ and compensating with an adjustment of $R$, $B,$ and the normalisation of the electron population. Here we are only 
interested in the SSC model with the aim of comparing the different hadronic solutions (Sect.~\ref{sec:signatures}) and do not discuss these models in any detail.

Mixed lepto-hadronic scenarios, in which both proton-induced cascades and SSC emission make up the high-energy bump in the SED, do not present a possible solution for the two HBLs studied
here in the framework of our model and with the constraints we have imposed. \citet{Mastichiadis2013} and \citet{Dimitrakoudis2014} show that it is possible to find models for an SED of 
Mrk\,421 for low magnetic fields and relatively small emission regions in which the high-energy bump is dominated by cascade emission. Their study is based on a dataset from observations 
in 2001 that is less complete than the more recent multi-wavelength (MWL) data used in our study, especially due to the absence of data from {\it Fermi}-LAT. Another important difference lies in the additional 
constraints we impose on our model parameters and on the restriction of our solutions to low jet powers and equipartition ratios. By imposing small values of $\gamma_{p;max}$, which is 
considered a free parameter in their model and is chosen to be roughly three orders of magnitude below the value we would get from our constraints, and by adjusting  the slopes 
of the electron and proton distributions separately, a distinct set of ``leptohadronic-pion'' solutions can be found. These solutions are marked by very steep injection spectra with indices below 1.5, a large 
deviation from equipartition that is typically $u_p / u_B$ of the order 10$^3$, and very large jet powers of the order 10$^{48}$ erg s$^{-1}$.

%______________________________________________________________

\section{Comparison to the parameter space of models for ultra-high-frequency peaked BL Lac objects}
\label{sec:UHBL}

\begin{figure}
\centering
 \includegraphics[width=0.9\columnwidth]{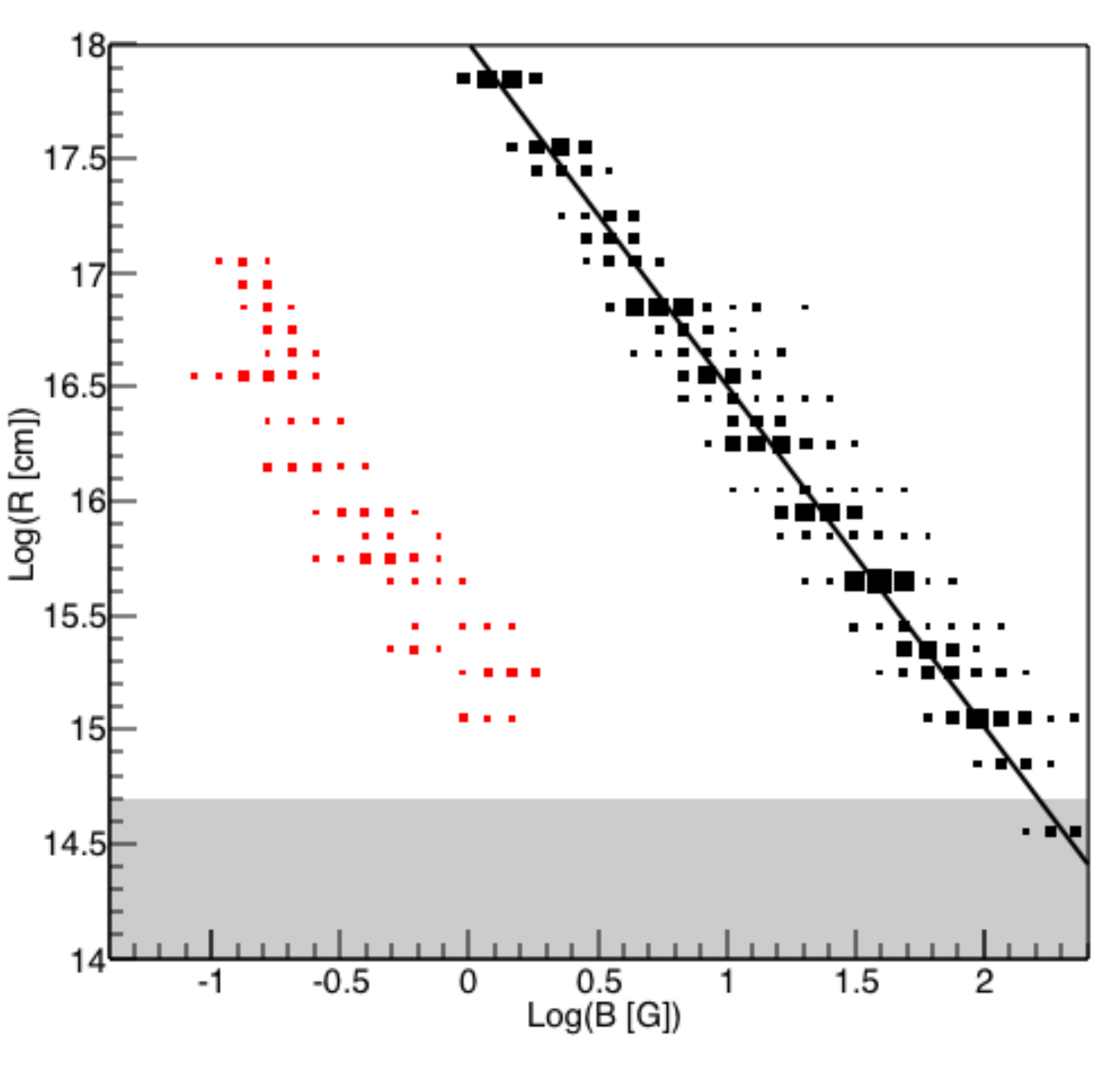}
 \caption{Source extension vs. magnetic field for the UHBL models discussed by \citetalias{Cerruti2015}. The locations for the two different types of models are shown in black for the proton-synchrotron scenario and in red 
 for the mixed lepto-hadronic scenario. The grey band indicates the range of Schwarzschild radii for the five UHBLs considered. }
  \label{fig:rb_uhbl}
\end{figure}

\begin{figure}
\centering
 \includegraphics[width=0.9\columnwidth]{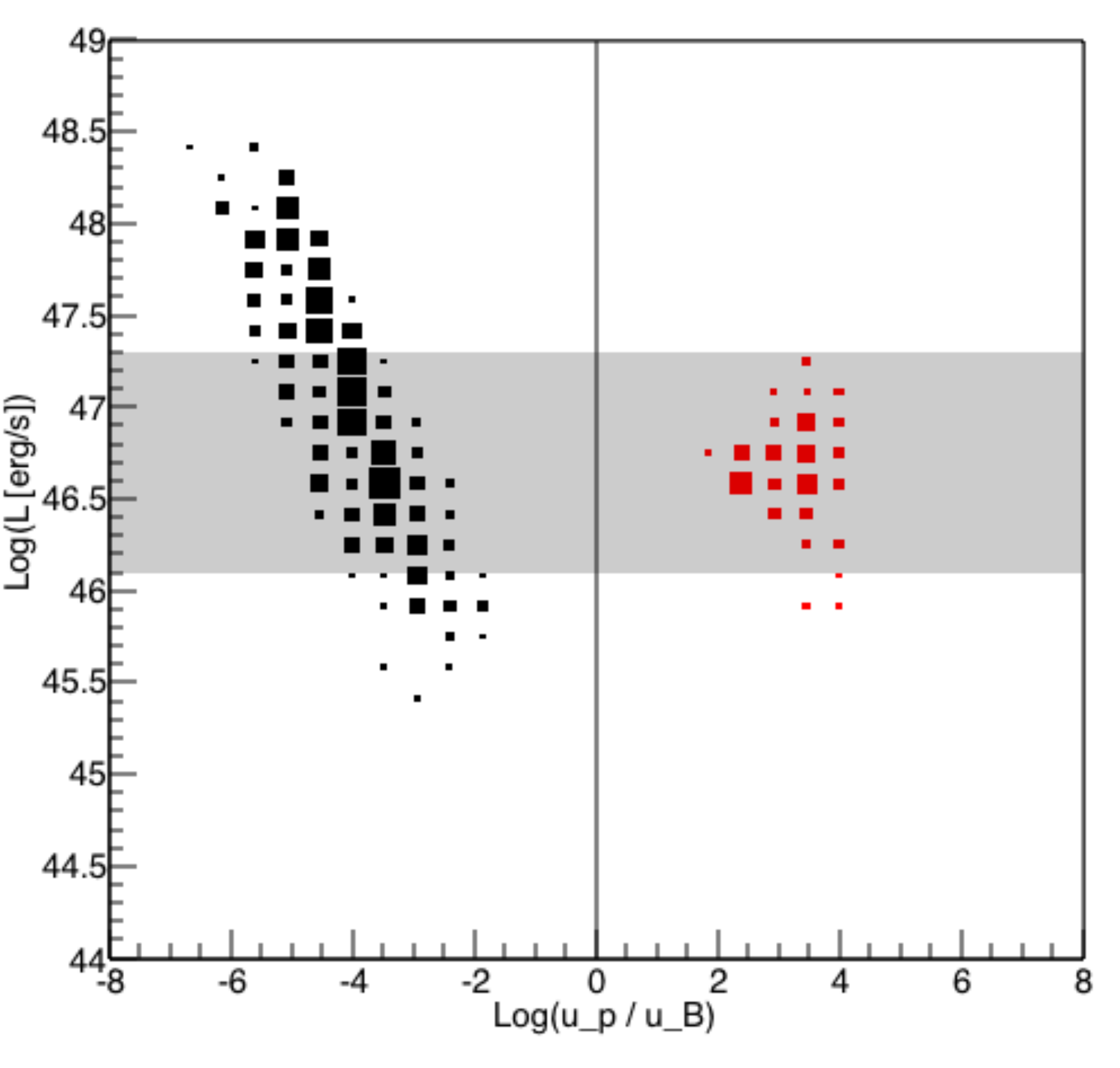}
 \caption{Jet power vs. equipartition ratio for the UHBL models discussed by \citetalias{Cerruti2015}. The locations for the two different types of models are shown in black for the proton-synchrotron scenario and in red for the mixed lepto-hadronic scenario. The grey band indicates the range of Eddington luminosities for the UHBLs considered.}
 \label{fig:Leta_uhbl}
\end{figure}

The application of our lepto-hadronic code to UHBLs is described in~\citetalias{Cerruti2015}, where solutions are given for five UHBLs: 1ES\,0229+200, 1ES\,0347-121, RGB\,J0710+591, 
1ES\,1101-232, and 1ES\,1218+304. The ensemble of the lepto-hadronic solutions found for those sources in $\log{B}$-$\log{R}$ space and $L_j$-$\eta$ space are shown in Figs.~\ref{fig:rb_uhbl} and~\ref{fig:Leta_uhbl} for comparison with the HBL solutions~\footnote{Compared to~\citetalias{Cerruti2015}, the jet powers shown here are larger by a factor of two to account for double-sided jets.}. It should be noted that solutions with $L_j < L_{edd}$ can be found for each source separately.
The same bulk Doppler factor of $\delta = 30$ was assumed in that study.

Proton-synchrotron radiation dominates the high-energy spectral bump for a large set of solutions on both sides of the dividing line between different cooling regimes, as shown in Fig.\ref{fig:rb_uhbl}. 
Because UHBLs are defined by higher peak frequencies of the high-energy bump compared to HBLs, solutions are shifted to the right, parallel to the diagonal band that
marks the HBL solutions in Fig.~\ref{fig:rb_hbl}. For these models, contributions from muon-synchrotron and cascade emission are negligible. The energy budget is largely dominated 
by the energy density of the magnetic field (cf. Fig.~\ref{fig:Leta_uhbl}) and the jet power decreases approximately as $1 / B$.

It should be noted that whether adiabatic or radiative cooling dominates at the highest proton energies depends on the assumptions made about the acceleration timescale. This explains why \citet{Muecke2003},  for example, find solutions for HBLs in which both timescales are comparable at the highest energies by assuming highly efficient particle acceleration. Independently of the actual behaviour of the acceleration timescale, if one assumes that the same particle acceleration mechanism is at play in HBLs and UHBLs, it can be seen that UHBLs correspond to a more radiatively-efficient regime in which radiative losses are more important than for HBLs.

The very steep spectral shape required to fit the {\it Fermi}-LAT data in UHBLs implies that jet powers are still acceptable even with larger values of $B$ and $R$ compared to HBLs. It also
leads to smaller values of $\eta$. 
For this reason, proton-photon interactions in UHBLs are largely dominated by proton-synchrotron emission and there are no ``cascade bumps'' even for small source extensions.
The muon-synchrotron scenario does not allow us to represent the SEDs of UHBLs because the proton-synchrotron peak of the model is strongly constrained by the high-energy 
bump and cannot be shifted to lower energies without violating the constraints from the {\it Fermi}-LAT data.

For UHBLs, a set of mixed lepto-hadronic solutions can be found in a more compact region in $\log{B}$-$\log{R}$ space for values of $B$ between approximately $\ 0.1$\,G and  approximately$\ 2$\,G, in which 
a combination of both SSC and proton-induced cascade emission is responsible for the high-energy spectral bump. For these models, proton-synchrotron emission occurs at intermediate energies and is very weak compared to the emission from SSC and cascades. The energy budget is dominated by the kinetic energy of the relativistic protons (cf. Fig.~\ref{fig:Leta_uhbl}).
Such mixed lepto-hadronic solutions cannot be found for less extreme HBLs when we impose our usual constraints, as was discussed in Sect.~\ref{sec:SSC}.
In both scenarios applicable to UHBLs, the proton synchrotron and mixed lepto-hadronic, contribution from muon-synchrotron emission is small and there is no 
spectral feature in the TeV range arising from cascades.

%______________________________________________________________

\section{Expected ``cascade bump'' signatures for the Cherenkov Telescope Array}
\label{sec:signatures}

The hadronic signatures in the VHE spectrum we are interested in can be seen at some level over the whole range of solutions 
discussed in Sect.~\ref{sec:application}. For these models, we want to test the possibility of detecting the ``cascade bump'' with the future CTA
and thus to distinguish the hadronic scenarios from a basic one-zone SSC model. 

\subsection{Method}
The expected spectra for CTA were simulated using the publicly available instrument response functions~\footnote{\url{https://portal.cta-observatory.org/Pages/CTA-Performance.aspx}} 
for the full southern array in the case of PKS\,2155-304 and for the full northern array in the case of Mrk\,421. The current layout for the southern array led to roughly a  two times better 
differential sensitivity in the range around a few TeV compared to the northern array. 

\begin{figure}[htb!]
\centering
\includegraphics[width=\columnwidth]{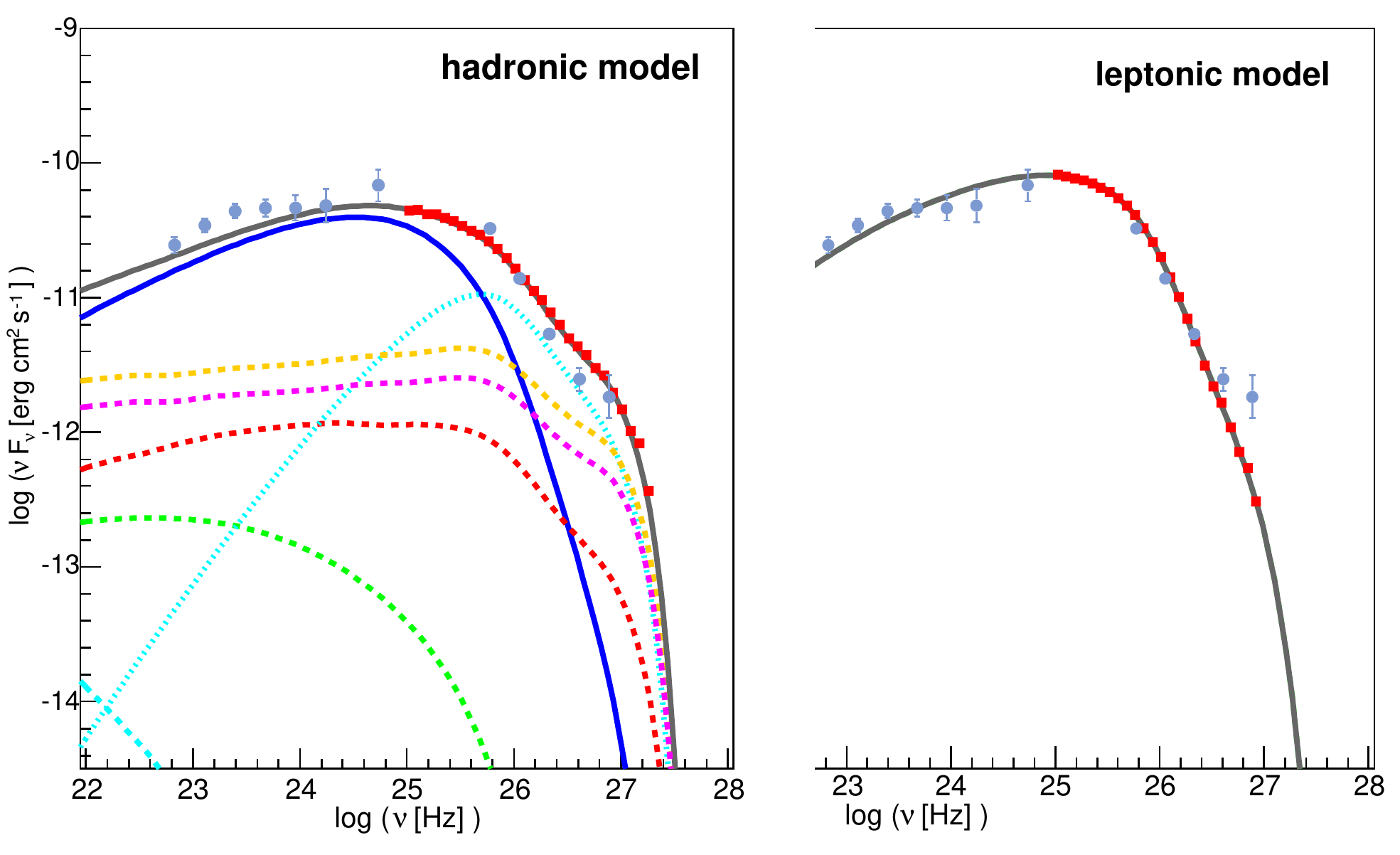}
%\vspace{4cm}
\caption{Example of testing emission scenarios with CTA: comparison of the expected CTA spectra for two specific emission models for PKS\,2155-304. 
A hadronic scenario is shown on the left, in which $n_1=2.1$, $\log{B [G]}=1.1$. and $\log{R [cm]}=15.6$, and a standard leptonic SSC model on the right. 
The exposure time assumed for the simulations, 33\,hr, is the same as the live time for the H.E.S.S. observations, which are represented by grey data points above 3$\times$10$^{25}$\,Hz. 
H.E.S.S.\ upper limits are not shown. Uncertainties in the CTA data points are smaller than the red squares. 
}
 \label{fig:kspAGNmodels}
\end{figure}

For each modelled spectrum, the integrated flux of expected $\gamma$-rays and cosmic-ray background events per energy bin was determined and weighted 
with the effective area from the performance file. The resulting event rate was then multiplied by the observation time, yielding the total number of detected events per bin. 
The number of excess events was determined by assuming that the $\gamma$-rays and cosmic rays follow a Poisson distribution and are subject to the simulated energy resolution of the instrument. 
The background rate was assumed to be extracted from a region on the sky that is five times larger than the ``on-source'' region, corresponding to a standard situation for observations
in ``wobble-mode''. The uncertainty in the number of excess events was calculated using the method by \citet{Li1983}. The excess and its uncertainty were then converted into a spectral point with 
a statistical uncertainty.
An example for two simulated CTA spectra is shown in Fig.~\ref{fig:kspAGNmodels}. For the chosen model, one can clearly see the difference in the shape of the simulated CTA spectra
for an SSC and a hadronic scenario.

To take into account statistical fluctuations of the $\gamma$-ray rate and background rate, and the effect of a limited energy resolution, several realisations of the simulated spectra were 
generated. For each hadronic model we simulated one hundred spectra and compared them to one hundred spectra simulated for the SSC model for the same source. We verified that a larger number 
of realisations is not necessary because it does not significantly change our results.

To compare the simulated hadronic and SSC spectra, first the simulated SSC spectra were fitted with a simple logparabolic function, where the lower limit of the fitted energy range was 
adjusted to optimise the reduced $\chi^2$ of the fit. The form of this function, characterised with three parameters (P1, P2, P3), is given by:
\begin{equation}
\log{f (\log{E}) } = P0 + ( -P1 - P2 * ( \log{E} - \log{E0} ) ) * (\log{E} - \log{E0} )  ;
\label{equ:logparabola}
\end{equation}
energies are in units of TeV and the reference energy is fixed with $\log{E0}$ = -1.

The real shape of the overall SSC spectrum bump, which is absorbed on the EBL, is generally not well represented by a logparabola. However, if the energy range is restricted to energies above about 100\,GeV the logparabolic function provides a satisfactory characterisation with typical fit probabilities above 10\%. 
The same fit function is then applied to a second SSC model for each source to verify that it does indeed provide a general description of the SSC scenario and not only of a single model. 

\begin{figure}[htb!]
\centering
\includegraphics[width=0.9\columnwidth]{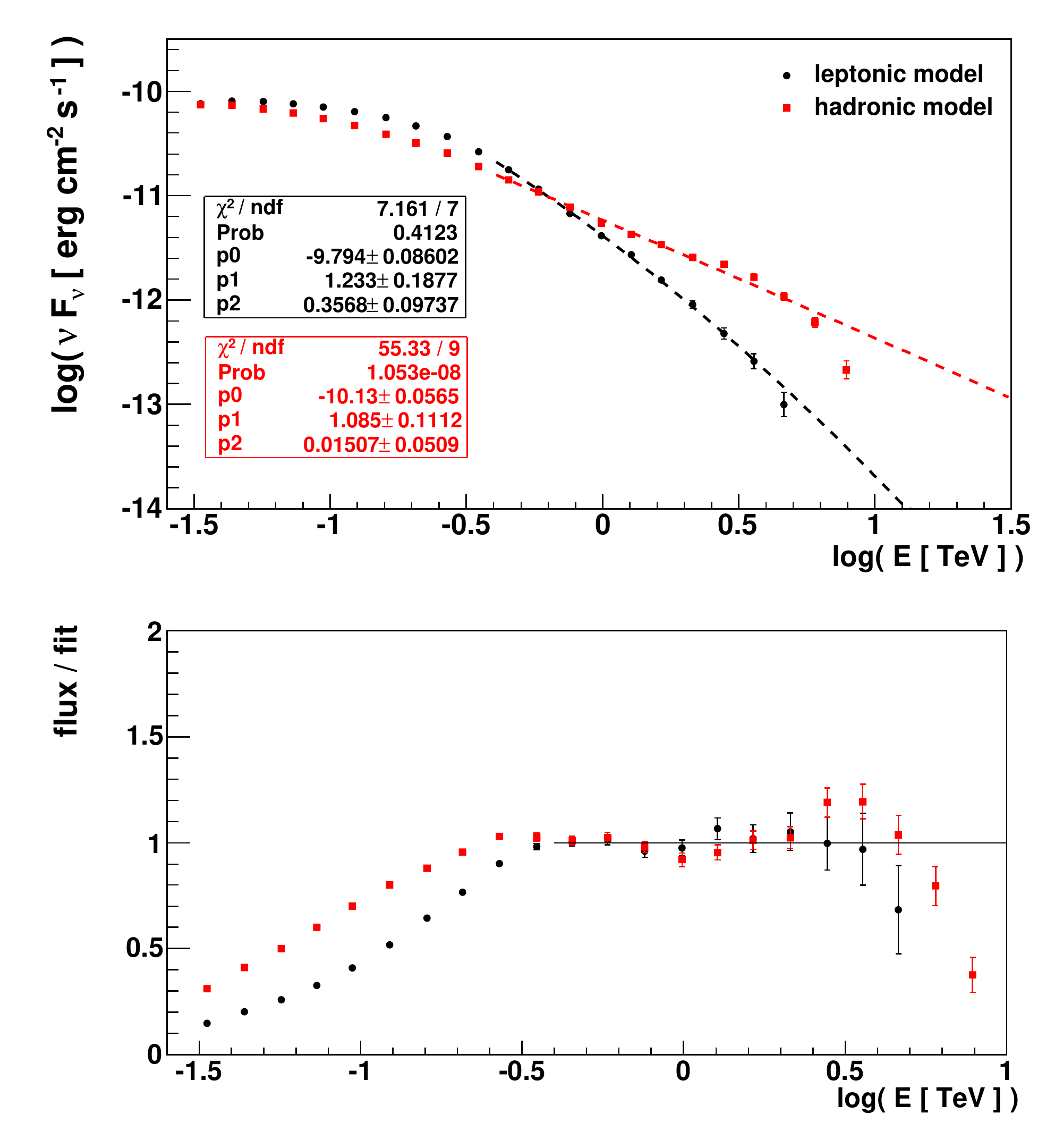}
%\vspace{4cm}
\caption{Example of a logparabolic fit to realisations of simulated spectra, upper panel, for an SSC model (black circles, upper fit parameters) and a 
hadronic model (red squares, lower fit parameters). The spectra correspond to a 50\,hr exposure time on the source PKS\,2155-304. The lower panel shows the ratio of the simulated fluxes over the fit function.
}
 \label{fig:sedfit}
\end{figure}

The same fit function applied to hadronic models over the same energy range usually results in worse reduced-$\chi^2$ values. An example for a logparabolic fit to a realisation of a 
hadronic model and of an SSC model is shown in Fig.~\ref{fig:sedfit}. 

It should be noted that a direct comparison between the simulated SSC and hadronic spectra, using for example a Kolmogorov-Smirnov test, risks being inconclusive. This is  because we are only interested in
comparing features in the spectral shape and not the absolute values of the spectral distributions between two models, which could be significantly different even between two SSC models that are acceptable
solutions for the given dataset.

All spectra simulated for the SSC and hadronic models for a given source were fitted with the same function and the values of the resulting reduced $\chi^2$ were recorded. 
The quality of the logparabolic fit was thus used to discriminate between spectra with SSC and with hadronic shapes.
The distributions of the reduced $\chi^2$ were characterised by their mean and standard deviation. As a simple ad hoc criterion, which can be refined in future studies, we consider that two models start being 
distinguishable if there is no overlap in their reduced $\chi^2$ distributions within one standard deviation from the mean values, meaning that\ if 
\begin{equation}
\left \langle \chi^2_{hadronic} \right \rangle - \sigma_{hadronic} \quad > \quad \left \langle \chi^2_{SSC} \right \rangle + \sigma_{SSC},
\label{equ:chi2}
\end{equation}the results in the following section do not change significantly when the mean and width of a Gaussian fit to the reduced $\chi^2$ distribution is used to characterise the different models,
instead of the mean and standard deviation of the distribution itself.

\subsection{Detectability as a function of exposure time}

The detectability of the ``cascade bump'' in the different hadronic models for PKS\,2155-304 and Mrk\,421 was tested for three different observation times with CTA: 20\,hr, 50\,hr, and 100\,hr. These are typical exposure times used with current IACT arrays on a single source when considering that the low-state flux can be summed
over several observation periods. Both blazars chosen for this study will be observed regularly with CTA, as calibration sources, but also within the AGN Key Science
Programme \citep{CTA2016}. 

The logparabolic function provides a good fit for two independent SSC models for each of the two sources. In the case of PKS\,2155-304, we compare an SSC model that includes the 
optical points against an alternative model that is not constrained by the optical emission, supposed in this case to stem from a different emission region, for example an extended jet. 
The reduced $\chi^2$ distributions for one hundred realisations of each of the two SSC models are in good agreement for 20\,hr and 50\,hr exposure times and still have a large overlap for 100\,hr.
The 1-$\sigma$ envelope of the resulting reduced $\chi^2$ distributions for fits above approximately$\ 400$\,GeV is below 1.3 for observations of 20\,hr or 50\,hr and below 2.4 for observations of 100\,hr at which point the shapes of the simulated SSC spectra do not always match the logparabolic function very well. 
 
In the case of Mrk\,421, we compare two SSC models with Doppler factors of $\delta = 30$ and $50$, with logparabolic fits above approximately $\ 600$ \,GeV. The 1-$\sigma$ envelope is below 1.5 for 20\,hr and 50\,hr exposure times and increases to 2.0 for 100\,hr.

\begin{figure}[htb!]
\centering
\includegraphics[width=\columnwidth]{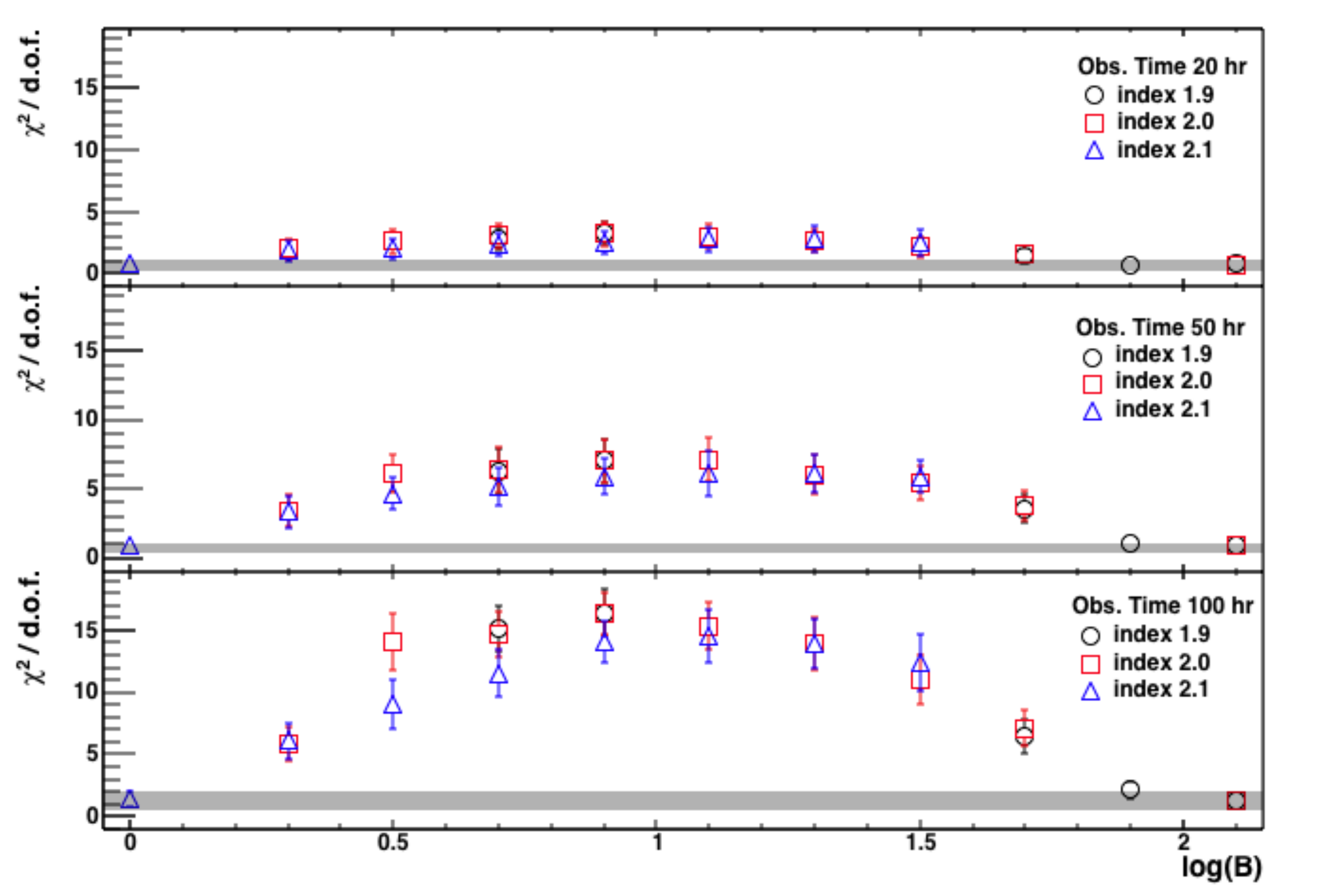}
%\vspace{4cm}
\caption{Mean values of the reduced $\chi^2$ distributions of logparabolic fits to one hundred realisations of each model for the SEDs of PKS\,2155-304. 
The abscissa shows the logarithm of the magnetic field strength to distinguish different hadronic models. The error bars show the standard deviation of the distributions. The energy range of 
the fit was chosen to provide a good reduced $\chi^2$ for two different SSC models. The grey band is the union of the 1-$\sigma$ envelopes of those SSC spectrum fits. }
 \label{fig:chi2_2155}
\end{figure}

\begin{figure}[htb!]
\centering
\includegraphics[width=\columnwidth]{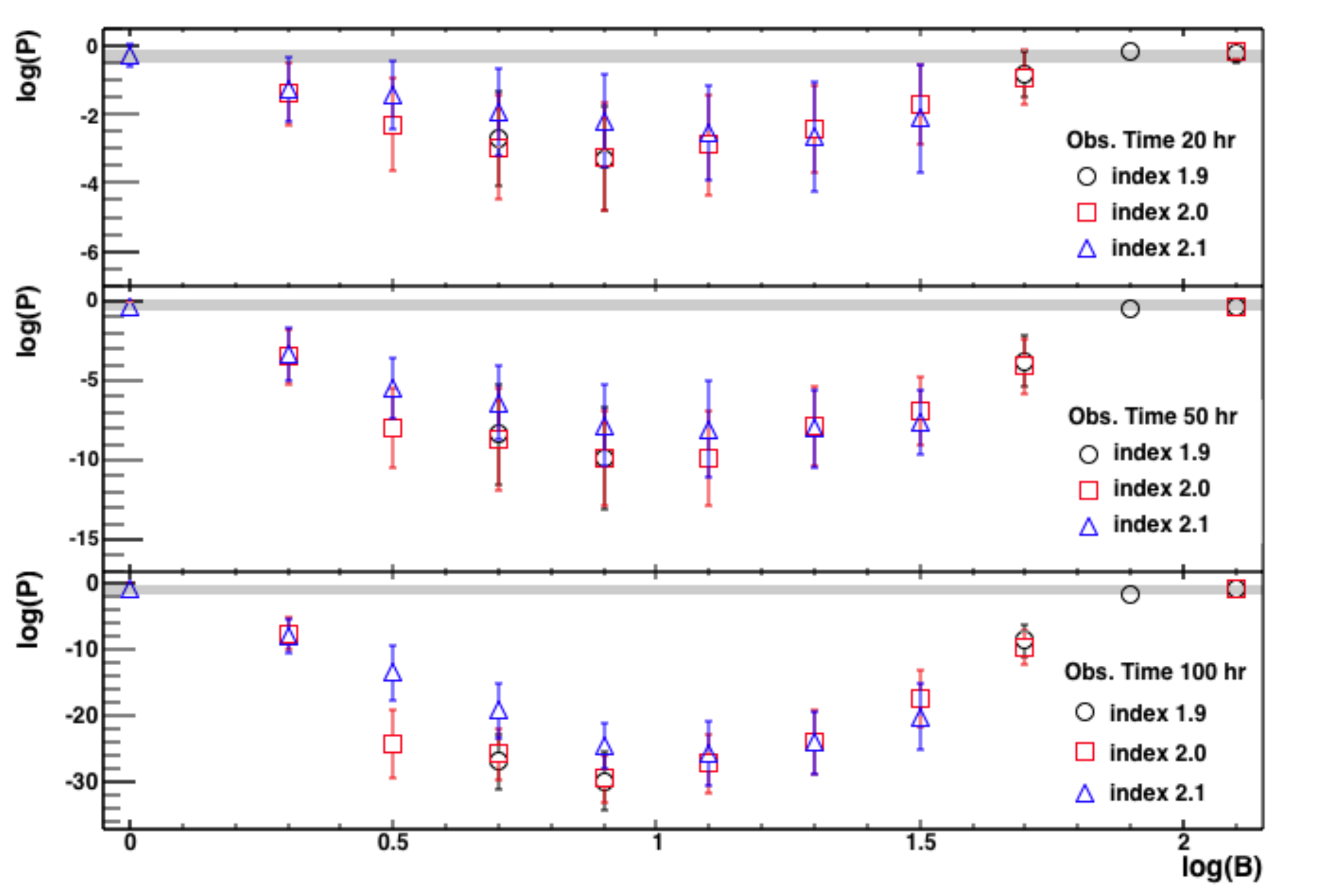}
%\vspace{4cm}
\caption{Mean logarithmic values of the probabilities for the same fits as Fig.~\ref{fig:chi2_2155} for PKS\,2155-304. 
The abscissa shows the logarithm of the magnetic field strength to distinguish different hadronic models. The error bars show the standard deviation of the distributions. The grey band is the union 
of the 1-$\sigma$ envelopes of the SSC spectrum fits. 
}
 \label{fig:prob_2155}
\end{figure}

\begin{figure}[htb!]
\centering
\includegraphics[width=\columnwidth]{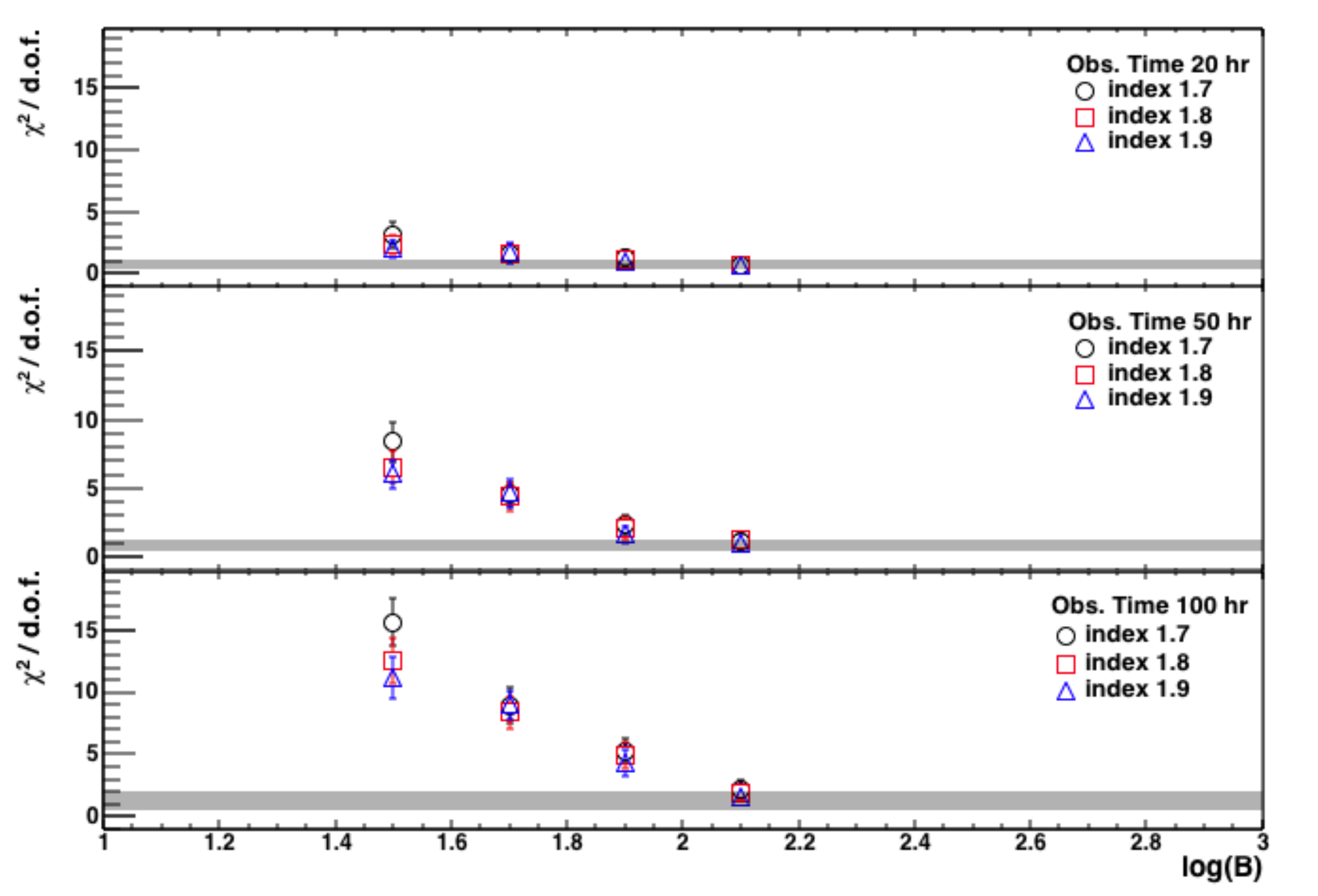}
%\vspace{4cm}
\caption{Mean values of the reduced $\chi^2$ distributions of logparabolic fits to one hundred realisations of each model for the SEDs of Mrk\,421.
The caption of Fig.~\ref{fig:chi2_2155} contains more details.
}
 \label{fig:chi2_421}
\end{figure}

\begin{figure}[htb!]
\centering
\includegraphics[width=\columnwidth]{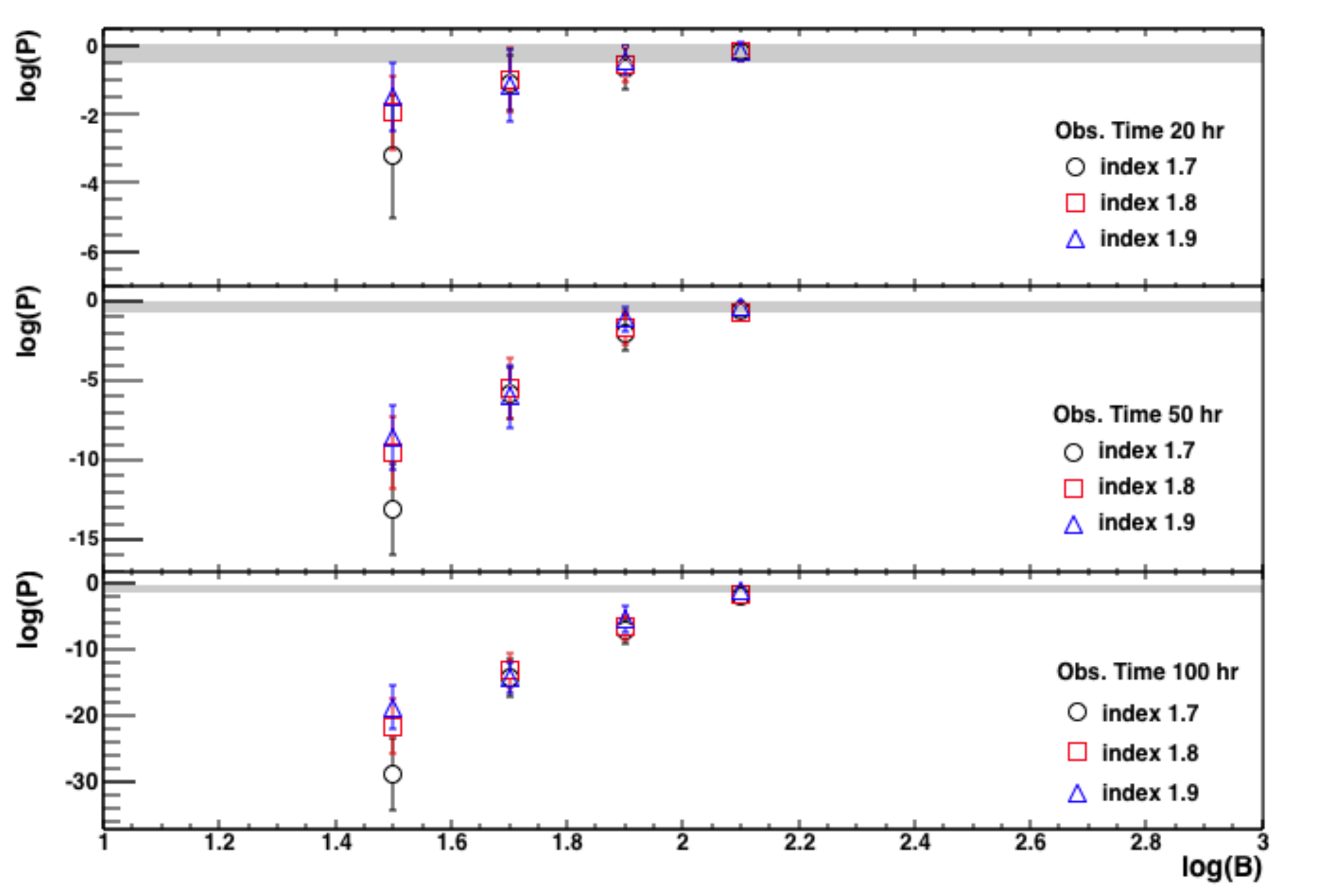}
%\vspace{4cm}
\caption{Mean logarithmic values of the probabilities for the same fits as Fig.~\ref{fig:chi2_421} for Mrk\,421. 
The abscissa shows the logarithm of the magnetic field strength to distinguish different hadronic models. The error bars show the standard deviation of the distributions. The grey band is the union 
of the 1-$\sigma$ envelopes of the SSC spectrum fits. 
}
 \label{fig:prob_421}
\end{figure}

When the same logparabolic fits are applied to the hadronic models, in most cases the reduced $\chi^2$ of the fit is different from that obtained with the SSC model by more than one
standard deviation for 50\,hr of observation time, and in several cases already by 20\,hr (cf. Figs. \ref{fig:chi2_2155} and~\ref{fig:chi2_421}). The difference between SSC and hadronic models becomes clearer as the exposure time increases.
When investing 100\,hr of observation time, the 
large majority of our hadronic models for PKS\,2155-304 and for Mrk\,421 are clearly distinguishable from the SSC models. Figures~\ref{fig:prob_2155} and \ref{fig:prob_421} show the 
corresponding fit probabilities (p-values) in terms of the mean and standard deviation for the distributions of the logarithmic probabilities. For 50\,hr of observation time, for example, most hadronic
models for PKS\,2155-304 have probabilities less than $1 \times 10^{-5}$, whereas the SSC models have probabilities greater than 0.1.

For PKS\,2155-304, the model with a spectral index of 2.1 and a very small magnetic field of 1\,G corresponds to a pure proton-synchrotron scenario, in which contributions from muon-synchrotron 
and cascades are very small and do not lead to significant features. This model is also the one with the largest jet power. On the other hand, models with very large magnetic 
fields of $B \gtrsim 70$\,G and dense emission regions also present a challenge, because the contribution from cascades and muon synchrotron are strongly absorbed at the highest energies, 
leading to very steep and smooth spectra without detectable features. These models are also disfavoured on physical grounds, because the radius of the emission region is very close to
the Schwarzschild radius of the source. 

Pure proton-synchrotron models for Mrk\,421 do not exist, as discussed above. Models with very large magnetic fields above $\gtrsim 100$\,G suffer from the same problems for detection
of the ``cascade bump'' as in the case of PKS\,2155-304. They are again disfavoured due to the smallness of the emission region.

\subsection{Detectability as a function of flux level and redshift}

The two sources selected for this case study are particularly bright HBLs, even in their low states. To test how the detectability of the ``cascade bump'' depends
on the absolute flux level, we scaled the fluxes of an SSC model and of a well distinguishable hadronic model by factors between 0.5 and 0.1 for each of the sources. 

For a model for PKS\,2155-304 with a ``cascade bump'' detectable in 20\,hr exposure time, at least 50\,hr are needed for a detection of the signature when the flux is reduced to 40\%
of its initial value. For a reduction to 30\%, more than 100\,hr are needed.  
When selecting a well-distinguishable model for Mrk\,421, in which the ``cascade bump'' is detected in 20\,hr, and reducing its flux to 40\% of its initial value, again the signature is only
detectable in 50\,hr of exposure time or more. Around 100\,hr are needed for a flux reduced to 30\%. 
Thus only sources with relatively high flux levels during their low states may be considered for searches of this spectral signature with CTA. 

The HBLs PKS\,2155-304 and Mrk\,421 are nearby sources with redshifts of $z=0.116$ and $z=0.031,$ respectively. To test the dependence of the ``cascade bump'' 
signature on the source distance, we artificially redshifted an SSC model and a hadronic model with well-detectable ``cascade bumps'' to values of $z=0.15, 0.2$ for PKS\,2155-304 
and to $z=0.1, 0.15, 0.2$ for Mrk\,421, and evaluated the fit results for these shifted fluxes. Unsurprisingly, the source redshift is a very important factor for the detectability of the 
hadronic signature. 
For PKS\,2155-304, moving the redshift to 0.15 increases the minimum exposure time for a detection of the signature from 20\,hr to 50\,hr. 
In the case of Mrk\,421, putting the source at a redshift of 0.1 already increases the required observation time from 20\,hr to 100\,hr. 
Increasing the source distance leads not only to a reduction of the observed flux, but the stronger absorption on the EBL rapidly washes out the spectral structure at the highest 
energies. 

It should be stressed that the detectability of the hadronic signature does not only depend on the flux level and redshift of the source, but also on its intrinsic spectral shape. 
The above estimates based on our two sources are meant to provide some general trends, although actual numbers might vary significantly for other sources. Sources
with large high-energy bumps and hard spectra in the TeV range might prove especially more accessible, even at larger redshifts and for lower fluxes.

%______________________________________________________________

\section{Discussion}
\label{sec:discussion}

\subsection{Impact of our simplifying assumptions}

%% Influence of model assumptions
To arrive at the selection of models for the two HBL sources, several physically-motivated simplifying assumptions were made to reduce the space
of the model parameters. They are discussed in the following paragraphs.
\newline

% co-acceleration 
Assuming an identical index for the injection spectra of protons and primary electrons significantly reduces the acceptable values of the slope for both spectra
and excludes pure proton-synchrotron scenarios for the SED of Mrk\,421, given the constraints from the optical data. It also restricts the minimum jet power for
the solutions for PKS\,2155-304, by excluding solutions with $n_1 < 1.9$. This assumption ignores the complexity of the actual acceleration mechanisms that may
be at play inside the source and might lead to different particle spectra for protons and leptons, which are subject to diffusion on different length scales. In the same
sense, the assumption of a pure synchrotron cooling break in the electron spectrum might be an over-simplification, as is generally assumed in one-zone SSC models.
Nevertheless, the parameter space of our solutions should not change significantly for particle indices before cooling that are close to the value of 
around two, which is expected for Fermi-like acceleration. The stronger constraint comes from the fact that primary electrons are completely cooled whereas protons are not, resulting in a significantly steeper stationary spectrum for the electrons. This constraint is a solid consequence of the assumed co-acceleration scenario.
\newline

% maximum proton energy 
The maximum proton energy is determined by an equilibrium between acceleration and loss timescales, and thus in our framework only by the magnetic field $B$ 
and source radius $R$. A more sophisticated treatment of particle acceleration, diffusion, and energy loss may lead to a wider range of maximum possible proton energies for a given 
parametrisation of the source conditions. More diverse combinations between the proton-synchrotron component and the muon-synchrotron and cascade components might occur.
As an order-of-magnitude estimate, however, the current approach is sufficient.
\newline

% Doppler factor
We chose to fix the value of the bulk Doppler factor to $\delta = 30$, but the present study may be extended to a range of values. We  verified the effect on the resulting 
models for a given $B$ when varying the Doppler factor to $\delta = 20$ and to $\delta = 40$ while keeping the peak positions and peak fluxes of the modelled SEDs constant. Increasing 
$\delta$ by a factor $f_\delta$ leads to an increase in the observed energy flux level ``$\nu F_{\nu}$''that is proportional to $f_\delta^4$. At the same time, the frequencies of the electron-synchrotron
and proton-synchrotron peak positions increase by a factor $f_\delta$. A reduction in $R$, which leads to a reduction in $\gamma_{p; max}$ given our constraints, and an adjustment of the 
maximum electrons Lorentz factor $\gamma_{e;max}$, are necessary to keep the positions of the peaks fixed at their initial frequencies. 

Reducing $R$ leads to a linear reduction of $\gamma_{p;max}$ 
and thus to a quadratic reduction in the proton-synchrotron peak frequency. When compensating for the shift in the peak frequency by reducing $R$ by a factor $f_\delta^{-1/2}$, the flux
level decreases by a factor $f_\delta^{-3/2}$. It is thus still a factor $f_\delta^{5/2}$ higher than before the change in $\delta$. This remaining flux increase needs to be offset by reducing the particle densities by a factor $f_\delta^{-5/2}$. 
For the proton population, this reduced particle density leads to a smaller contribution from the muon-synchrotron and cascade components. Inversely, these components become more significant 
when lowering the value of $\delta$ while keeping the flux level and peak frequencies seen by the observer fixed. 
The low-energy turnover of the modelled SED in the optical range is kept fixed by decreasing the value of $\gamma_{min, e}$ when increasing $\delta$, which can lead to more standard values 
for certain models. 

The choice of $\delta$ has also an effect on the jet power. 
Given the small-angle approximation (Eq.~\ref{equ:delta}), the jet power is directly proportional to $\delta^2$, as can be seen from Eq.~\ref{equ:Ljet}. When neglecting the small 
contributions from the electron population and from the radiation fields, the jet power takes the following form in our approximation:
\begin{equation}
L_{j} \propto R^2  \delta^2 ( c_1 B^2 + c_2 K_p \gamma_{p,max} ),
\end{equation} 
with $c_1$ and $c_2$ constant. For a given $B$, when keeping flux level and peak positions fixed as discussed above, the jet power $L_{j}$ increases linearly with $\delta$ if the jet is dominated by the magnetic energy density $u_B$, due to the impact of 
the factors $R^2$ and $\delta^2$. For a jet dominated by the kinetic energy density of the proton population $u_p$ and for a proton 
spectrum with index $n_1 \sim 2$, the jet power changes as $L_{j} \propto f_\delta^{-1}$ as a consequence of the additional changes in $K_p$ and $\gamma_{p,max}$. 

For a typical model for PKS\,2155-304, initially in the $u_p$ dominated regime, increasing $\delta$ from 20 to 40 while adjusting flux level and synchrotron peak positions leads to a decrease in 
$L_{j}$ by about 15\% and a drop in the flux level of the muon-synchrotron and the cascade components of roughly a factor of two in $\nu F_{\nu}$. For values of $\delta$ in the usual range assumed
for blazar emission models, the results presented here should thus not change fundamentally. 
\newline

% Limit on Ljet
To constrain the total jet power for our solutions, we compare it to the Eddington luminosity as a natural scale defined by the mass of the central black hole. This is clearly meant as
an order-of-magnitude estimate for the maximum acceptable jet power instead of a strict limit. A different approach would be to consider the disk luminosity as a natural scale to compare to the jet power.
However, the disk luminosity depends on the radiative efficiency and thus on the physical conditions of the accretion disk, which is in general not observationally accessible, especially for BL Lac objects, 
and for which a multitude of models exist. In blazars it is generally seen that the jet power is not limited by the disk luminosity, even for the ``standard'' leptonic models \citep[e.g.][]{Celotti2008,Ghisellini2010}.
In the case of Mrk\,421, one can estimate the disk luminosity from the detectable emission of the broad line region. \citet{Sbarrato2012} provide a value of $0.5 \times 10^{42}$ erg s$^{-1}$ for the luminosity of
the broad line region for this source, translating to roughly $10^{43}$ erg s$^{-1}$ for the disk luminosity \citep[see e.g.][]{Ghisellini2013}, meaning\ about $5 \times 10^{-4} L_{edd, Mrk421}$.    
For a typical jet power of ~0.03 $L_{edd, Mrk421}$ (cf. Fig.~\ref{fig:Leta_hbl}), the ratio of jet power over disk luminosity would be less than 100. Although these values still require a relatively inefficient accretion
process, the hadronic interpretation for HBLs does not suffer from the extreme energetic requirements derived by \citet{Zdziarski2015} for a set of hadronic models, which they applied to FSRQs and low- and intermediate-frequency-peaked  BL Lac objects.
\newline

% EBL model
The study presented here was carried out using the model of the EBL by~\citet{Franceschini2008}, which is frequently used for the interpretation of VHE 
data and has been shown to be consistent with a range of observed sources~\citep{Biteau2015}.
Although the dependency of our results on the chosen model is not very strong, given that the sources under study are not very distant, we note that it was shown that the ``cascade bump'' signature becomes less well defined when assuming a higher opacity of the EBL~\citep{Zech2013}. On the other hand, models with a more transparent EBL favour a detection of the signature. 
\newline

% performance curves
Any change in the CTA performance curves also has an impact on our results. When studying for example the detectability of the Mrk\,421 models with the southern performance 
files, one clearly sees a significant improvement due to the better coverage at the highest energies that comes from a larger number of medium-sized telescopes and the additional small-size
telescopes not foreseen for the northern site. Models with undetectable ``cascade bumps'' for 100\,hr exposures with the northern performance curves show clear detectability when using the 
southern performance curves instead. 
The public performance curves used here represent preliminary expectations of the instrumental response functions for preliminary array layouts. Once the actual performance of the final 
CTA arrays is established the results might thus differ from our current predictions. 
\newline

%% issues with data analysis

% data sets with variable fluxes
The long exposure times of up to 100\,hr for the study that we are proposing require multiple separate pointings during the visibility periods of the sources. These pointings will be taken over 
several months or even years and will very likely contain data from different flux states. When extracting a spectrum from these data, one will have to be very careful with the treatment of flux 
and spectral variability. As discussed by \citet{Abdo11c} in the case of a long-term campaign on Mrk\,501, it will be important to exclude periods of flaring activity to build the spectrum from 
data in a low, persistent flux level. Based on past observations of HBLs with the current generation of Cherenkov telescopes, it seems that significant spectral variability in the VHE band occurs 
only during flaring episodes. 
Although some flux variability was detected in the long-term low-state VHE lightcurve of Mrk\,421 \citep{Aleksic2015}, significant spectral variability was seen neither in the Fermi-LAT data from 0.1 to 400 GeV, 
nor in the MAGIC data over the 4.5 months of observations \citep{Abdo11c}. Long-term VHE data from PKS\,2155-304 from H.E.S.S. also indicate only very small spectral variations during 
non-flaring periods \citep{Abramowski2010a}, which has been confirmed with more recent data \citep{Chevalier2015}. 
The risk of introducing spectral features due to a combination of data from different periods should thus be small if the datasets are carefully selected and combined.
\newline

% analysis method
A final but important point concerns the method applied in this study to characterise spectral features in the simulated SEDs. The fit, by $\chi^2$ minimisation, of a simple function
to the simulated spectrum points represents a robust and rapid first approach, but more sophisticated techniques will certainly be applied to analyse CTA spectra. For the spectral analysis
of current IACT data, maximum likelihood methods applied to forward folding of different spectral shapes have become a standard technique. The direct analysis of VHE data using
realistic model shapes as underlying hypothesis, instead of power laws and logparabola, possibly through the use of libraries of modelled SEDs over a range of acceptable parameters, 
would be a next step and could be built on current techniques used for the analysis of X-ray data. Ideally, one would apply such methods to a full and simultaneous MWL dataset. It can thus 
be realistically expected that the search for spectral features in actual CTA data will be more effective than the simple approach proposed here for a first evaluation of the detectability of
the ``cascade bump''.

\subsection{Additional constraints from variability}

%% Additional constraints from the variability time scale

Even in the relatively low states the present study is focusing on, some flux variability was found during the MWL campaigns of Mrk\,421 and PKS\,2155-304. 
For the latter source, \citet{Aharonian09} report a small overall variability, characterised by fractional rms, at the 30\% level in the MWL light curves of the campaign that covered twelve nights.
The optical, X-ray, and VHE light curves show flux-doubling on timescales of days.
For Mrk\,421, \citet{Abdo11c} report only low flux variability during the 2009 campaign that covered 4.5 months.
A more in-depth study by \citet{Aleksic2015} finds significant variability at all wavelengths, which is highest in X-rays, with variations that are typically smaller than a 
factor of two. The authors find variability in the X-ray and VHE bands on day timescales, and in the optical and UV band on weekly or longer timescales. 

When requiring that the size of the emission region be sufficiently small for the assumed Doppler factor $\delta = 30$, 
to allow for variability of the order of one day one can use the usual light-crossing time argument to arrive at a limiting radius:
\begin{equation}
R \le  c \:  t_{\rm var} \:  \delta / (1 +  z).
\label{equ:var}
\end{equation}
This translates into a constraint on the source extension of $R \lesssim 7 \times 10^{16}$ cm for both sources. 
Although this kind of an additional constraint would be of no consequence for our selection of models for Mrk\,421, which already present small emission regions, a few models
for PKS\,2155-304 would be excluded. This concerns the models with the most important jet powers. 
\newline

% correlations PKS 2155
Apart from variability timescales, correlated behaviour between different wavelength bands is another observable characteristic that might help distinguish leptonic from
hadronic scenarios.
\citet{Aharonian09} find a correlation between the optical and VHE bands for PKS\,2155-304 on timescales of days, but no correlation between optical and {\it Fermi}-LAT data. Although our stationary 
model does not permit us to study variability, the scenarios that ascribe the {\it Fermi}-LAT flux to proton-synchrotron emission and the VHE flux mostly to muon-synchrotron 
emission might explain a difference in the variability patterns between these two domains, if confirmed. Muon-synchrotron emission is a consequence of proton-photon 
interactions and thus subject to correlated variations with the target photon field, meaning the electron-synchrotron emission responsible for the optical and X-ray spectrum 
in our interpretation, although time-lags between the high- and low-energy components would need to be evaluated for a given set of source parameters.

% correlations Mrk 421
According to \citet{Aleksic2015}, Mrk\,421 shows a positive correlation between the VHE and X-ray fluxes with zero time lag during the 2009 MWL campaign. 
The authors claim that as a consequence, the direct high-energy correlation supports leptonic models over hadronic ones. 
In general, correlated behaviour between the different energy bands in the lepto-hadronic model can be accounted for in a co-acceleration or co-injection scenario and via the proton-photon interactions. However,  the detailed behaviour can be complex and needs to be studied with time-dependent models~\citep[e.g.][]{Mastichiadis2013,Diltz2015}.

The authors also report an anticorrelation between the optical and UV band on the one hand, and the X-ray band on the other, whereas they report that there does not seem to be a correlation with the radio band.
If the emission in the optical and UV band and the X-ray emission come from different emission regions and the anticorrelation is a coincidence, we would have less constraints on our model on 
the $n_1$ parameter, but might require higher $\gamma_{e;min}$ values because the optical and UV data would still present upper limits.

\subsection{Non-uniqueness of the ``cascade bump'' signature}

%% False positives:
An obvious limitation of the present study is the fact that the hadronic models were only compared to the most standard one-zone SSC models. 
More complex scenarios might well produce comparable spectral hardening in the VHE spectra, but the higher degree of complexity that is generally accompanied by a 
larger number of free parameters would need to be well justified against the basic hadronic scenario proposed here.
A potentially similar spectral feature might arise in the scenarios discussed in the following paragraphs.
\newline

% second-order SSC
Emission from second-order SSC, meaning a second upscattering of a fraction of high-energy photons on the relativistic electrons in a standard SSC model, could in 
principle lead to an additional spectral component at high energies. In the case of HBLs however, this component is in general completely negligible compared 
to first-order SSC. Inverse-Compton upscattering of the bulk of the first-order SSC photons on the electron population around the break energy would take place
deeply within the Klein-Nishina regime.

%electron pile-up
A different effect that can lead to spectral hardening in the radiative emission from electrons is described by \citet{Moderski2005}. In sources with strong external 
photon fields, where electron cooling is dominated by the Inverse Compton process, a hardening or pile-up in the steady-state electron spectrum can form for energies
where cooling becomes ineffecient due to Klein-Nishina effects. This would result in an upturn or ''bump'' in the synchrotron spectrum. However, the effect is expected to be negligible 
for the high-energy Inverse Compton spectrum, where the hardening of the steady-state electrons is in competition with a softening of the emission due to the same Klein-Nishina 
effects. We would thus not expect an appreciable feature from this effect, particularly for HBLs, for which external photon fields can usually be neglected and synchrotron cooling 
dominates over Inverse Compton cooling for the electron population.

% multi-zone SSC
Multi-zone SSC models could in principle generate spectral hardening at TeV energies. One could image emission from a first zone dominating the SED
over almost the entire observable energy range, whereas SSC emission from a second, more compact zone would only only appear at VHE energies. Although the Klein-Nishina
effect would limit the high-energy reach of a second SSC component, this kind of a scenario could still be possible and might be difficult to distinguish from the hadronic 
``cascade bump''. Information on spectrally resolved variability might be needed to rule against or in favour of each scenario.

% EC
An additional spectral component in leptonic models might also arise from external Compton emission caused by the upscattering of photons from external fields, such as  broad
line region, disk emission, dust torus, stellar radiation field...etc. For HBLs, this kind of a component is not expected to contribute significantly to the SED due to the weakness
or absence of the non-detectable external photon fields. If such a component was present, potential VHE features would also be limited by the Klein-Nishina effect. 

% absorption features -> 0447 paper
Spectral hardening in the VHE band might arise for scenarios in which $\gamma$-rays in a certain energy range are absorbed on external photon fields in the source, for example from the 
broad line region~\citep{Senturk2013,Poutanen2010}, accretion disk, or torus~\citep{Donea2003}, and where the flux recovers at higher energies. However, these kinds of absorption features would
be expected to occur instead in the {\it Fermi}-LAT band with a flux recovery in the low VHE band. Besides, as discussed above, although external photon fields are very likely to play an important 
role for flat-spectrum radio quasars, they are usually considered negligible for HBLs~\citep[see also the discussion by][]{HESS2013}.

% external cascade models
Spectral hardening in the VHE range is also predicted by scenarios in which ultra-energetic photons or protons escape from the source and trigger particle cascades by 
interactions with the EBL and cosmic microwave background (CMB)~\citep[e.g.][]{Aharonian02a,Essey2011,Murase2012,Aharonian2013b,Taylor2011}. However, this kind of a component would be expected for sources with relatively 
high redshifts. It would also not exhibit the same spectral shape and temporal behaviour.

\subsection{Prospects for current and future Imaging Air Cherenkov telescope arrays}

%% Constraints from currently available data
The HBL\ PKS\,2155-304 is frequently observed with the H.E.S.S.\ IACT array. According to~\citet{Chevalier2015}, about 260\,hr of data were taken between 2004 and 2012, excluding the very 
luminous flares seen in 2006. Even so, the long-term light curve shows flux variations by a factor of a few. When assuming, very roughly, a ten times better sensitivity for CTA with respect
to H.E.S.S., the currently available VHE dataset on this source would correspond to maybe 40\,hr of an equivalent CTA exposure. This would  include different flux states and does not account for the
increasing gap between the current IACT sensitivities and the expected CTA performance at energies above a few TeV, where the additional performance of the small-size telescopes in 
the southern array becomes important~\citep{Bernlohr2013}. It would in any case be interesting to study the complete set of available data to probe the most prominent ``cascade bump''
features. A similarly rich dataset might also be available on Mrk\,421 from observations with the MAGIC and VERITAS IACTs. For this blazar, 840\,hr of TeV data had been collected with 
the Whipple IACT \citep{Acciari2014}, albeit with a lower sensitivity than current arrays.

%% Link with CTA KSP etc. , 
The two sources we have studied here are foreseen to be observed frequently with CTA. Given their brightness in the VHE range, they will be used as calibration sources and they are 
also prominently included in the proposed AGN Key Science Project~\citep{CTA2016}. Their long-term light curve will be studied with regular observations over an important fraction of
the lifetime of CTA. Deep exposures over shorter time ranges are also foreseen, to help derive high-quality spectra in a single state. Even if observation times of up to 100\,hr will be difficult to
achieve over a short time range, observations from a few seasons could be summed, as long as  the sources show sufficiently similar flux states. 

In general, nearby bright HBLs are expected to be the best candidates to search for the ``cascade bump'' signature. According to the TeVCAT 
catalogue (see footnote on page 2) there are currently twenty-two known VHE emitting HBLs with $z<0.15$ that 
could be considered for such a search, but only the sources with high fluxes during low states are promising targets. Apart from PKS\,2155-304 and
Mrk\,421, about twelve known HBLs seem promising targets when excluding UHBLs and sources with very faint fluxes, meaning below 2\% of the Crab flux above 200\,GeV, 
in low or average states: Mrk\,501, PKS 2005-489, RGB\,J0152+017, 1ES\,1959+650, 1ES\,2344+514, 1ES\,1727+502, IC\,310, 1ES\,0806+524, 1ES\,1215+303, 
B3\,2247+381, 1ES\,1741+196, and PKS\,1440-389 although it has an uncertain redshift.

Additional target candidates could be searched amongst the eight VHE HBLs currently known without redshifts and amongst a handful of nearby intermedium-frequency peaked BL Lac objects (IBL).
However, in the latter, external photon fields might no longer be negligible, which would further complicate the search for hadronic signatures.

In the case of a non-detection of the suggested spectral hardening with CTA after an accumulated observation time of about 100\,hr, hadronic interpretations of the SEDs of the sources in 
question would become strongly constrained. The most feasible models, when considering the required source extension and energetics, would be rejected. This would at least permit us to 
strongly reduce the available parameter space for one-zone hadronic solutions. Combined with future observational constraints on the variability timescale during the low state, one-zone hadronic solutions might be ruled out entirely. How such an assessment would be modified when loosening the constraints on co-acceleration and $\gamma_{p,max}$ will need to be investigated further.

\subsection{Ultra-high-energy cosmic rays and high-energy neutrinos}

% energy limit for protons in our models ? 
The maximum energy of the proton population is constrained by a comparison of the assumed acceleration timescale and the shortest cooling timescale, as explained in Sect.~\ref{sec:model}. 
For Mrk\,421, we find maximum proton Lorentz factors up to $\gamma_{p,max} \sim 1 \times 10^9$ for the models presented here that correspond to an energy of about $10^{18}$\,eV. 
Proton-synchrotron solutions for PKS\,2155-304 with large extensions of the emission region around $10^{17}$\,cm have $\gamma_{p,max}$ values up to about $10^{10}$, meaning proton energies up
to about $10^{19}$\,eV. When accounting for Doppler boosting of the particle energies, certain solutions permit us thus to reach maximum proton energies close to what is required to account for the 
most energetic UHECRs, based on our simplistic description of the acceleration and loss timescales. In addition, a rapidly decreasing fraction of protons with Lorentz factors above $\gamma_{p,max}$ 
is expected from the exponential tail of the particle distribution. The most energetic UHECRs might however be nuclei \citep[e.g.][]{Abbasi2015a}, which we do not consider in our code, instead of protons.

% expected neutrino spectra 
The use of the Monte Carlo code SOPHIA allows us to directly extract the neutrino spectra for a given model, which only needs to be transformed to the observer frame on Earth.
As detailed in \citet{Cerruti2015}, we carefully account for the radiative cooling of muons before extracting the neutrino spectra resulting from muon decay. The expected 
neutrino flux is directly related to the importance of proton-photon interactions inside the source, when ignoring potential neutrino production through neutron decay and UHECR interactions outside 
of the source. Scanning the parameter space for a given source instead of proposing only a single solution makes it possible to estimate a range of possible neutrino spectra for the
source within our lepto-hadronic framework. 

\begin{figure}[htb!]
\centering
\includegraphics[width=\columnwidth]{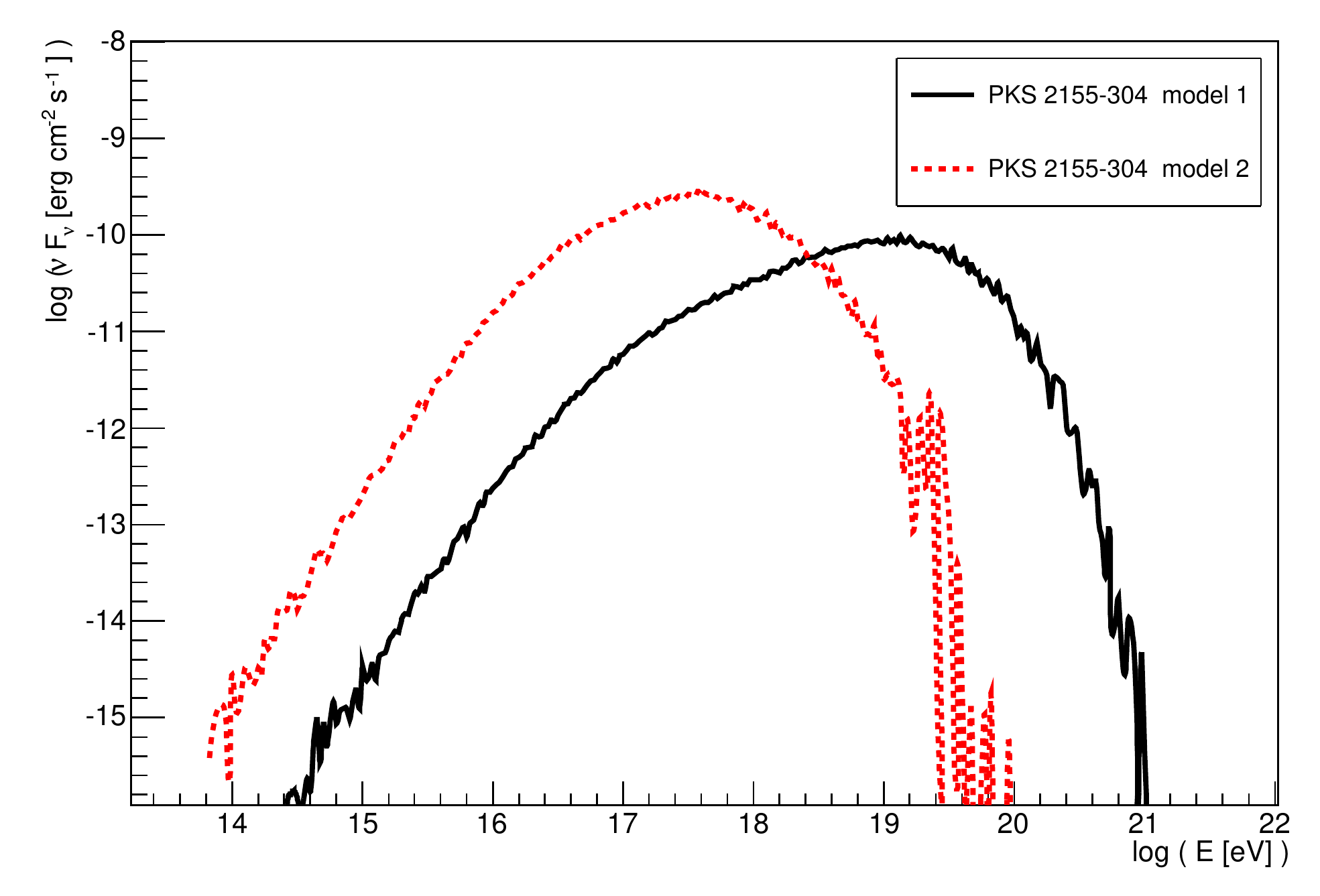}
%\vspace{4cm}
\caption{Expected neutrino spectra on Earth (all flavours combined) for two hadronic models for PKS\,2155-304. `Model 1' has a small $\log{B [G]}=0.3$, whereas 
`Model 2' has a much larger $\log{B [G]}=1.9$.
}
 \label{fig:neu2155}
\end{figure}

\begin{figure}[htb!]
\centering
\includegraphics[width=\columnwidth]{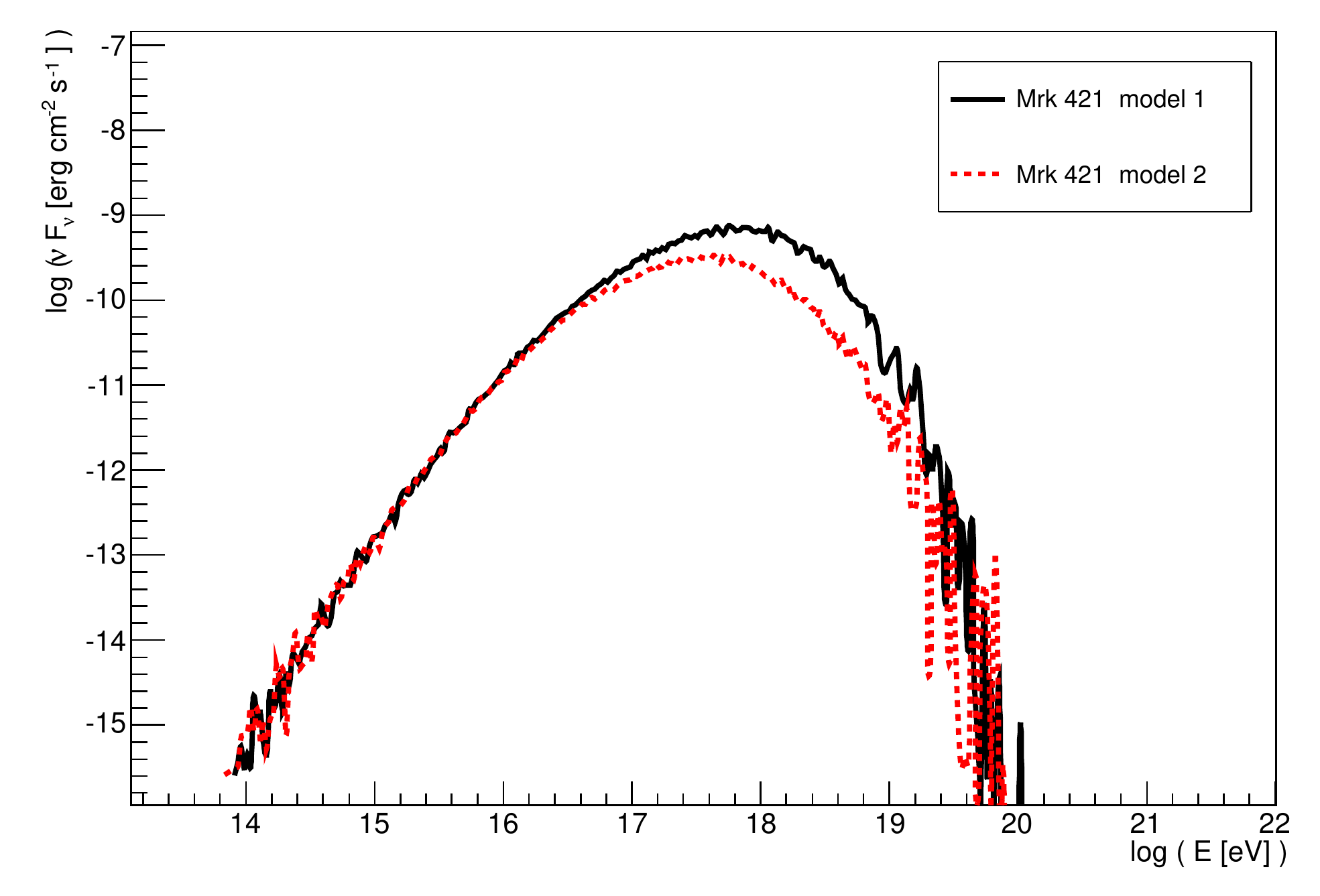}
%\vspace{4cm}
\caption{Expected neutrino spectra on Earth (all flavours combined) for two hadronic models for Mrk\,421. `Model 1' has a $\log{B [G]}=1.5$, whereas 
`Model 2' has a $\log{B [G]}=2.1$.
}
 \label{fig:neu421}
\end{figure}

Figures~\ref{fig:neu2155} and~\ref{fig:neu421} show two solutions each for the two HBLs under study, for models with different magnetic field strengths. Although the relatively small parameter space of hadronic 
solutions for Mrk\,421 is reflected in a small difference between the expected neutrino spectra for this source, the spread is much more significant in the case of 
PKS\,2155-304. For the latter, solutions close to the pure proton-synchrotron scenario lead to a low flux with a high energy peak, whereas solutions with a significant contribution from muon-synchrotron 
and cascade emission result in higher neutrino fluxes with lower peak energies. 

Contrary to the lepto-hadronic models discussed by~\citet{Petropoulou2015}, which have a small $\gamma_{p,max}$, large jet power, and are dominated by proton-photon interactions,
the models selected following our criteria lead to neutrino spectra that peak at higher energies. For the sources studied here, the expected neutrino emission seems out of reach for the IceCube
telescope \citep[cf.][]{Aartsen2014} when considering low $\gamma$-ray flux states. Flaring activity, if related to hadronic processes, can significantly increase the neutrino production, as discussed by \citet{Petropoulou2016}. Next-generation instruments with a higher energy reach, like ARA \citep{ARA2015} or GRAND \citep{Martineau2016}, might have a better chance of reaching the required high-energy sensitivity for a detection in this scenario. A more in-depth study of this topic will be treated in a dedicated publication.

%______________________________________________________________

\section{Conclusions}
\label{sec:conclusions}

We explore the parameter space of the one-zone hadronic blazar emission model for two bright VHE HBLs, PKS\,2155-304 and Mrk\,421, during low emission states. 
Satisfactory realisations of the model, where the high-energy emission is interpreted as a combination of proton-synchrotron, muon-synchrotron and proton-photon induced 
cascade emission, can be found for a range of parameters. The muon-synchrotron and cascade emission become more prominent as the density of the emission
region is increased while its size is reduced. The TeV spectrum can be dominated by proton-synchrotron or muon-synchrotron emission, depending on the chosen
solution.

The mixed lepto-hadronic models found for UHBLs~\citepalias{Cerruti2015} for magnetic fields of strengths intermediate between those for typical proton-synchrotron and 
SSC models do no provide satisfactory solutions for the HBLs under study, given our choice of simplifying assumptions about the particle populations and on the acceptable 
jet power. Muon-synchrotron emission, on the other hand, cannot dominate the TeV spectrum for UHBLs. Thus HBLs and UHBLs seem to populate distinct parameter spaces 
of the lepto-hadronic one-zone model.

All the hadronic models found for the two sources under study show a hardening in the multi-TeV spectrum at some level, due to the emission from synchrotron-pair cascades that are induced 
by proton-photon interactions. This adds to the dominant proton- or muon-synchrotron emission at these energies. Using a simple logparabolic fit to distinguish between the expected 
spectral shapes from pure SSC and from hadronic emission, we show that this characteristic ``cascade bump'' should be detectable for most models with CTA within 50\,hr, or in a few 
cases within 100\,hr, of observation time. This is especially the case for those models that present preferred solutions due to low jet powers and plausible extensions of the emission region.
The faint spectral feature is only expected to be detectable in nearby HBLs, of redshifts smaller than  approximately$\ 0.15$ and with flux levels at least a few tenths of those of the sources 
under study. We identify of the order of ten known TeV blazars that might be considered for future searches for such a signature.
Although the ``cascade bump'' permits the distinction between a standard one-zone SSC model and a simple one-zone hadronic model, confusion with signatures from more complex SSC 
scenarios has not been studied here and cannot be excluded a priori.

\begin{acknowledgements}
The authors wish to acknowledge discussions with C. Boisson, H. Sol, S. Inoue, D. Pelat, G. Henri, A. Reimer, J.-P. Ernenwein, and L. Stawarz that greatly helped improve this work.\\
This paper has gone through internal review by the CTA Consortium. 
\end{acknowledgements}

%-------------------------------------------------------------------

%\begin{thebibliography}{}
%
%
%\end{thebibliography}

\bibliographystyle{aa}
\bibliography{refAZ}

\clearpage

\appendix

\section{Examples of relevant timescales for models for PKS\,2155-304}
\label{app:2155times}
The following figures show the relevant acceleration and cooling timescales for the different particle populations considered in the code as a function of the 
Lorentz factor of the particles. The vertical dashed line labelled ``p gyroradius'' indicates the proton Lorentz factor $\gamma_{p,max}$ that corresponds to a 
gyro-radius of the size of the radius of the emission region. The results shown here correspond to the exemplary models for PKS\,2155-304 shown in Fig.~\ref{fig:sed_2155}.

  \begin{figure}[h]
   \centering
               \includegraphics[width=\columnwidth]{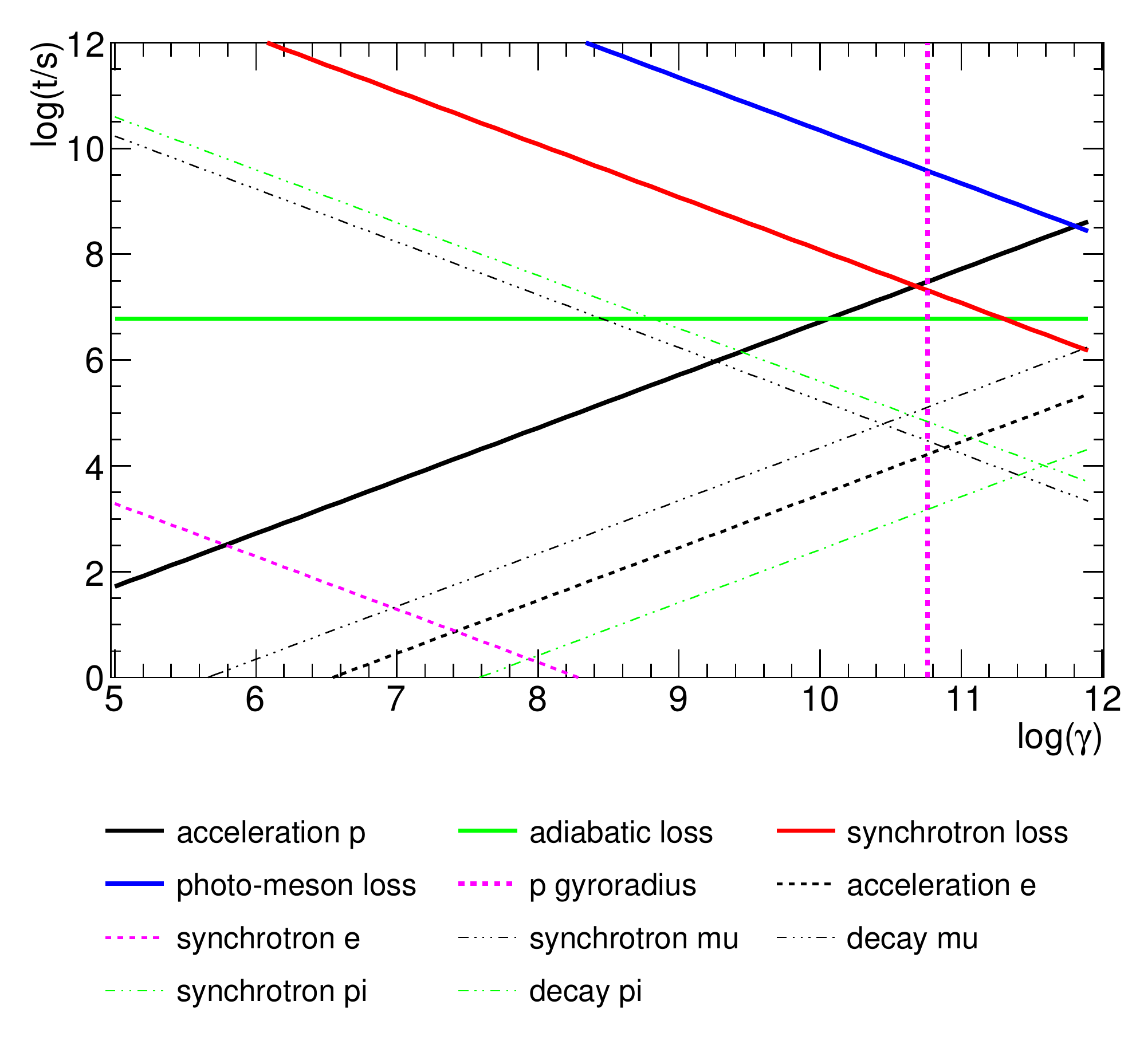}
      \caption{Relevant timescales for PKS\,2155-304 for the model shown in Fig.~\ref{fig:sed_2155} in the left panel. }
      \label{fig:times_2155_1}
  \end{figure}

 \begin{figure}[h]
   \centering
               \includegraphics[width=\columnwidth]{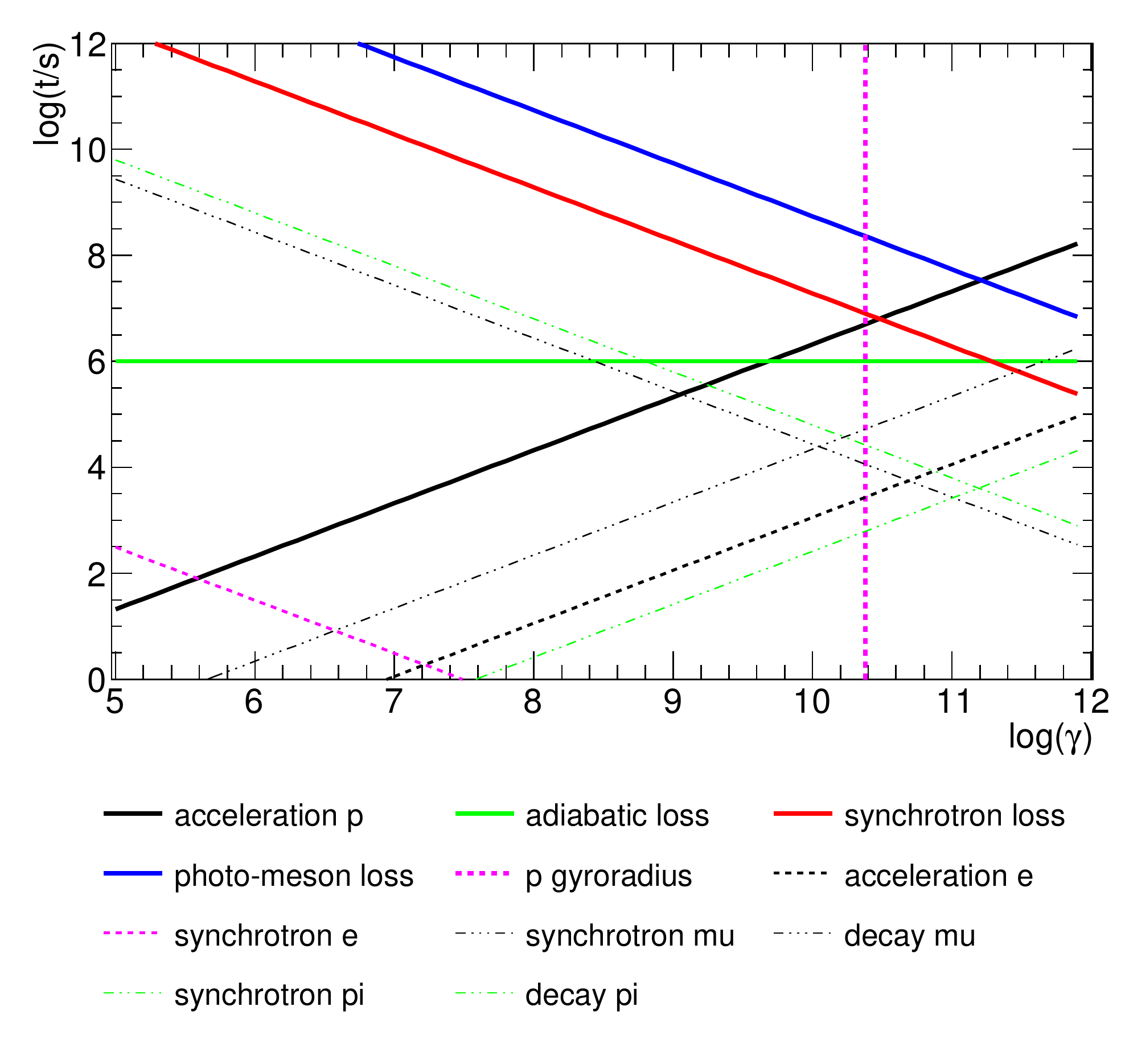}
      \caption{Relevant timescales for PKS\,2155-304 for the model shown in Fig.~\ref{fig:sed_2155} in the right panel.}
      \label{fig:times_2155_2}
  \end{figure}

\section{Examples of relevant timescales for models for Mrk\,421}
\label{app:421times}
The following figures show the relevant acceleration and cooling timescales for the different particle populations considered in the code as a function of the 
Lorentz factor of the particles. The vertical dashed line labelled ``p gyroradius'' indicates the proton Lorentz factor $\gamma_{p,max}$ that corresponds to a 
gyro-radius of the size of the radius of the emission region. The results shown here correspond to the exemplary models for Mrk\,421 shown in Fig.~\ref{fig:sed_421}.\newline

  \begin{figure}[h]
   \centering
               \includegraphics[width=\columnwidth]{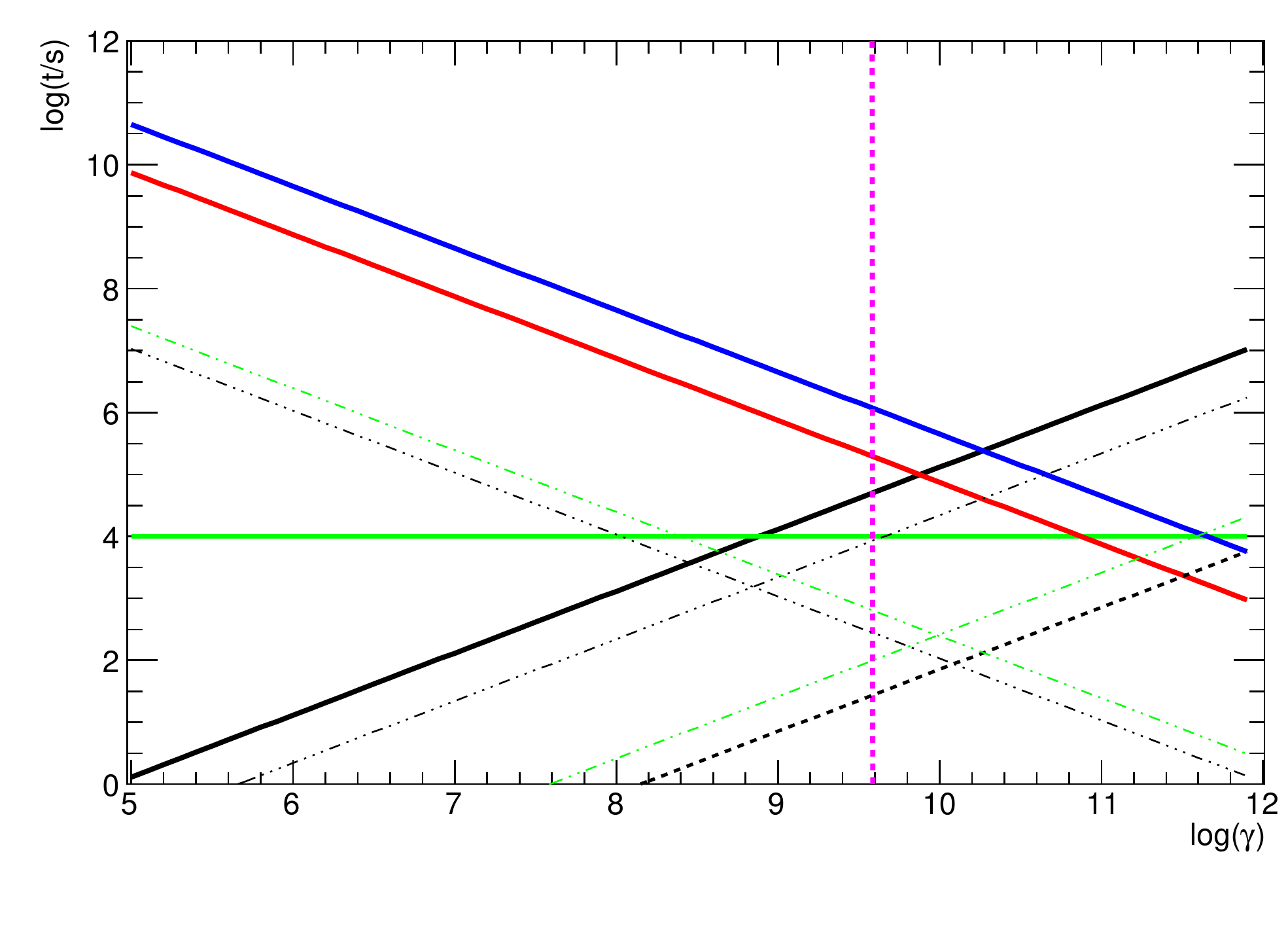}
      \caption{Relevant timescales for Mrk\,421 for the model shown in Fig.~\ref{fig:sed_421} in the left panel. 
       For a description of the different lines, see Sect.~\ref{app:2155times} above.}
      \label{fig:times_421_1}
  \end{figure}

 \begin{figure}[h]
   \centering
               \includegraphics[width=\columnwidth]{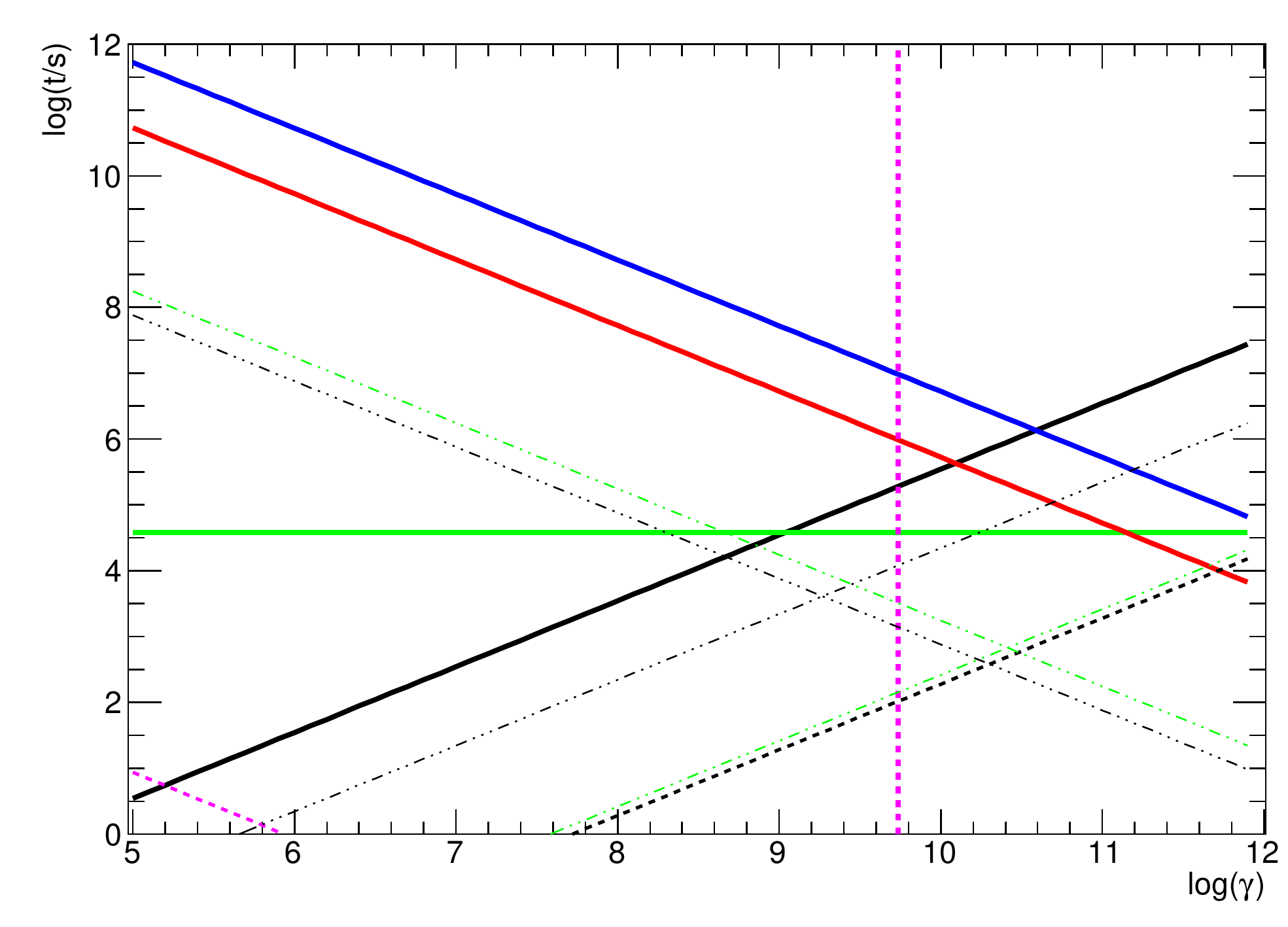}
      \caption{Relevant timescales for Mrk\,421 for the model shown in Fig.~\ref{fig:sed_421} in the right panel. 
      For a description of the different lines, see Sect.~\ref{app:2155times} above.}
      \label{fig:times_421_2}
  \end{figure}

\newpage

\section{Examples of models for PKS\,2155-304}
\label{app:2155}
The following figures show all the hadronic models for PKS\,2155-304 for an intermediate index of the proton spectrum $n_1 = 2.0$ that passed our selection.
The models for $n_1 = 1.9$ and $n_1 = 2.1$ are not included here for sake of brevity. Figure~\ref{fig:ps_diagonal} contains the interpretation of the different curves.

  \begin{figure}[h]
   \centering
               \includegraphics[width=\columnwidth]{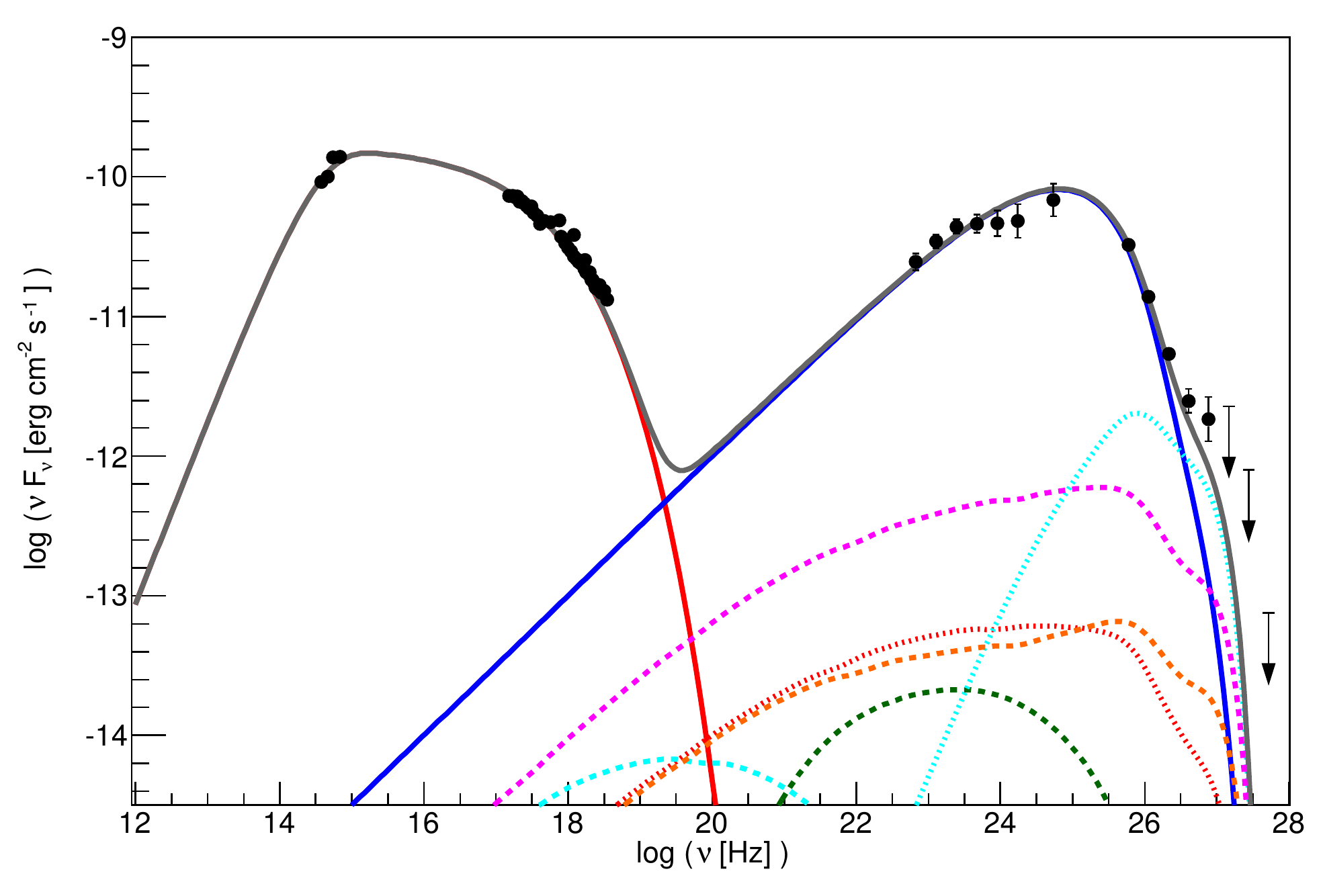}
      \caption{SED for PKS\,2155-304 with a hadronic model with
       a magnetic field of $\log{B [G]}=0.3$ and an emission region of size $\log{R [cm]}=16.9$. }
  \end{figure}

  \begin{figure}[h]
   \centering
               \includegraphics[width=\columnwidth]{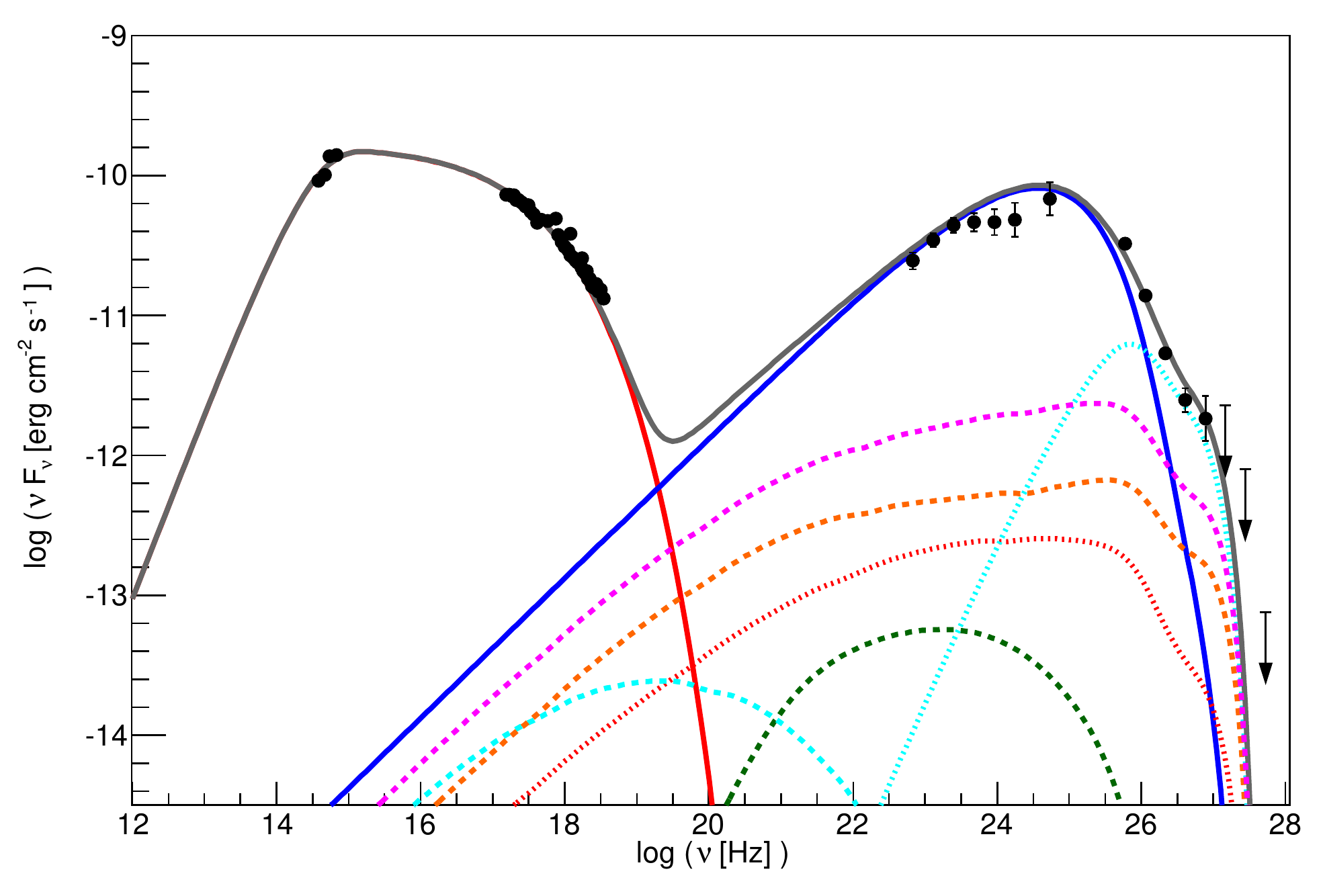}
      \caption{SED for PKS\,2155-304 with a hadronic model with
       a magnetic field of $\log{B [G]}=0.5$ and an emission region of size $\log{R [cm]}=16.5$. }
  \end{figure}

  \begin{figure}[h]
   \centering
               \includegraphics[width=\columnwidth]{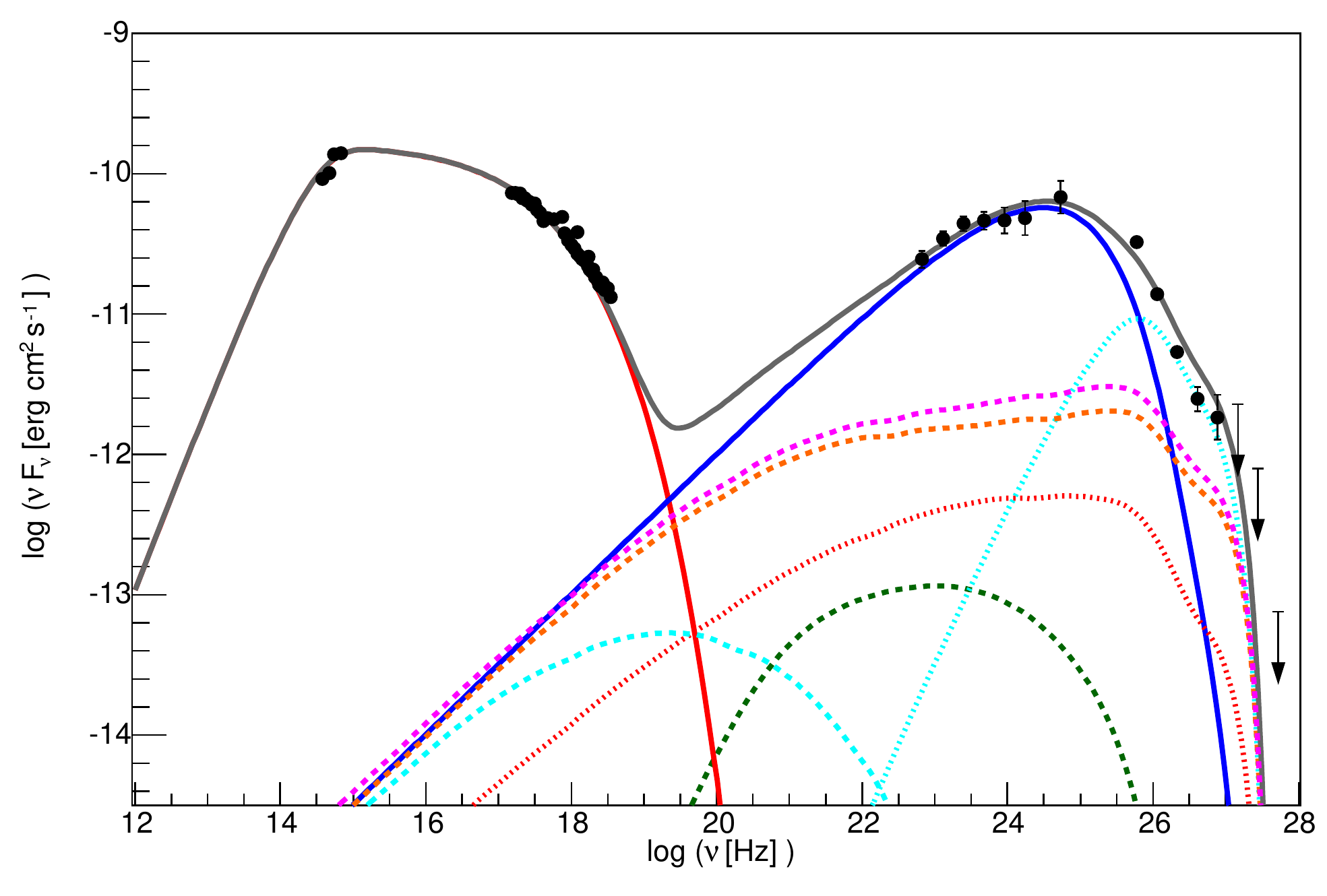}
      \caption{SED for PKS\,2155-304 with a hadronic model with
       a magnetic field of $\log{B [G]}=0.7$ and an emission region of size $\log{R [cm]}=16.2$. }
  \end{figure}

  \begin{figure}[h]
   \centering
               \includegraphics[width=\columnwidth]{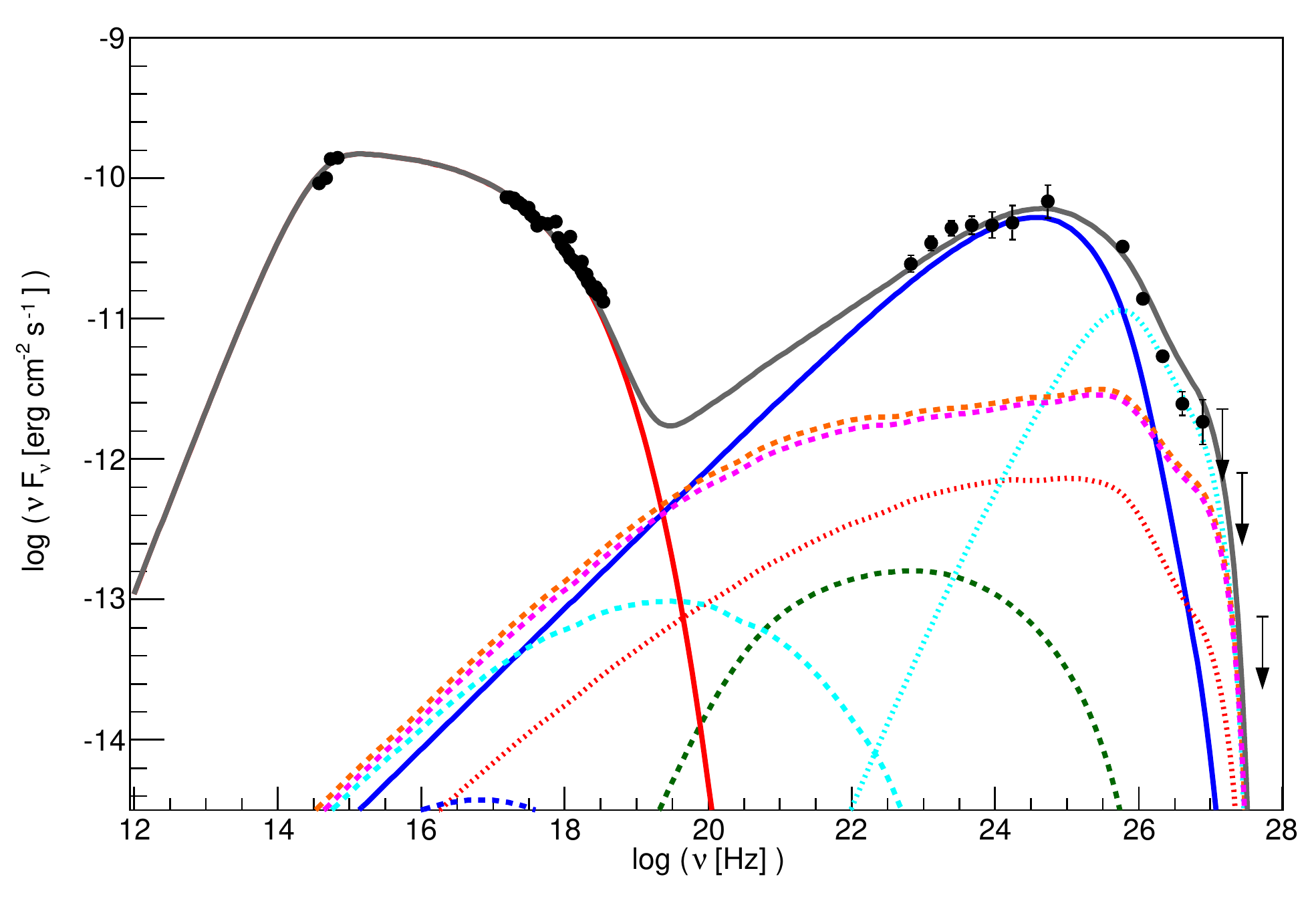}
      \caption{SED for PKS\,2155-304 with a hadronic model with
       a magnetic field of $\log{B [G]}=0.9$ and an emission region of size $\log{R [cm]}=15.9$. }
  \end{figure}

  \begin{figure}[h]
   \centering
               \includegraphics[width=\columnwidth]{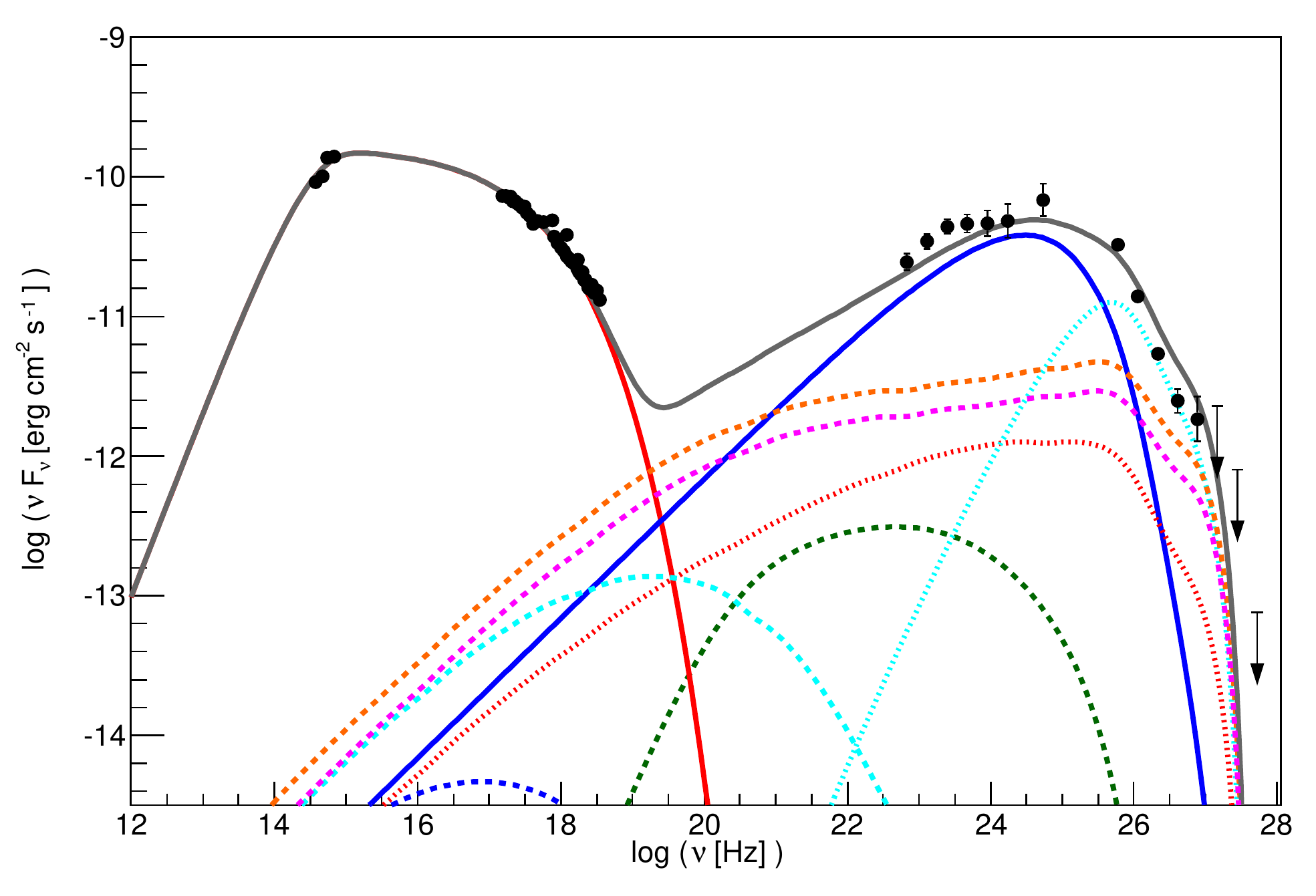}
      \caption{SED for PKS\,2155-304 with a hadronic model with
       a magnetic field of $\log{B [G]}=1.1$ and an emission region of size $\log{R [cm]}=15.5$. }
  \end{figure}

  \begin{figure}[h]
   \centering
               \includegraphics[width=\columnwidth]{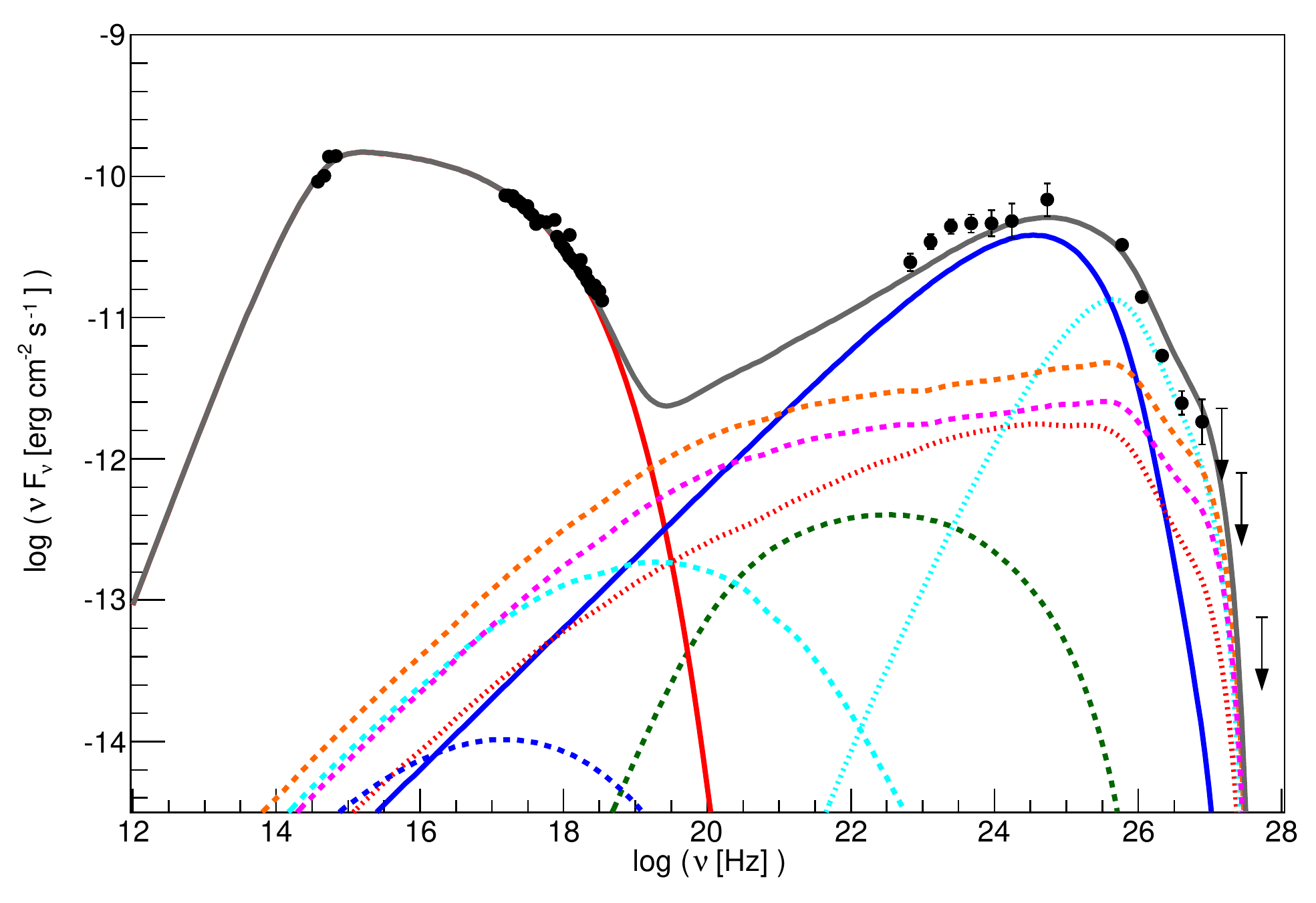}
      \caption{SED for PKS\,2155-304 with a hadronic model with
       a magnetic field of $\log{B [G]}=1.3$ and an emission region of size $\log{R [cm]}=15.3$. }
  \end{figure}

  \begin{figure}[h]
   \centering
               \includegraphics[width=\columnwidth]{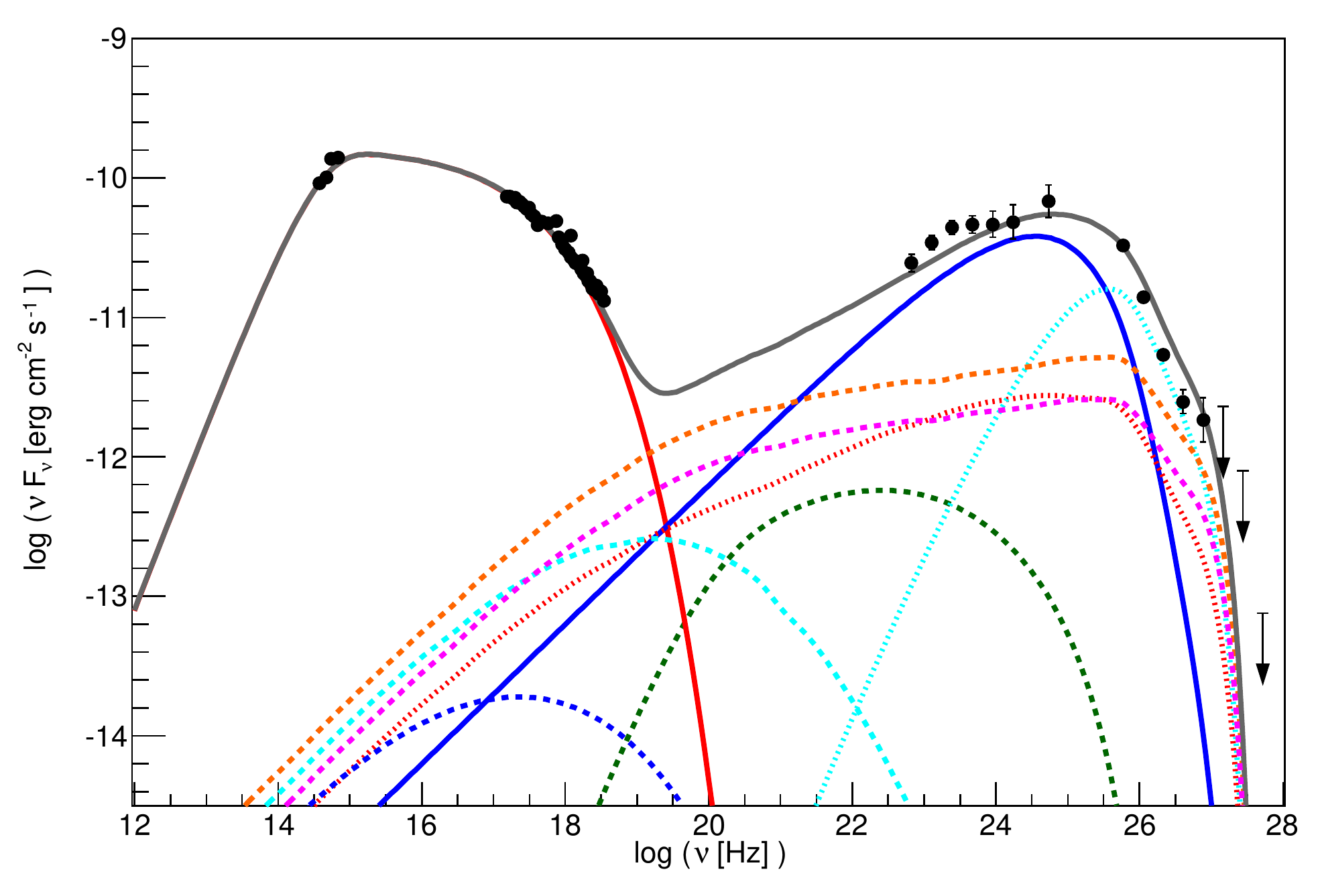}
      \caption{SED for PKS\,2155-304 with a hadronic model with
       a magnetic field of $\log{B [G]}=1.5$ and an emission region of size $\log{R [cm]}=15.0$. }
  \end{figure}

  \begin{figure}[h]
   \centering
               \includegraphics[width=\columnwidth]{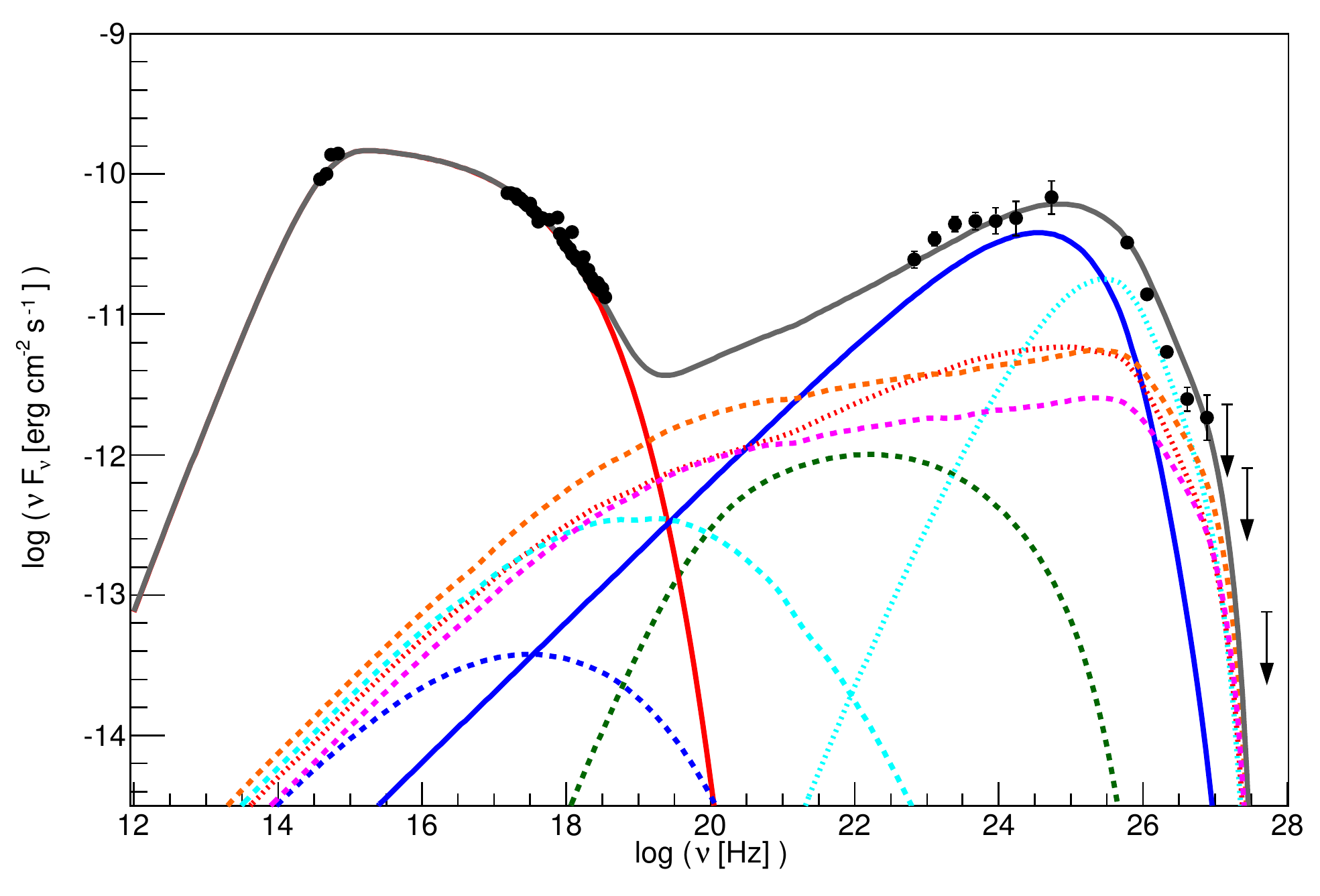}
      \caption{SED for PKS\,2155-304 with a hadronic model with
       a magnetic field of $\log{B [G]}=1.7$ and an emission region of size $\log{R [cm]}=14.7$. }
  \end{figure}

  \begin{figure}[h]
   \centering
               \includegraphics[width=\columnwidth]{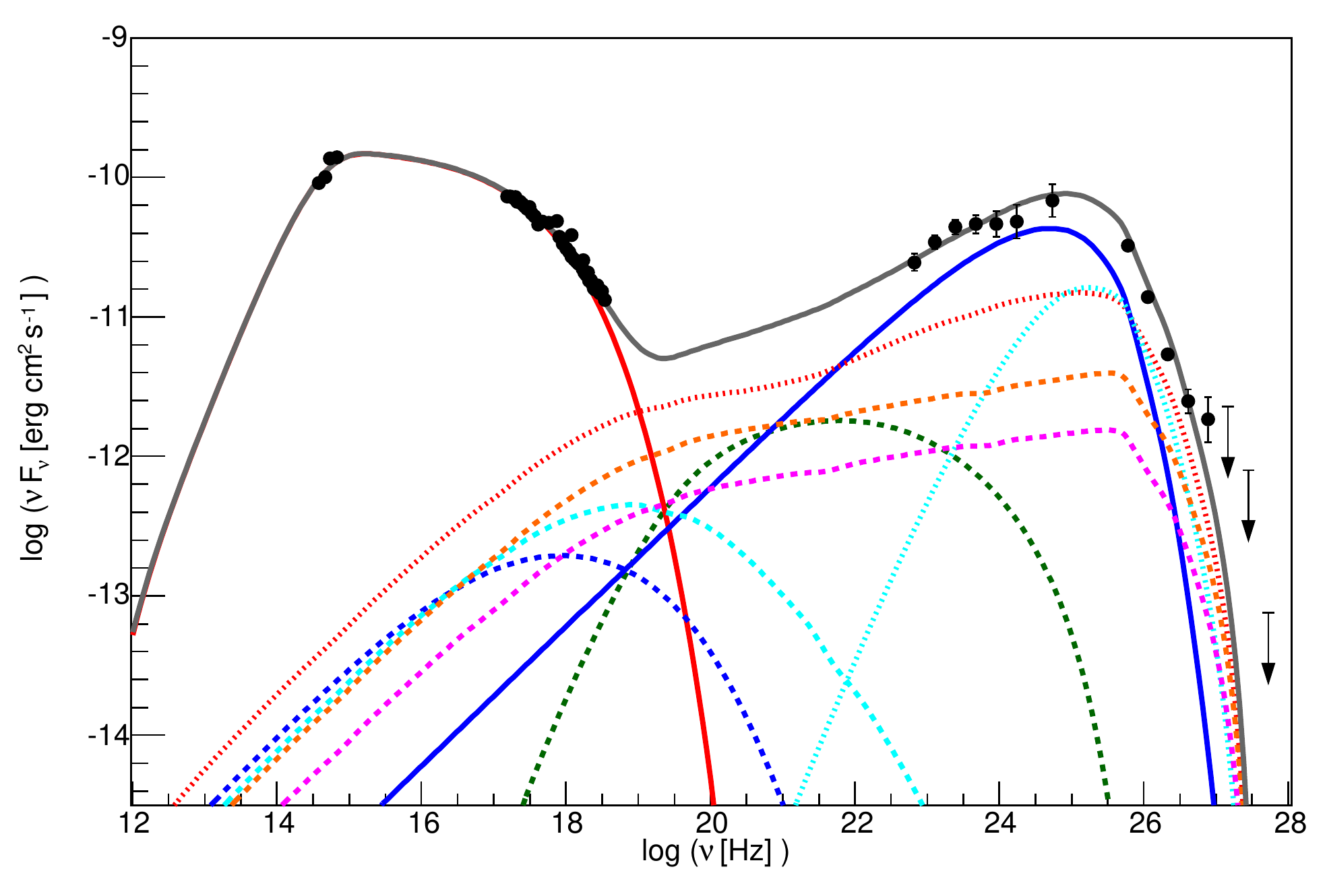}
      \caption{SED for PKS\,2155-304 with a hadronic model with
       a magnetic field of $\log{B [G]}=2.1$ and an emission region of size $\log{R [cm]}=14.2$. }
  \end{figure}

\clearpage
\section{Selected models for Mrk\,421}
\label{app:421}
The following figures show all the hadronic models for Mrk\,421 for an intermediate index of the proton spectrum $n_1 = 1.8$ that passed our selection.
The models for $n_1 = 1.7$ and $n_1 = 1.9$ are not included here for sake of brevity. Figure~\ref{fig:ps_diagonal} contains the interpretation of the different curves.

 \begin{figure}[h]
   \centering
              \includegraphics[width=\columnwidth]{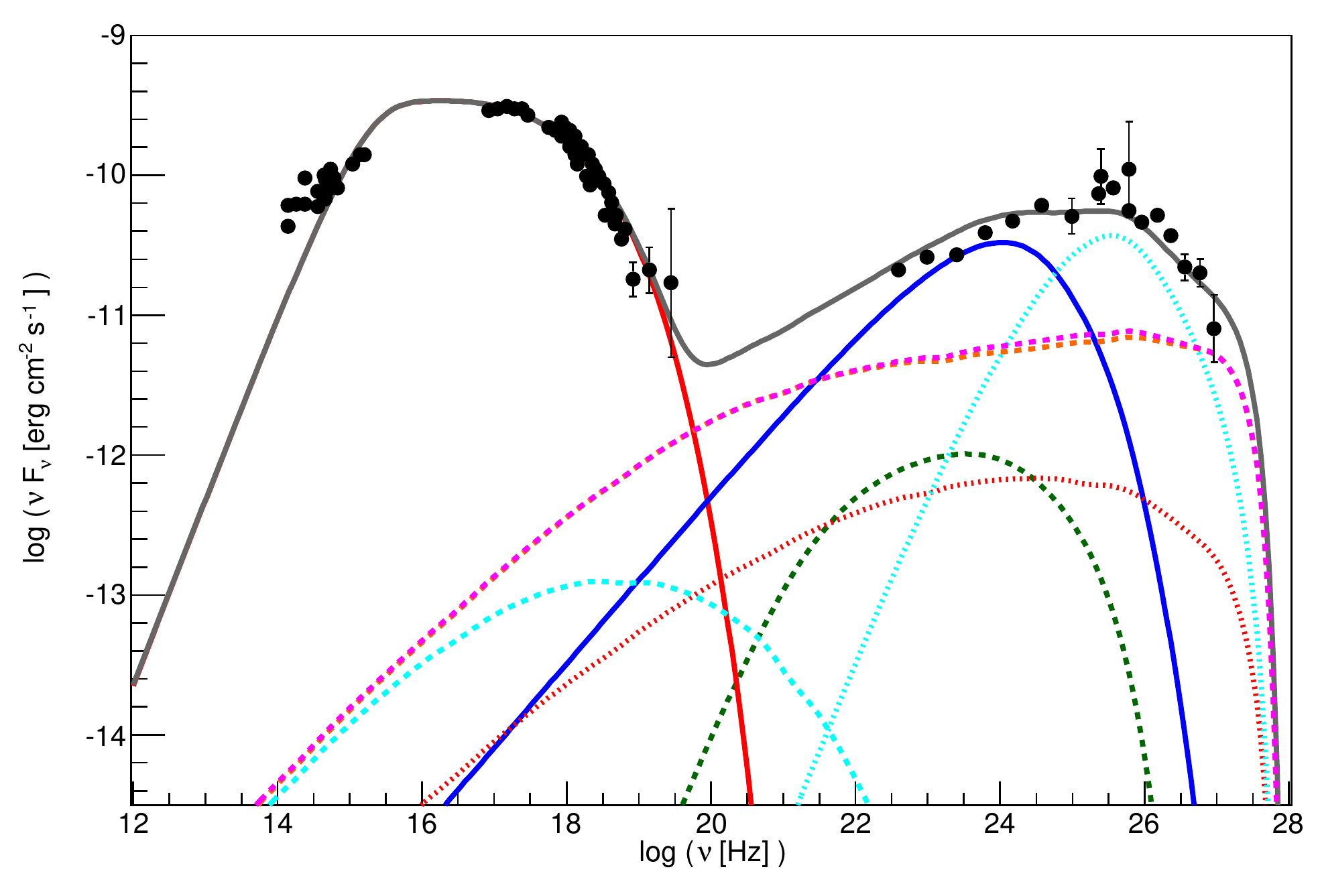}
      \caption{SED for Mrk\,421 with a hadronic model with
        a magnetic field of $\log{B [G]}=1.5$ and an emission region of size $\log{R [cm]}=14.7$.}
  \end{figure}

 \begin{figure}[h]
   \centering
              \includegraphics[width=\columnwidth]{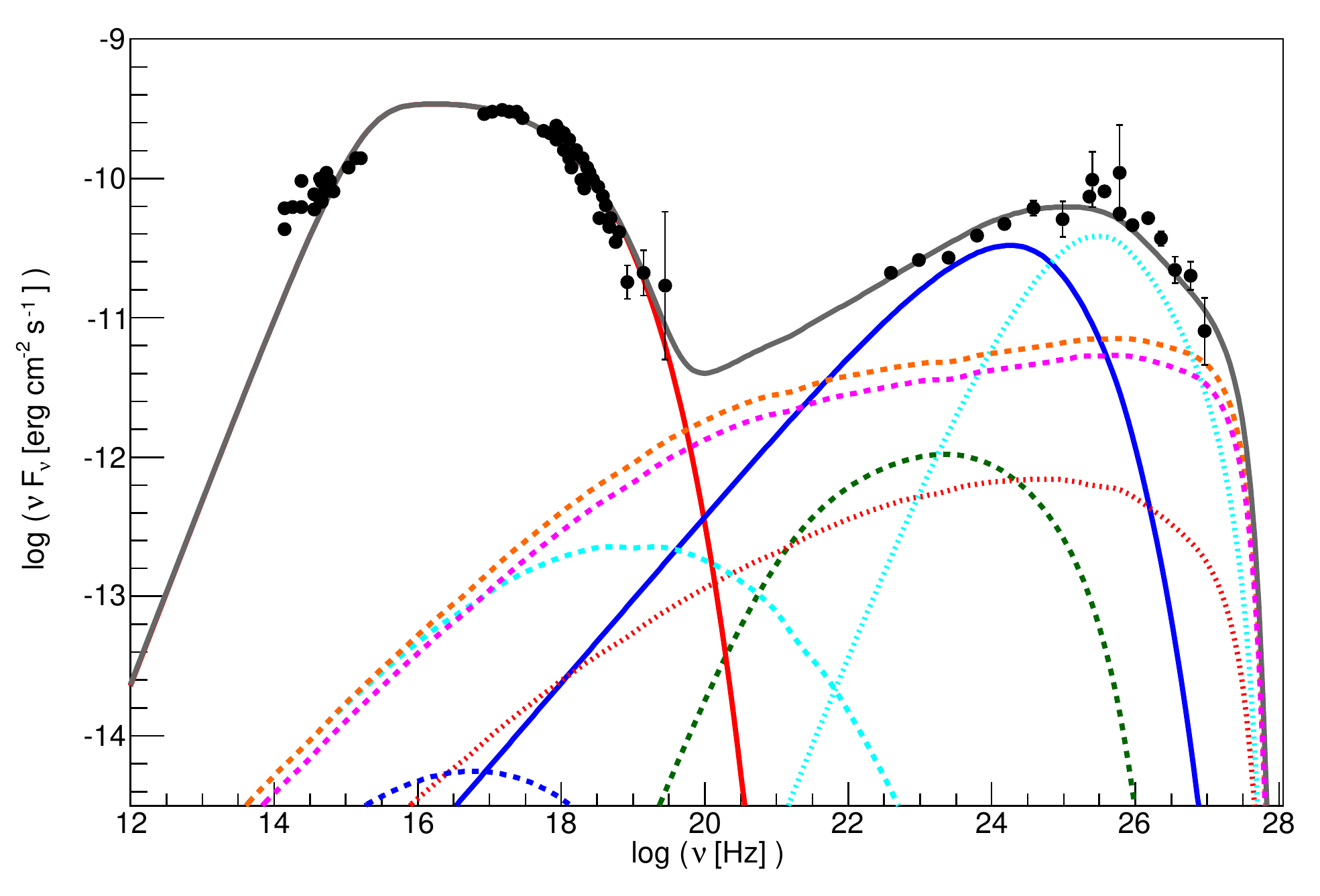}
      \caption{SED for Mrk\,421 with a hadronic model with
        a magnetic field of $\log{B [G]}=1.7$ and an emission region of size $\log{R [cm]}=14.5$.}
  \end{figure}

 \begin{figure}[h]
   \centering
              \includegraphics[width=\columnwidth]{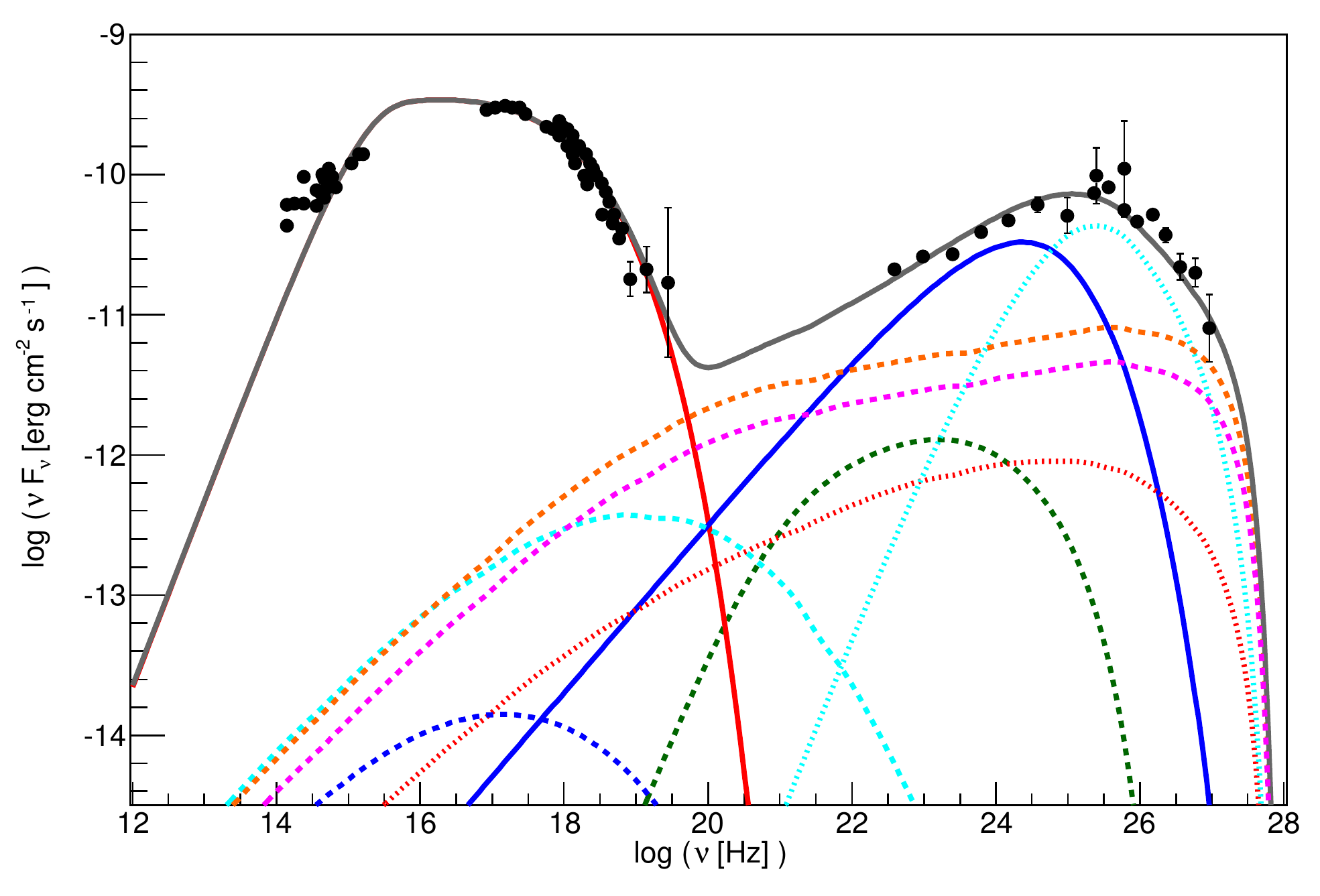}
      \caption{SED for Mrk\,421 with a hadronic model with
        a magnetic field of $\log{B [G]}=1.9$ and an emission region of size $\log{R [cm]}=14.2$.}
  \end{figure}

 \begin{figure}[h]
   \centering
              \includegraphics[width=\columnwidth]{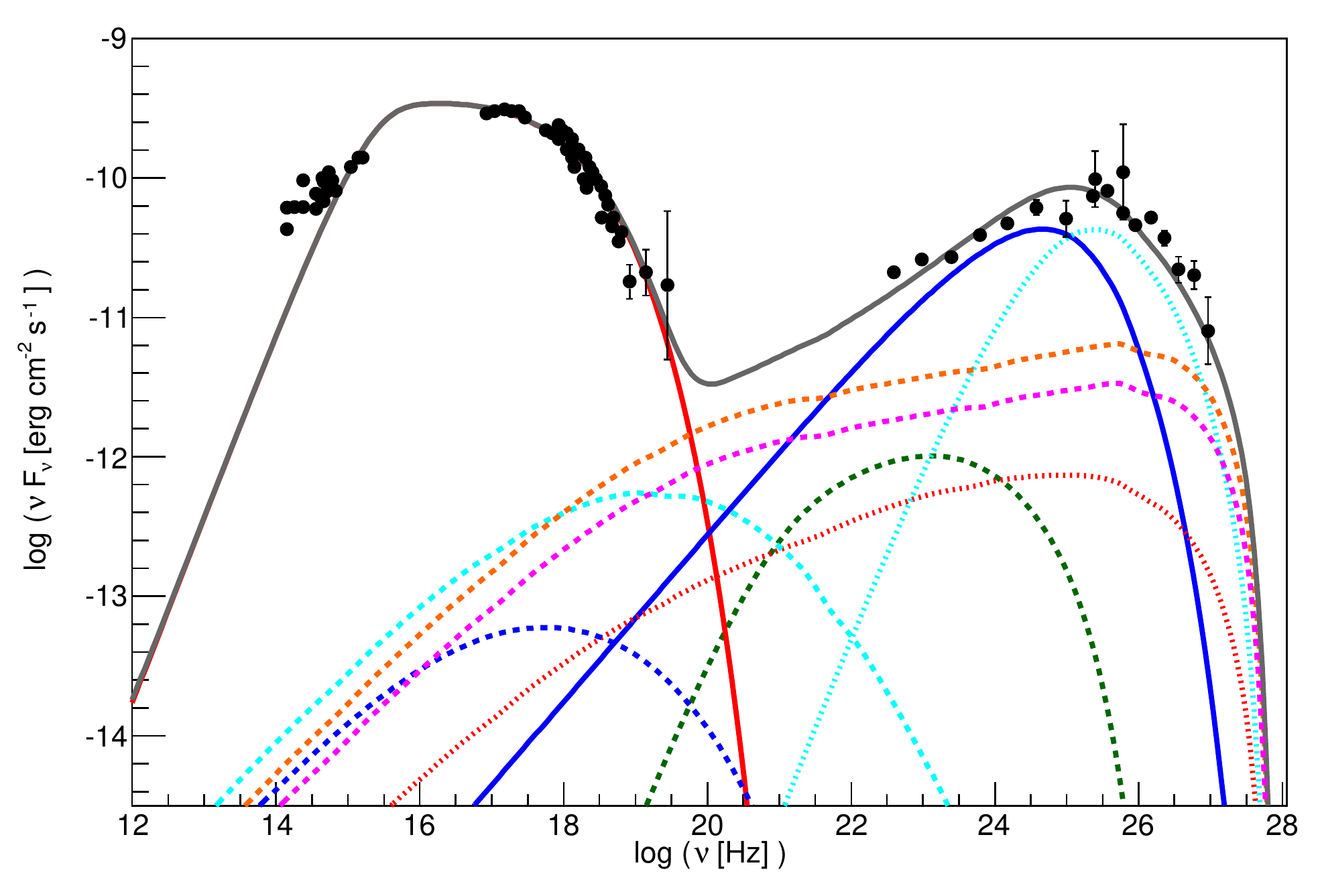}
      \caption{SED for Mrk\,421 with a hadronic model with
        a magnetic field of $\log{B [G]}=2.1$ and an emission region of size $\log{R [cm]}=14.1$.}
  \end{figure}

\end{document}